\documentclass[prx,aps,superscriptaddress,twocolumn]{revtex4}
\usepackage{amsmath}
\usepackage{amsthm}
\usepackage{amsfonts}
\usepackage{listings}
\usepackage{graphicx,wasysym}
\usepackage{verbatim}
\usepackage{pdflscape,amssymb}
\usepackage{float,braket}
\usepackage{booktabs}
\usepackage{mathrsfs}
\usepackage{diagbox}
\usepackage[colorlinks,
linkcolor=blue,
anchorcolor=blue,
citecolor=blue
]{hyperref}

\usepackage{physics}

\usepackage{enumerate}
\usepackage{latexsym}
\usepackage{color}
\usepackage{bm}
\usepackage{hyperref}
\hypersetup{
	pdfnewwindow=true, colorlinks=true,
	linkcolor=blue, anchorcolor=blue,
	citecolor=blue, filecolor=blue,
	menucolor=blue, urlcolor=blue}

\usepackage{psfrag}

\usepackage{subfigure}

\RequirePackage[normalem]{ulem} %DIF PREAMBLE
\RequirePackage{color}\definecolor{RED}{rgb}{1,0,0}\definecolor{BLUE}{rgb}{0,0,1} %DIF PREAMBLE
\newcommand{\webirvsp}{\href{https://github.com/zjwang11/irvsp}{\texttt{IRVSP}} }

\newcommand{\I}{\mathrm{\uppercase\expandafter{\romannumeral1}}}
\newcommand{\II}{\mathrm{\uppercase\expandafter{\romannumeral2}}}
\newcommand{\III}{\mathrm{\uppercase\expandafter{\romannumeral3}}}
\newcommand{\IV}{\mathrm{\uppercase\expandafter{\romannumeral4}}}

\def\ie{{\it i.e.},\ }
\def\eg{{\it e.g.}\ }
\def\ea{{\it et al.}}

\begin{document}
	
	\title{{\ttfamily VASP2KP}: $ k\cdot p $ models and Land\'e $g$-factors from \textit{ab initio} calculations}

	\author{Sheng Zhang}
	\thanks{These authors contributed equally to this work.}
	\affiliation{Beijing National Laboratory for Condensed Matter Physics,
		and Institute of Physics, Chinese Academy of Sciences, Beijing 100190, China}
	\affiliation{University of Chinese Academy of Sciences, Beijing 100049, China}

	\author{Haohao Sheng}
	\thanks{These authors contributed equally to this work.}
	\affiliation{Beijing National Laboratory for Condensed Matter Physics,
		and Institute of Physics, Chinese Academy of Sciences, Beijing 100190, China}
	\affiliation{University of Chinese Academy of Sciences, Beijing 100049, China}

	\author{Zhi-Da Song}
	\email{songzd@pku.edu.cn}
	\affiliation{International Center for Quantum Materials, School of Physics, Peking University, Beijing 100871, China}
    \affiliation{Hefei National Laboratory, Hefei 230088, China}
    \affiliation{Collaborative Innovation Center of Quantum Matter, Beijing 100871, China}

        \author{Chenhao Liang}
	\affiliation{Beijing National Laboratory for Condensed Matter Physics,
		and Institute of Physics, Chinese Academy of Sciences, Beijing 100190, China}
	\affiliation{University of Chinese Academy of Sciences, Beijing 100049, China}

	\author{Yi Jiang}
	\affiliation{Beijing National Laboratory for Condensed Matter Physics,
		and Institute of Physics, Chinese Academy of Sciences, Beijing 100190, China}
	\affiliation{University of Chinese Academy of Sciences, Beijing 100049, China}

	\author{Song Sun}
	\affiliation{Beijing National Laboratory for Condensed Matter Physics,
		and Institute of Physics, Chinese Academy of Sciences, Beijing 100190, China}
	\affiliation{University of Chinese Academy of Sciences, Beijing 100049, China}

        \author{Quansheng Wu}
	\affiliation{Beijing National Laboratory for Condensed Matter Physics,
		and Institute of Physics, Chinese Academy of Sciences, Beijing 100190, China}
	\affiliation{University of Chinese Academy of Sciences, Beijing 100049, China}
 
	\author{Hongming Weng}
	\affiliation{Beijing National Laboratory for Condensed Matter Physics,
		and Institute of Physics, Chinese Academy of Sciences, Beijing 100190, China}
	\affiliation{University of Chinese Academy of Sciences, Beijing 100049, China}
	
	\author{Zhong Fang}
	\affiliation{Beijing National Laboratory for Condensed Matter Physics,
		and Institute of Physics, Chinese Academy of Sciences, Beijing 100190, China}
	\affiliation{University of Chinese Academy of Sciences, Beijing 100049, China}

	\author{Xi Dai}
	\affiliation{Department of Physics, Hong Kong University of Science and Technology, Hong Kong 999077, China}

	\author{Zhijun Wang}
	\email{wzj@iphy.ac.cn}
	\affiliation{Beijing National Laboratory for Condensed Matter Physics,
		and Institute of Physics, Chinese Academy of Sciences, Beijing 100190, China}
	\affiliation{University of Chinese Academy of Sciences, Beijing 100049, China}
	
	\begin{abstract}
        The $k\cdot p$  method is significant in condensed matter physics for the compact and analytical Hamiltonian. 
        %With obtained irreducible representations (irreps) of the $\vb*k$-little group  [Gao \textit{et al.}, \href{https://journals.aps.org/prb/abstract/10.1103/PhysRevB.106.035150}{Comput. Phys. Commun., \textbf{261}, 107760 (2021)}], the low-energy $k\cdot p$ effective Hamiltonian can be automatically generated under the standard basis of the irreps [Jiang \textit{et al.}, \href{https://iopscience.iop.org/article/10.1088/0256-307X/38/7/077104/meta}{Chin. Phys. Lett., \textbf{39}, 077104 (2021)}], but the parameters are usually obtained by fitting experimental or theoretical band dispersions. In the presence of magnetic field, it is described by effective Zeeman's coupling Hamiltonian with Land\'e $ g $-factors~[Song \textit{et al.}, \href{https://doi.org/10.1142/9789811231711_0013}{Memorial Volume for Shoucheng Zhang, 263-281  (2021)}].
        In the presence of magnetic field, it is described by the effective Zeeman's coupling Hamiltonian with Land\'e $ g $-factors.
		Here, we develop an \emph{open-source} package {\ttfamily VASP2KP} (including two parts: {\ttfamily vasp2mat} and {\ttfamily mat2kp}) to compute $k\cdot p$ parameters and Land\'e $g$-factors directly from the wavefunctions provided by the density functional theory (DFT) as implemented in Vienna \textit{ab~initio} Simulation Package (VASP).
		First, we develop a VASP patch {\ttfamily vasp2mat} to compute matrix representations of the generalized momentum operator $ \hat{\vb* \pi}=\hat{\vb* p}+\frac{1}{2mc^2}\left(\hat{{\vb* s}}\times\nabla V({\vb* r})\right) $, spin operator $\hat{\vb*s}$, time reversal operator $\hat{T}$ and crystalline symmetry operators $\hat{R}$ on the DFT wavefunctions.
		Second, we develop a python code {\ttfamily mat2kp} to obtain the unitary transformation $U$ that rotates the degenerate DFT basis towards the standard basis, and then automatically compute the $k\cdot p$ parameters and $g$-factors.
		The theory and the methodology behind {\ttfamily VASP2KP} are described in detail. The matrix elements of the operators are derived comprehensively and computed correctly within the projector augmented wave method. We apply this package to some materials, \eg Bi$_2$Se$_3$, Na$_3$Bi, Te, InAs and 1H-TMD monolayers. The obtained effective model's dispersions are in good agreement with the DFT data around the specific wave vector, and the $g$-factors are consistent with experimental data. The {\ttfamily VASP2KP} package is available at \href{https://github.com/zjwang11/VASP2KP}{https://github.com/zjwang11/VASP2KP}.
	\end{abstract}

	%By inputing the standard corepresentation, {\ttfamily VASP2KP} can construct the k⋅p k\cdot p  Hamiltonian of the certain order and calculate the undetermined parameters based on the VASP calculation. 
	%Similarly, {\ttfamily VASP2KP} can also construct the effective Zeeman's coupling with calculated Land\'e $ g  -factors. 

	\maketitle
	\section{Introduction}
	Electronic band structures hold immense importance in the field of condensed matter physics and materials science, providing crucial insights into the behavior of electrons within crystalline materials. They reveals information about distributions of energy levels, energy gaps, and density of states, which are essential for determining material electrical conductivities, optical properties, thermal behavior, and so on. In this background, the density functional theory (DFT)~\cite{PhysRev.136.B864,PhysRev.140.A1133} was developed, which gave birth to many first-principles calculation softwares or codes based on them, including VASP~\cite{VASP1,VASP2}, Quantum Espresso~\cite{Giannozzi_2009,Giannozzi_2017}, CASTEP~\cite{Clark}, ABINIT~\cite{Gonze2020,Romero2020}, and so on~\cite{Jose_M_Soler_2002,10.1063/5.0005077,10.1063/1.5143061,10.1063/1.5143190}. 
	
	However, DFT bands are quite complex, which contain quite a large number of energy bands that hardly affect the desired physical properties thus making the physical pictures difficult to understand. In order to construct a model that only involves a few bands with a greater impact on the physical properties of materials, the $ k\cdot p $ method is proposed, which is used to construct an effective model to describe the quasiparticles near the specific wave vector in the reciprocal space~\cite{LUQUE201582}. The $ k\cdot p $ models constructed through the theory of invariants are analytical and only contain a few important bands, thus making the physical pictures quite clear. There are some arbitrary parameters in $ k\cdot p $ models, which could be determined by fitting to the corresponding experimental data or the DFT calculation data such as band structures. The $ k\cdot p $ method has been successfully applied to many condensed-matter systems, including metals~\cite{Gresch_2017,PhysRev.97.869}, semiconductors~\cite{doi:10.1021/acs.nanolett.5b00101,KANE1957249}, topological insulators and superconductors~\cite{HZhang,zhang2009topological,PhysRevLett.103.266801,PhysRevLett.107.186806}, spin-lasers~\cite{PhysRevB.92.075311,PhysRevB.83.125309}, nanostructured solids~\cite{MARQUARDT2021110318}, two-dimensional van der Waals materials~\cite{PhysRevB.100.115203,Kormanyos_2015} and so on~\cite{PhysRevLett.124.226402,Junior,PhysRevResearch.2.033256,10.1063/1.4945112}. 
	
	When a magnetic field is applied to a condensed matter system, we can use effective Zeeman's coupling Hamiltonian to describe the effects of the magnetic field. Effective Zeeman's coupling determines the split of Kramers states under magnetic field, leading to outcomes like the Pauli paramagnetism of the metals and the van Vleck paramagnetism of insulators. The important parameters in Zeeman's coupling are Land\'e $ g $-factors. Land\'e $ g $-factors of materials have been widely studied in 
	quantum wires~\cite{PhysRevB.76.045324,ZAMANI2017243,nano13091461,ZAMANI2018145}, quantum dots~\cite{PhysRevLett.96.026804,GHARAATI201717}, semiconductors nanostructures~\cite{PhysRevB.63.085310,PhysRevB.58.16353,PhysRevLett.119.037701,PhysRevB.86.195302,PhysRevLett.97.236402}, topological materials~\cite{C5NR05250E,PhysRevB.104.085302} and so on~\cite{10.1063/1.1423328,SEMENOV201657,FISCHER200155}. Moreover, the theory of invariants can also be used to construct Zeeman's coupling Hamiltonian conveniently. However, there is no available code to compute Land\'e $ g $-factors and $k\cdot p$ parameters directly from DFT wavefunctions.
	
    First of all, Gao \ea ~developed the program \webirvsp to determine the irreducible representations (irreps) of bands at any $\vb*k$-point in VASP calculations~\cite{GAO2021107760}. Then, Jiang \ea \  developed the python package {\ttfamily kdotp-generator} to generate the $k\cdot p$ effective Hamiltonian for the given irreps automatically~\cite{Jiang_2021}. Furthermore, Song \ea \ derived effective Zeeman's coupling and $g$-factors in DFT calculations~\cite{song-gfactor}. Thus, it is straightforward for us to generate the {\ttfamily VASP2KP} code to construct the effective models and to compute the model parameters from VASP wavefunctions directly. Although many similar codes are developed subsequently for Quantum Espresso, such as IR2PW~\cite{RHZhang}, IrRep~\cite{IeRep}, and DFT2kp~\cite{cassiano2023dft2kp}, their functions do not go beyond the above mentioned codes. The DFT2kp can not generate the Zeeman's coupling Hamiltonians. Additionally, the matrix elements of symmetry operators are not obtained properly, because the projector augmented wave (PAW) corrections are neglected in DFT2kp.

	In this work, we have developed an \emph{open-source} package {\ttfamily VASP2KP}, which can generate the effective $ k\cdot p $ model and Zeeman's coupling, and obtain the values of the $ k\cdot p $ parameters and $g$-factors.
	This package contains two parts: a VASP patch {\ttfamily vasp2mat} and a post-processing python code {\ttfamily mat2kp}.
	First, we use {\ttfamily vasp2mat} to generate matrix representations for generalized momentum $\hat{\vb*\pi}=\hat{\vb*p}+\frac{1}{2mc^2}(\hat{\vb*s}\times\nabla V(\vb* r))$, spin $\hat{\vb*s}$, time reversal $ \hat{T} $, and crystalline symmetry operators $\hat{R}$ in DFT calculations. The matrix elements of the space group operators are derived in detail and computed correctly in the PAW wavefunctions.
	Second, {\ttfamily mat2kp} can obtain the unitary transformation $ U $ from the degenerate wavefunctions to the $k\cdot p$ standard basis, and then compute these parameters automatically. The obtained effective masses and $g$-factors are important and comparable with experimental observations.

	The paper is organized as the following. In Sec. \ref{th}, the theoretical foundations behind {\ttfamily VASP2KP} are introduced. In Sec. \ref{capability-sec} the main algorithm steps are described in detail. The results obtained by {\ttfamily VASP2KP} are shown and analysed for some typical materials in Sec. \ref{exa}, and we have some discussions in Sec. \ref{dis}.
	
	\section{Theory and methodology}\label{th}
	In this section, we first introduce the $ k\cdot p $ method in Sec. \ref{th.A}. The method to obtain Zeeman's coupling is presented in Sec. \ref{th.Zeeman}. Then the theory of invariants is reviewed and the invariant $ k\cdot p $ Hamiltonian as well as Zeeman's coupling are obtained in Sec. \ref{th.B}. In Sec. \ref{th.C}, we propose a general routine
    to get the unitary transformation that change degenerate DFT basis to the $k\cdot p$ standard basis. Last, the method to calculate $k\cdot p$ parameters and $g$-factors is introduced in Sec. \ref{th.solve}.
	
	\subsection{\texorpdfstring{$ k\cdot p $}{} effective Hamiltonian}\label{th.A}
	When we only care about wave vector $ {\vb* K} $ around a specific wave vector $ {\vb* k}_0 $ in the Brillouin zone, it can be well described by using the $ k\cdot p $ effective model. Suppose that $ \psi_{n{\vb* K}}({\vb* r}) $ is the wavefunction of $ n $-th band which satisfies the Schrödinger equation with spin-orbit coupling (SOC), \ie $ \hat{H}_B\psi_{n{\vb* K}}({\vb* r})=\epsilon_{n}({\vb* K})\psi_{n{\vb* K}}({\vb* r}) $, where
	\begin{equation}
		\hat{H}_B=\frac{\hat{p}^2}{2m}+V({\vb* r})+\frac{1}{2m^2c^2}\left(\hat{{\vb* s}}\times\nabla V({\vb* r})\right)\cdot\hat{{\vb* p}}
	\end{equation}
	is the Bloch Hamiltonian operator with SOC, and $ \hat{\vb* p} $ is the momentum operator, $ V({\vb* r}) $ is the potential in crystal, $ \hat{{\vb* s}} $ is the spin momentum operator, $ m $ is the electron mass, and $ c $ is the light speed in vacuum. To expand the Hamiltonian at $ \vb*k_0 $, introducing a transformation $ \phi_{n{\vb* k}}({\vb* r})=\psi_{n{\vb* K}}({\vb* r})e^{-i\vb* k\cdot\vb* r} $, where $ {\vb* k}={\vb* K}-{\vb* k}_0 $ is the deviation from $ {\vb* k}_0 $, the Schrödinger equation which $ \phi_{\vb* k}({\vb* r}) $ obeys is 
	\begin{equation}
		\left(\hat{H}_B+\frac{\hbar^2k^2}{2m}+\hat{H}'({\vb* k})\right)\phi_{n{\vb* k}}({\vb* r})=\epsilon_n({\vb* K})\phi_{n{\vb* k}}({\vb* r})
		\label{eq2}
	\end{equation}
	where $ \hat{H}'({\vb* k})=\frac{\hbar}{m}{\vb* k}\cdot{\vb* \pi} $ is the first-order term, and $ \hat{\vb* \pi}=\hat{\vb* p}+\frac{1}{2mc^2}\left(\hat{{\vb* s}}\times\nabla V({\vb* r})\right) $ is the generalized momentum operator with SOC. We can take $ \hat{H}^{kp}=\hat{H}_B+\frac{\hbar^2k^2}{2m}+\hat{H}'({\vb* k}) $ as the equivalent Hamiltonian of $ \hat{H}_B $ since the eigenvalues of them are all equal.
	
	Suppose that we have got the eigenenergies $\{ \epsilon_n({\vb* k}_0) \}$ and the eigenstates $ \{\ket{n({\vb* k}_0)}\} $ of $ \hat{H}_B $. Then we can obtain the generalized momentum elements by $ {\vb* \pi}_{mn}=\mel{m({\vb* k}_0)}{\hat{\vb* \pi}}{n({\vb* k}_0)} $. Thus the matrix elements of $ \hat{H}^{kp} $ can be obtained by
	\begin{equation}
		H^{\text{kp-all}}_{mn}({\vb*k})=\left(\epsilon_n({\vb*k}_0)+\frac{\hbar^2k^2}{2m}\right)\delta_{mn}+\frac{\hbar}{m}{\vb*\pi}_{mn}\cdot{\vb*k}
		\label{kp-initial}
	\end{equation}
    Usually, we aimed at several low-energy bands (the set of which is denoted as $\mathcal{A}$). The set of other bands is denoted as $ \mathcal{B} $. Then we can fold down the $H^{\text{kp-all}}_{mn}({\vb*k})$ Hamiltonian into a subspace $\mathcal{A}$ via Löwdin partitioning. After two-order Löwdin partitioning, the effective $ k\cdot p $  Hamiltonian is transformed as (see Appendix \ref{lowdin-app} for details)
	\begin{equation}
		\begin{aligned}
			H^{kp}_{\alpha\beta}({\vb* k})=&\left(\epsilon_\alpha({\vb* k}_0)+\frac{\hbar^2k^2}{2m}\right)\delta_{\alpha\beta}+\frac{\hbar}{m}{\vb* \pi}_{\alpha\beta}\cdot{\vb* k}\\
			&+\frac{\hbar^2}{2m^2}\sum_{l\in \mathcal{B}}\sum_{ij}\pi^{i}_{\alpha l}\pi^{j}_{l\beta}k^ik^j\\
			&\times \left(\frac{1}{\epsilon_\alpha({\vb* k}_0)-\epsilon_l({\vb* k}_0)}+\frac{1}{\epsilon_\beta({\vb* k}_0)-\epsilon_l({\vb* k}_0)}\right)
		\end{aligned}
		\label{eq-kp-origin}
	\end{equation}
    with $ i,j\in \{x,y,z\} $. Hereafter we use $\alpha, \beta\in\mathcal{A}$, $l\in\mathcal{B}$ and $m, n\in \mathcal{A}\cup \mathcal{B}$. Moreover, the $ k\cdot p $ Hamiltonian of the third order can be constructed similarly in Appendix \ref{lowdin-app}.
	
	\subsection{Zeeman's coupling}\label{th.Zeeman}
	When a magnetic field is applied to the system, the momenta $ \hbar k^i $ should be replaced by $ (-i\hbar\partial^i+eA^i) $ according to Peierls substitution, where $ {\vb* A} $ is the vector potential and $e$ (positively valued) is the elementary charge. Therefore, $ \hbar^2k^ik^j $ in the summation will be replaced by the sum of the gauge dependent term $ \frac{1}{2}\left\{-i\hbar\partial^i+eA^i,-i\hbar\partial^j+eA^j\right\} $ and gauge independent term $ \frac{1}{2}\left[-i\hbar\partial^i+eA^i,-i\hbar\partial^j+eA^j\right]=-\frac{i\hbar e}{2}\sum_k\epsilon^{ijk}B_k $, where $ {\vb* B} $ is the magnetic induction intensity. In this case, the total Hamiltonian can be written by the summation of the $ k\cdot p $ effective Hamiltonian $ H^{kp}_{\alpha\beta} $ and Zeeman's coupling $ H^{Z}_{\alpha\beta} $, which are gauge dependent and gauge invariant, respectively. The $ k\cdot p $ effective Hamiltonian $ H^{kp}_{\alpha\beta} $ can be expressed as
	\begin{equation}
		\begin{aligned}
		H^{kp}_{\alpha\beta}=&\epsilon_\alpha({\vb* k}_0)\delta_{\alpha\beta}+\frac{\hbar}{m}{\vb*\pi}_{\alpha\beta}\cdot\left(-i\nabla+\frac{e}{\hbar}{\vb* A}\right)\\
		&+\sum_{ij}M_{\alpha\beta}^{ij}\left(-i\partial^i+\frac{e}{\hbar}A^i\right)\left(-i\partial^j+\frac{e}{\hbar}A^j\right)
        \end{aligned},
        \label{2kp-1}
	\end{equation}
	where
	\begin{equation}
		\begin{aligned}
	M_{\alpha\beta}^{ij}=&\frac{\hbar^2}{2m}\delta_{\alpha\beta}\delta_{ij}+\frac{\hbar^2}{4m^2}\sum_{l\in\mathcal{B}}\left(\pi^{i}_{\alpha l}\pi^{j}_{l\beta}+\pi^{j}_{\alpha l}\pi^{i}_{l\beta}\right)\\
     &\times
     \left(\frac{1}{\epsilon_\alpha({\vb *k}_0)-\epsilon_l({\vb *k}_0)}+\frac{1}{\epsilon_\beta({\vb *k}_0)-\epsilon_l({\vb *k}_0)}\right)
		\end{aligned}.
		\label{2kp-2}
	\end{equation}
	Zeeman's coupling term can be expressed by 
	\begin{equation}
		H^{Z}_{\alpha\beta}=\frac{\mu_B}{\hbar}\left({\vb *L}_{\alpha\beta}+2{\vb *s}_{\alpha\beta}\right)\cdot{\vb *B},
		\label{Zeeman}
	\end{equation}
	where
	\begin{equation}
		\begin{aligned}
			L_{\alpha\beta}^{k}=& - \frac{i\hbar}{2m}\sum_{l\in\mathcal{B}}\sum_{ij}\epsilon^{ijk}\pi^{i}_{\alpha l}\pi^{j}_{l\beta}\\
			&\times\left(\frac{1}{\epsilon_\alpha({\vb *k}_0)-\epsilon_l({\vb *k}_0)}+\frac{1}{\epsilon_\beta({\vb *k}_0)-\epsilon_l({\vb *k}_0)}\right)
		\end{aligned}
		\label{Zeeman-L}
	\end{equation}
	can be considered as the orbital contribution, $ {\vb *s}_{\alpha\beta}=\mel{\alpha({\vb* k}_0)}{\hat{{\vb* s}}}{\beta({\vb* k}_0)} $ are the spin elements, and $ \mu_B $ is the Bohr magneton. The detailed derivation is shown in Appendix \ref{Zeeman-app}. 
	
	\subsection{Theory of invariants}\label{th.B}
	Suppose that the little group at $ {\vb* k}_0 $ is $ L $, and $ D(R) $ is a representation (rep) for $ \forall R\in L $, whose dimension is $ N $. There are a total of $ N^2 $ independent Hermitian matrices $ {\vb*X}={X_i\ (i=1,2,\cdots,N^2)} $ in $ N $ dimensions that constitute an $ N $-dimensional matrix space. It is simple to get a new rep $ D^{(X)} $ by
	\begin{equation}
		D(R)X_nD(R)^{-1}=\sum_{m}X_mD^{(X)}_{mn}(R)
		\label{m_b}
	\end{equation}
	for $ \forall R\in L $. Also, we can construct polynomial space of order $ p\in\mathbb{N} $, whose basis can be chosen as $ {\vb*g}({\vb*k})=\{k_x^{i}k_y^{j}k_z^{l}|i+j+l\leq p,i,j,l\in \mathbb{N}\} $. It is simple to get a new rep $ D^{(g)} $ by
	\begin{equation}
		\hat{R}g_n(\vb*k)=g_n(R^{-1}{\vb*k})=\sum_m g_m({\vb*k})D^{(g)}_{mn}(R).
	\end{equation}
	Spatial inversion $ \hat{P} $ and time reversal $ \hat{T} $ may also be the elements of the little group $ L $, which transform the wave vector as $ \vb*k \xrightarrow{\hat{P}} -\vb*k $ and $ \vb*k \xrightarrow{\hat{T}} -\vb*k $, respectively.
	
	Decomposing $ D^{(X)} $ and $ D^{(g)} $ into irreps, we have
	\begin{equation}
		D^{(X)}(R)\simeq\bigoplus_\lambda a_\lambda D^{\lambda}(R)
	\end{equation}
	\begin{equation}
		D^{(g)}(R)\simeq\bigoplus_\lambda b_\lambda D^{\lambda}(R)
	\end{equation}
	where $D^{\lambda}$ represents irreps and $ a_\lambda $ or $ b_\lambda $ denotes the multiplicities of $ D^{\lambda} $. Note that the irreps decomposed from the $ D^{(X)} $ and $ D^{(g)} $ are not exactly the same. It is simple to get the matrix basis $ X^{\lambda,\mu} $ and polynomial basis $ \vb*g^{\lambda,\nu}(\vb*k) $ of $D^{\lambda}$-irrep. An irrep may correspond to more than one sets of matrix basis or polynomial basis, thus enabling us to add an indicator $ \mu/\nu $ to distinguish them. After obtaining matrix basis and polynomial basis, the $ k\cdot p $ Hamiltonian can be written as
	\begin{equation}
		H^{kp}({\vb*k})=\sum_{\lambda}\sum_{\mu=1}^{a_\lambda}\sum_{\nu=1}^{b_\lambda}c_{\lambda\mu\nu}\vb*X^{\lambda,\mu}\cdot\vb*g^{*\lambda,\nu}({\vb*k}),
		\label{kp-opt}
	\end{equation}
	where $ c_{\lambda\mu\nu} $ are undetermined $k\cdot p$ parameters, which must be real~\cite{Jiang_2021}. It can be proved that the $ p $-th order $k\cdot p$ Hamiltonian expressed by Eq. (\ref{kp-opt}) satisfies the relation
	\begin{equation}
		H^{kp}(\hat{R}{\vb*k})=D(R)H^{kp}({\vb*k})D(R)^{-1}
		\label{kp-relation}
	\end{equation}
	for $ \forall R\in L $~\cite{Jiang_2021}. 
	
	Moreover, in the presence of magnetic field, Zeeman's coupling can also be constructed in the same way. Unlike the wave vector $ \vb* k $, the magnetic field $ \vb*B $ is a pseudovector, thus making it satisfy the relation $ \vb*B \xrightarrow{\hat{P}} \vb*B $ and $ \vb*B \xrightarrow{\hat{T}} -\vb*B $ under spatial inversion and time reversal, respectively.
	Effective Zeeman's coupling satisfies the relation
	\begin{equation}
		H^Z(\hat{R}\vb*B) = D(R)H^Z(\vb*B)D(R)^{-1}
	\end{equation}
	
	By the way, given the matrix representations of the generators of the little group $ L $, the python package {\ttfamily kdotp-generator} constructs the standard $ k\cdot p $ Hamiltonian or Zeeman's coupling with undetermined $ k\cdot p $ parameters as Eq. (\ref{kp-opt}). The standard matrices $ D^{\text{std}}(R) $ are given on the Bilbao Crystalline Server (BCS) for the assigned irreps by \webirvsp.
  %If D(R) D(R)  is the standard corep Dstd(R) D^{\text{std}}(R) , then the standard k⋅p k\cdot p  Hamiltonian Hkp−std H^{kp-\text{std}}  and Zeeman's coupling Hkp−std H^{kp-\text{std}}  can be obtained by this method.

	\subsection{Unitary transformation}\label{th.C}
	In DFT calculations, when the irrep is $n$-fold ($n\geq 2$) and the eigenstates are degenerate, there is $U(n)$ ambiguity in the DFT eigenstates.  In this section, we propose a general routine to get the unitary transformation that changes degenerate DFT wavefunctions to the $ k\cdot p $ standard basis. 
	We can obtain the eigenenergies $\{ \epsilon_n({\vb*k}_0)\} $ and eigenstates $ \{\ket{n({\vb* k}_0)}\} $ in VASP. 
    The matrix representation of $\hat{R}$ under the VASP basis set would be calculated by {\ttfamily mat2kp}, which is denoted as $ D^{\text{num}}(R) $. 
	
	However, $ D^{\text{num}}(R) $ is usually not in the standard form; they are related by a unitary transformation, \ie
	\begin{equation}
		D^{\text{std}}(R) = U^{\dagger}D^{\text{num}}(R)U
		\label{eq-transform}
	\end{equation}
	for $ \forall R \in L $, where $ U $ is a unitary matrix. To find $ U $, it is obvious that only the generators of the $\vb*k_0$-little group $ L $ should be taken into account. They are space group operators $\{R_t|\vb*v_t\}$, consisting of rotational part and translational part.  The translational part is usually expressed by a phase factor. The Eq. (\ref{eq-transform}) can be rewritten as 
	\begin{equation}
		D^{\text{num}}(R)U-UD^{\text{std}}(R)=\vb*0
		\label{U-all}
	\end{equation}
	where $\vb*0$ is a zero matrix. The real part and the imaginary part of each elements of $ U $ are independent variables, which are denoted as $ U_{r11}, U_{r12}, \cdots U_{rnn} $ and $ U_{i11}, U_{i12}, \cdots U_{inn} $, respectively. From Eq. (\ref{U-all}), it is clear to find that all these variables satisfy a linear equation set. Using a column vector $ {\vb* u}=(U_{r11}, U_{r12},\cdots,U_{rnn},U_{i11},U_{i12},\cdots,U_{inn})^T $, the Eq. (\ref{U-all}) can be rewritten as
	\begin{equation}
		Q{\vb* u}=\vb*0
	\end{equation}
	where $ Q $ is a real coefficient matrix defined by Eq. (\ref{U-all}). The details are shown in Appendix \ref{Q-app}.
	
	Therefore, all vectors in the null space of the matrix $ Q $ is the solution to Eq. (\ref{eq-transform}). Perform singular value decomposition on matrix $ Q $, we can obtain $ Q=V_1\Sigma V_2^T $ thus $ QV_2=V_1\Sigma $, where $ V_1 $ and $ V_2 $ are orthogonal matrices and $ \Sigma $ is a diagonal matrix. It can be proved that the column vectors in $ V_1 $ correspond to the singular value 0 form the basis of the null space of the matrix $ Q $, which are denoted as $ \{{\vb*u_1},{\vb*u_2},\cdots,{\vb*u_m}\} $. After reshaping these vectors, the basis of the solution space of $ U $ of Eq. (\ref{eq-transform}) are denoted as $ \{U_1,U_2,\cdots,U_n\} $. Any solution can be written as
	\begin{equation}
		U=\sum_{\alpha=1}^{m}\lambda_\alpha U_\alpha
	\end{equation}
	where $ \lambda_\alpha $ must be real.
	
	Thus, one has to get one set of the parameters $ \{\lambda_\alpha\} $ to satisfy $ U^{\dagger}U=\mathbb{I} $, where $\mathbb{I} $ is an identity matrix. However, this relation is nonlinear, thus making it hard to solve by treating it as an equation directly. In {\ttfamily mat2kp}, the unitary $ U $ is obtained via the sequential least squares programming to find optimal parameter set $ \{\lambda_\alpha\} $, which can minimize the error $ \varepsilon=\sum_{ij}\left|(U^{\dagger}U-\mathbb{I})_{ij}\right| $.

    \subsection{Calculations of parameters}\label{th.solve}
    Applying Eqs.~(\ref{eq-kp-origin}-\ref{Zeeman-L}) in DFT calculations, one can get the numerical Hamiltonian $H^{\text{num}}$, one have to solve the equation $H^{\text{std}}=U^{-1}H^{\text{num}}U $ to get all the parameters in $H^{\text{std}}$. 
 	By the way, from Eq. (\ref{kp-opt}), it is easy to find that the equations are all linear equations. However, there are more equations than parameters. The coefficients of each linear equation are not accurate because of numerical errors generated from VASP calculations, making it impossible to solve by the traditional Gaussian elimination method. We can write all equations in a matrix form 
	\begin{equation}
		A\bm{x}=\bm{b},
		\label{equset}
	\end{equation}
	where $ A $ is a constant matrix, and $ \bm{x} $ is the column vector comprised of all undetermined real parameters. To eliminate the constraint of the real $ \bm{x}$, suppose that $ A=A_r+iA_i $ and $ \bm{b}=\bm{b}_r+i\bm{b}_i $, where the subscript $ r $ represents the real part and the subscript $ i $ represents the imaginary part. Then Eq. (\ref{equset}) can be transformed as
	\begin{equation}
		\left(\begin{aligned}
			A_r\\A_i
		\end{aligned}\right)\bm{x}=\left(\begin{aligned}
			\bm{b}_r\\ \bm{b}_i
		\end{aligned}\right).
	\end{equation}
	In this case, all the matrices become real (the real constraint is automatically satisfied). Therefore, the linear least square method can be used to calculate the parameters and to give the error as well.

    \section{Capability of {\ttfamily VASP2KP}}\label{capability-sec}
    
    The {\ttfamily VASP2KP} package contains two parts: the VASP patch {\ttfamily vasp2mat} and the post-processing python code {\ttfamily mat2kp}.
    The workflow is presented in Fig.~\ref{fig:vasp2kp}.
    The methodology of obtaining the matrix elements of the generalized momentum, spin, time reversal and crystalline rotational operators in VASP is introduced in Sec. \ref{th.D}. The main algorithm steps of {\ttfamily VASP2KP} are described in Sec. \ref{th.E}. Lastly, main steps to construct $ k\cdot p $ models and Zeeman's coupling via {\ttfamily VASP2KP} are shown in Sec. \ref{th.G}.
    
	\subsection{{\ttfamily vasp2mat}: to compute matrix elements from DFT wavefunctions}\label{th.D}
	The matrix elements of $\hat{\vb*\pi}$, $\hat{\vb*s}$, $\hat{T}$ and $\hat{R}$ are required in constructing the $ k\cdot p $ Hamiltonian and Zeeman's coupling under the DFT wavefunctions as shown in Sec. \ref{th.A}. The computation method of  $\hat{\vb*\pi}$ matrix elements was first introduced in Ref.~\cite{song-gfactor}. Here we generalize it to nonlocal operators. 
    In VASP, the PAW potential is used, which is a combination and generalization of the linear augmented plane wave method. Therefore, we derive these matrix elements under the PAW wavefunctions, which are implemented in {\ttfamily vasp2mat} to compute the matrix elements. 
	
	The relation of all electron wavefunction ($ \ket{n(\vb*k_0)} $) and pseudo wavefunction ($ \ket{\widetilde{n}(\vb*k_0)} $) in PAW is
	\begin{equation}
		\ket{n(\vb*k_0)}=\mathcal{T}\ket{\widetilde{n}(\vb*k_0)},
	\end{equation} where the linear transformation $ \mathcal{T} $ can be expressed as~\cite{PhysRevB.50.17953,blochl2003projector}
	\begin{equation}
		\mathcal{T}=1+\sum_{a\mu\zeta}\left(\ket{\phi^a_{\mu}\zeta}-\ket{\widetilde{\phi}^a_{\mu}\zeta}\right)\bra{\widetilde{p}^a_{\mu}\zeta}.
		\label{T-trans}
	\end{equation}
	The $ \ket{\phi^a_{\mu}\zeta} $ is the direct product of the real space all electron partial wavefunction $ \phi^a_{\mu}(\vb* r) $ at the $ a $-th atom with a spinor wavefunction $ \ket{\zeta} (\zeta=\uparrow \text{or}\downarrow)  $, 
	$ \ket{\widetilde{\phi}^a_{\mu}\zeta} $ is the direct product of the real space pseudo partial wavefunction $ \widetilde{\phi}^a_{\mu}(\vb* r) $ with a spinor wavefunction $ \ket{\zeta} $, and $ \ket{\widetilde{p}^a_{\mu}\zeta} $ is a projector wavefunction comprised of the direct product of a real space projector wavefunction $ \widetilde{p}^a_{\mu}({\vb* r}) $ and a spinor wavefunction $ \ket{\zeta}$. 
    Here, $ \phi^{a}_{\mu}(\vb* r) $'s are obtained by all-electron calculation for the reference atom. 
    The pseudo partial wavefunctions $ \widetilde{\phi}^a_{\mu}(\vb* r) $ are identical to $ \phi^{a}_{\mu}(\vb* r) $ outside the augmentation sphere of the corresponding atom and are much softer than $ \phi^{a}_{\mu}(\vb* r) $ inside the augmentation sphere;
    $ \widetilde{\phi}^{a}_{\mu}(\vb* r) $'s provide a complete basis for the pseudo wavefunction $\ket{\widetilde{n}(\boldsymbol{k}_0)}$ inside the augmentation sphere.
    The projector wavefunctions are defined in such a way, {\it i.e.}, $\langle \widetilde{p}^a_{\mu} | \widetilde{\phi}^{a}_{\mu'} \rangle=\delta_{\mu\mu'}$, that $\langle \widetilde{p}_{\mu}^a\zeta | \widetilde{n}(\boldsymbol{k}_0)\rangle$ gives the expanding coefficients of $| \widetilde{n}(\boldsymbol{k}_0)\rangle$ on $ \widetilde{\phi}^a_{\mu}(\vb* r) $. 
    Usually the projector wavefunctions are chosen to be zero outside the core radius. 
    Therefore, outside the augmentation sphere the third and second terms in $\mathcal{T}$ cancel each other exactly, and inside the augmentation sphere the first and third terms cancel each other exactly. 
    $\mathcal{T}$ leaves $| \widetilde{n}(\boldsymbol{k}_0)\rangle$ unchanged outside the augmentation sphere and maps it to the all-electron wavefunction inside the augmentation sphere. 
    Both ${\phi}^a_{\mu}(\vb* r) $ and $ \widetilde{\phi}^a_{\mu}(\vb* r) $ are stored as a radial part times an angular part, which can be expressed as
	\begin{equation}
		\left\{
		\begin{aligned}
			\phi^{a}_{\mu}(\vb* r)=Y^{m_\mu}_{l_\mu}(\widehat{\vb*r-{\vb*\tau}_a})R^a_{\mu}(\left|\vb*r-{\vb*\tau}_a\right|)\\
			\widetilde{\phi}^{a}_{\mu}(\vb* r)=Y^{m_\mu}_{l_\mu}(\widehat{\vb*r-{\vb*\tau}_a})\widetilde{R}^a_{\mu}(\left|\vb*r-{\vb*\tau}_a\right|)
		\end{aligned}\right.
	\end{equation}
	where $ {\vb*\tau}_a $ is the site of the $ a $-th atom, $Y^{m}_l$ are sphere harmonics, and $R^{a}_\mu$ are real functions. The hatted vectors represent the unit vector along the corresponding directions. 
	
	According to Eq. (\ref{T-trans}), the PAW matrix form of a local operator $ \hat{F} $ can be expressed by
	\begin{equation}
		\begin{aligned}
			F_{mn}=&\mel{m(\vb*k_0)}{\hat{F}}{n(\vb*k_0)}=\mel{\widetilde{m}(\vb*k_0)}{\mathcal{T}^{\dagger}\hat{F}\mathcal{T}}{\widetilde{n}(\vb*k_0)}\\
			=&\mel{\widetilde{m}(\vb*k_0)}{\hat{F}}{\widetilde{n}(\vb*k_0)}\\
			&+\sum_{a\mu\nu}\sum_{\zeta\zeta'}\braket{\widetilde{m}(\vb*k_0)}{\widetilde{p}^a_{\mu}\zeta}F^{a}_{\mu\zeta,\nu\zeta'}\braket{\widetilde{p}^a_{\nu}\zeta'}{\widetilde{n}(\vb*k_0)}
		\end{aligned},
		\label{matrix-f}
	\end{equation}
	where $ F^{a}_{\mu\zeta,\nu\zeta'} $ is the projection matrix of the operator $ \hat{F} $ in the $ a $-th atom's augmentation sphere, which is defined as
	\begin{equation}
		F^{a}_{\mu\zeta,\nu\zeta'}=\mel{\phi^a_{\mu}\zeta}{\hat{F}}{\phi^{a}_{\nu}\zeta'}-\mel{\widetilde{\phi}^a_{\mu}\zeta}{\hat{F}}{\widetilde{\phi}^{a}_{\nu}\zeta'}.
		\label{ele-f}
	\end{equation}
    The first term in $F^{a}_{\mu\zeta,\mu'\zeta'}$ gives the contribution from the all-electron wavefunction in the augmentation sphere, and the second term cancels the contribution from the pseudo wavefunction in the augmentation sphere that is counted in the first term of Eq.~(\ref{matrix-f}). 
    
	In PAW, the plane wave is used to span the pseudo wavefunction $ \widetilde{\psi}_{n \vb*k_0}(\vb*r,\zeta)=\braket{\vb*r,\zeta}{\widetilde{n}(\vb*k_0)} $, which is expressed by
	\begin{equation}
		\widetilde{\psi}_{n\vb*k_0}(\vb*r, \zeta)=\sum_{\vb*G}c^{n\vb*k_0}_{\zeta\vb*G}e^{i({\vb*k_0+\vb*G})\cdot{\vb*r}}
		\label{plane-wave}
	\end{equation}
	where $ c^{n\vb*k_0}_{\zeta\vb*G} $ are the plane wave coefficients. 
	
	The projection coefficients $ \braket{\widetilde{p}^a_{\mu}\zeta}{\widetilde{n}(\vb*k_0)} $ in the second term of Eq. (\ref{matrix-f}) have already been calculated in VASP. Therefore, to calculate the matrix of the local operator $ \hat{F} $, we need to calculate $\mel{\widetilde{m}(\vb*k_0)}{\hat{F}}{\widetilde{n}(\vb*k_0)}$ and $F^{a}_{\mu\zeta,\nu\zeta'}$ in Eq. (\ref{matrix-f}).
	
	\subsubsection{Generalized momentum matrix}
	The matrix of the generalized momentum $ \hat{\vb*\pi} $ can be expressed by substituting $ \hat{\vb*\pi} $ into $ \hat{F} $ in Eq. (\ref{matrix-f}) as follows:
	\begin{equation}
		\begin{aligned}
			&\vb*\pi_{mn}=\mel{\widetilde{m}(\vb*k_0)}{\hat{\vb*p}}{\widetilde{n}(\vb*k_0)}+\frac{1}{2mc^2}\mel{\widetilde{m}(\vb*k_0)}{\hat{\vb*s}\times \nabla V}{\widetilde{n}(\vb*k_0)}\\
			+&\sum_{a\mu\nu}\sum_{\zeta\zeta'}\braket{\widetilde{m}(\vb*k_0)}{\widetilde{p}^a_{\mu}\zeta}\vb*p^{a}_{\mu\zeta,\nu\zeta'}\braket{\widetilde{p}^a_{\nu}\zeta'}{\widetilde{n}(\vb*k_0)}\\
			+&\frac{1}{2mc^2}\sum_{a\mu\nu\zeta\zeta'}\braket{\widetilde{m}(\vb*k_0)}{\widetilde{p}^a_{\mu}\zeta}\mel{\phi^a_{\mu}\zeta}{\hat{\vb*s}\times\nabla V}{\phi^{a}_{\nu}\zeta'}\braket{\widetilde{p}^a_{\nu}\zeta'}{\widetilde{n}(\vb*k_0)}\\
			-&\frac{1}{2mc^2}\sum_{a\mu\nu\zeta\zeta'}\braket{\widetilde{m}(\vb*k_0)}{\widetilde{p}^a_{\mu}\zeta}\mel{\widetilde{\phi}^a_{\mu}\zeta}{\hat{\vb*s}\times\nabla V}{\widetilde{\phi}^{a}_{\nu}\zeta'}\braket{\widetilde{p}^a_{\nu}\zeta'}{\widetilde{n}(\vb*k_0)}
			\label{pi-1}
		\end{aligned}
	\end{equation}
	Since the SOC effect is considered only within the augmentation spheres in VASP, where $ \ket{\widetilde{n}(\vb*k_0)}=\sum_{a\mu\zeta}\ket{\widetilde{\phi}^a_{\mu}\zeta}\braket{\widetilde{p}^a_\mu\zeta}{\widetilde{n}(\vb*k_0)} $, we can make the second term and the fifth term in Eq. (\ref{pi-1}) cancel out. Therefore, the generalized momentum matrix can be simplified to 
	\begin{equation}
		\begin{aligned}
			\vb*\pi_{mn}=&\mel{\widetilde{m}(\vb*k_0)}{\hat{\vb*p}}{\widetilde{n}(\vb*k_0)}\\
			&+\sum_{a\mu\nu}\sum_{\zeta\zeta'}\braket{\widetilde{m}(\vb*k_0)}{\widetilde{p}^a_{\mu}\zeta}\vb*\pi'^{a}_{\mu\zeta,\nu\zeta'}\braket{\widetilde{p}^a_{\nu}\zeta'}{\widetilde{n}(\vb*k_0)}
		\end{aligned}
		\label{pi-simp}
	\end{equation}
	where 
	\begin{equation}
		\begin{aligned}
			\vb*\pi'^{a}_{\mu\zeta,\nu\zeta'}=&\delta_{\zeta\zeta'}\left(\mel{\phi^a_\mu}{\hat{{\vb* p}}}{\phi^a_\nu}-\mel{\widetilde{\phi}^a_\mu}{\hat{{\vb* p}}}{\widetilde{\phi}^a_\nu}\right)\\
			&+\frac{\hbar^2}{4mc^2}\vb*\sigma_{\zeta\zeta'}\times\mel{\phi^a_\mu}{\nabla V}{\phi^a_\nu}
		\end{aligned}
		\label{pi'-matrix}
	\end{equation}
	The first term of $ \vb*\pi_{mn} $ in Eq. (\ref{pi-simp}) can be calculated by
	\begin{equation}
		\mel{\widetilde{m}(\vb*k_0)}{\hat{\vb*p}}{\widetilde{n}(\vb*k_0)}=\sum_{\vb*G}\hbar(\vb*k_0+\vb*G)c^{*m\vb*k_0}_{\zeta\vb*G}c^{n\vb*k_0}_{\zeta\vb*G}
		\label{p-matrix}
	\end{equation}
	where $ \hat{\vb*p}=-i\hbar\nabla $. The integrals in $ \vb*\pi'^a_{\mu\zeta,\nu\zeta'} $ can be calculated by separating the radial part and angular part, which are expressed as
	\begin{equation}
		\left\{\begin{aligned}
			&\begin{aligned}
				\mel{\phi^a_\mu}{\hat{{\vb* p}}}{\phi^a_\nu}&=-i\hbar\int\mathrm{d}\Omega Y^{m_\mu*}_{l_\mu}\nabla Y^{m_\nu}_{l_\nu}\int\mathrm{d}rr^2R^{a*}_\mu R^a_\nu\\
				&-i\hbar\int\mathrm{d}\Omega Y^{m_\mu*}_{l_\mu}\frac{\vb*r}{r}Y^{m_\nu}_{l_\nu}\int\mathrm{d}rr^2R^{a*}_{\mu}\partial_rR^{a}_\nu
			\end{aligned}\\
			&\begin{aligned}
				\mel{\widetilde{\phi}^a_\mu}{\hat{{\vb* p}}}{\widetilde{\phi}^a_\nu}=&-i\hbar\int\mathrm{d}\Omega Y^{m_\mu*}_{l_\mu}\nabla Y^{m_\nu}_{l_\nu}\int\mathrm{d}rr^2\widetilde{R}^{a*}_\mu \widetilde{R}^a_\nu\\
				&-i\hbar\int\mathrm{d}\Omega Y^{m_\mu*}_{l_\mu}\frac{\vb*r}{r}Y^{m_\nu}_{l_\nu}\int\mathrm{d}rr^2\widetilde{R}^{a*}_{\mu}\partial_r\widetilde{R}^{a}_\nu
			\end{aligned}\\
			&\mel{\phi^a_\mu}{\nabla V}{\phi^a_\nu}\approx\int\mathrm{d}\Omega Y^{m_\mu*}_{l_\mu}\frac{\vb*r}{r}Y^{m_\nu}_{l_\mu}\int\mathrm{d}rr^2R^{a*}_{\mu}\partial_rVR^{a}_\nu
		\end{aligned}\right.
		\label{pi'-matrix-ele}
	\end{equation}
Finally, by substituting Eqs. (\ref{pi'-matrix})-(\ref{pi'-matrix-ele}) into Eq. (\ref{pi-simp}), the matrices of the generalized momentum $ \hat{\vb*\pi} $ can be obtained.

        \begin{figure*}[t!]
 		\includegraphics[width=0.81\linewidth]{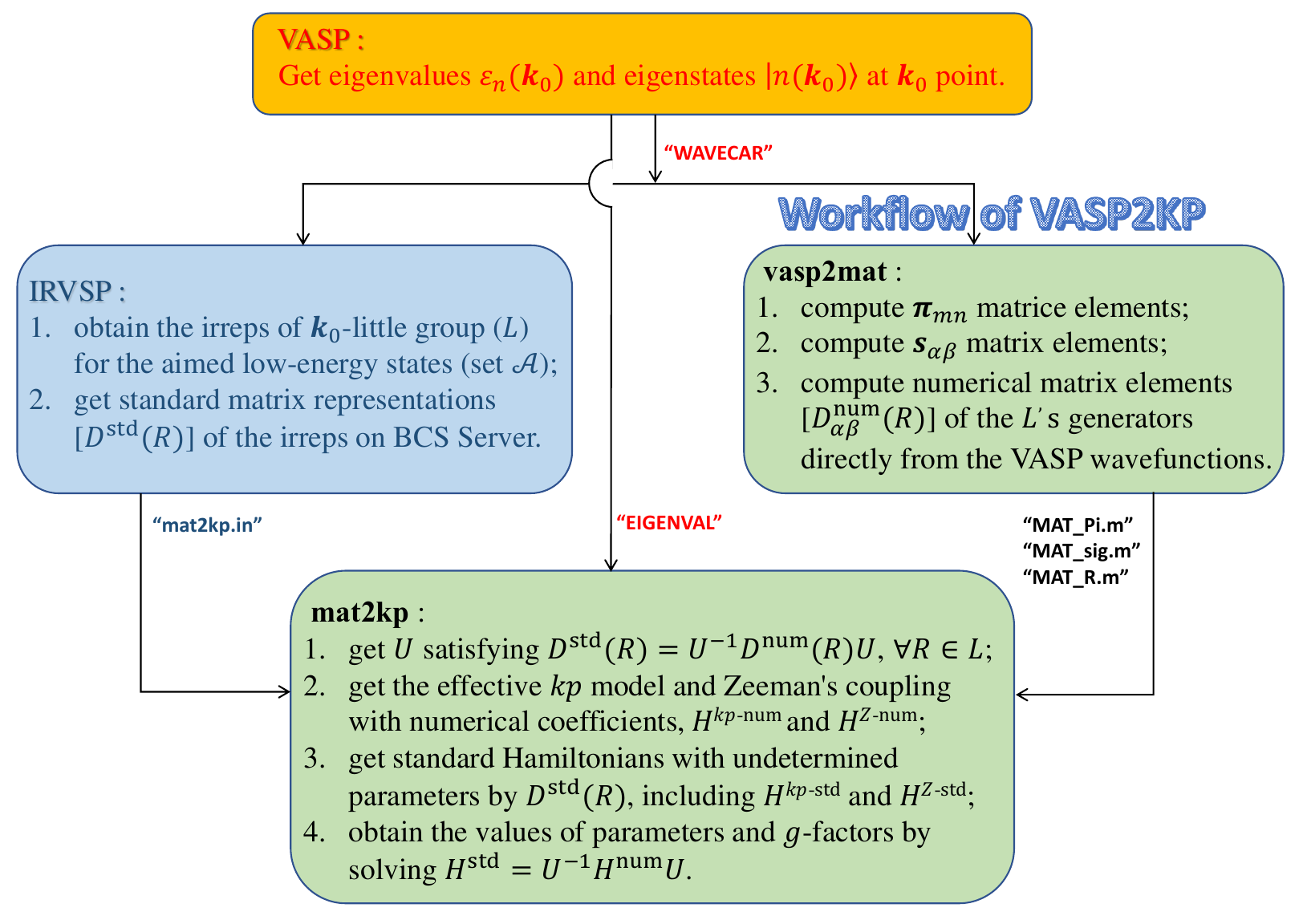}
        \caption{The workflow of {\ttfamily VASP2KP} to compute the $k \cdot p$ parameters and Land\'e $g$-factors directly from the VASP wavefunctions.}
        \label{fig:vasp2kp}
        \end{figure*}

	\subsubsection{Spin matrices}
	By substituting the local operator $ \hat{\vb*s} $ for $ \hat{F} $ in Eq. (\ref{matrix-f}), the corresponding matrix elements are given explicitly.
	The first term in Eq. (\ref{matrix-f}) can be calculated by
	\begin{equation}
        \begin{aligned}
		\mel{\widetilde{\alpha}(\vb*k_0)}{\hat{\vb*s}}{\widetilde{\beta}(\vb*k_0)}=&\frac{\hbar}{2}\sum_{\vb*G\vb*G'\zeta\zeta'}\vb*\sigma_{\zeta\zeta'}c^{*\alpha\vb*k_0}_{\zeta\vb*G}c^{\beta\vb*k_0}_{\zeta'\vb*G'} \delta_{\vb*G,\vb*G'}
		\label{spin-T-R-1}
        \end{aligned}
	\end{equation}
	where $ \hat{\vb*s}=\frac{\hbar}{2}\hat{\vb*\sigma} $ and the projection matrix can be obtained by
	\begin{equation}
        \begin{aligned}		\vb*s^{a}_{\mu\zeta,\nu\zeta'}=&\frac{\hbar}{2}\delta_{l_\mu l_\nu}\delta_{m_\mu m_\nu}\vb*\sigma_{\zeta\zeta'}\int\mathrm{d}rr^2\left(R^{a*}_\mu R^{a}_\nu-\widetilde{R}^{a*}_\mu \widetilde{R}^{a}_\nu\right)
        \end{aligned}
        \label{spin-T-R-2}
	\end{equation}
    %Here, $D^{(2l+1)}(R(\vec n , w))= e^{-i w \vec n \cdot \vec L} $ is the standard rotation matrix in $\ket{L=l;m=-l,\cdots,l}$ basis~\cite{GAO2021107760}. When the operation $R$ has a fractional translation, the detailed derivation can be found in Ref.~\cite{GAO2021107760}. 
    Finally, by substituting Eqs. (\ref{spin-T-R-1},\ref{spin-T-R-2}) into Eq. (\ref{matrix-f}), the matrices of the spin $ \hat{\vb*s} $ can be obtained in the PAW wavefunctions.

    \subsubsection{Space group operator matrices}
    The general (magnetic) space group operator (SGO) is expressed by $F=\{R|\vb*v\}$ or $F=\{TR|\vb*v\}$, which is usually a nonlocal (NL) operator. 
    One should notice that the SGO commutes with $\mathcal{T}$ in Eq.~(\ref{T-trans}). 
    Thus the matrix element can be written as 
    \begin{align}
        F^{NL}_{mn}=& \mel{m(\vb*k_0)}{\hat{F}}{n(\vb*k_0)}=\mel{\widetilde{m}(\vb*k_0)}{\mathcal{T}^{\dagger} \mathcal{T}}{\hat{F} \cdot \widetilde{n}(\vb*k_0)} \nonumber\\
        =& \langle {\widetilde{m}(\vb*k_0)} | \hat{F} \cdot \widetilde{n}(\vb*k_0)\rangle \nonumber\\
        & + \sum_{a\mu\nu\zeta}\braket{\widetilde{m}(\vb*k_0)}{\widetilde{p}^a_{\mu}\zeta}C^{a}_{\mu,\nu}\braket{\widetilde{p}^{a}_{\nu}\zeta}{ \hat{F} \cdot \widetilde{n}(\vb*k_0)}, 
    \end{align}
    where
    \begin{equation}
    \begin{aligned}
        &C^a_{\mu,\nu}=\delta_{l_\mu l_\nu}\delta_{m_\mu m_\nu}\int\mathrm{d}rr^2\left(R^{a*}_\mu R^{a}_\nu-\widetilde{R}^{a*}_\mu \widetilde{R}^{a}_\nu\right)\\
        &\mel{\vb*r,\zeta}{\left\{R|\vb*v\right\}}{\widetilde{\beta}(\vb*k_0)}=\sum_Gc^{\beta\vb*k_0}_{\zeta\vb*G} e^{i({\vb*k_0}+{\vb*G})\cdot R^{-1}({\vb*r}-{\vb*v})}\\
        &\mel{\vb*r,\zeta}{\left\{TR|\vb*v\right\}}{\widetilde{\beta}(\vb*k_0)}=\eta_\zeta\sum_Gc^{\beta\vb*k_0*}_{\overline{\zeta}\vb*G} e^{-i({\vb*k_0}+{\vb*G})\cdot R^{-1}({\vb*r}-{\vb*v})}
    \end{aligned}
    \end{equation}
    Here, $\eta_\uparrow=1$, $\eta_\downarrow=-1$, and
    $\bar\zeta$ indicates the opposite of spin $\zeta$.
    In our convention, the translation {$\vb*L$} is expressed by a phase factor of $e^{-i \vb*k_0\cdot \vb*L}$. More functions of the patch {\ttfamily vasp2mat} are presented in Appendix \ref{vasp_kp_all}.

	\subsection{The python code {\ttfamily mat2kp}}\label{th.E}
	In this subsection, we give a brief introduction of the main algorithm steps of {\ttfamily mat2kp}. This code needs the inputs of $\varepsilon_n(\vb*k_0)$, $\vb*{\pi}_{mn}$, $\vb*{s}_{\alpha\beta}$, $D^{\text{num}}_{\alpha\beta}(R)$ and $D^{\text{std}}_{\alpha\beta}(R)$, which correspond to the {\ttfamily EIGENVAL}, {\ttfamily MAT\_Pi.m}, {\ttfamily MAT\_sig.m}, {\ttfamily MAT\_R.m} and {\ttfamily mat2kp.in} files, respectively.
	The main steps of {\ttfamily mat2kp} are as follows:
	\begin{enumerate}
		\item Following Sec. \ref{th.C}, calculate the unitary transformation matrix $ U $ satisfying $ D^{\text{std}}(R) = U^\dagger D^{\text{num}}(R)U $ for $\forall R\in L$.
		\item By downfolding processing and using Eqs. (\ref{2kp-1}-\ref{Zeeman-L}), get the numerical $ k\cdot p $  Hamiltonian and Zeeman's coupling and compute the coefficients numerically under DFT basis, \ie $H^{kp-\text{num}}$ and $ H^{Z-\text{num}}$.
		%\item Transform the numerical k⋅p k\cdot p   Hamiltonian with Zeeman's coupling under VASP basis into the standard form via U U , which are denoted as Hkp−std−N H^{kp-\text{std}-N}  and HZ−std−N H^{Z-\text{std}-N} , respectively.
		\item Through theory of invariants in Sec. \ref{th.B}, import the package {\ttfamily kdotp-generator} and generate the standard $ k\cdot p $ Hamiltonian and Zeeman's coupling with a set of undetermined parameters, $ H^{kp-\text{std}} $ and $ H^{Z-\text{std}} $, respectively.
		\item Obtain the values of $ k\cdot p $ parameters and Land\'e $ g $-factors by solving $ H^{kp-\text{std}}=U^{-1}H^{kp-\text{num}}U $ and $ H^{Z-\text{std}}=U^{-1}H^{Z-\text{num}} U$ in Sec. \ref{th.solve}.
		%\item Present a report containing the k⋅p k\cdot p  Hamiltonian and Zeeman's coupling with the values for the k⋅p k\cdot p  parameters and Land\'e g g -factors in the eV and \AA \ units, respectively.
	\end{enumerate}

	\subsection{General steps to get the \texorpdfstring{$ k\cdot p $}{} model and parameters automatically}\label{th.G}
     The general workflow is given in Fig.~\ref{fig:vasp2kp}. We will take  Bi$ _2 $Se$ _3 $ as an example for illustration.
	
	\begin{enumerate}
		\item Run VASP to output the eigenstates $ \ket{n(\vb*k_0)} $ ($\ket{\widetilde{n}(\vb*k_0)}$ in {\ttfamily WAVECAR}) and  eigenvalues $ \varepsilon_n(\vb*{k}_0)$ ({\ttfamily EIGENVAL}) at $\vb*k_0$ point.
  
		\item Run {\ttfamily IRVSP} to get the irreps of the aimed low-energy bands (set $ \mathcal{A} $), and then obtain the standard matrix representations [$ D^{\text{std}}(R) $] of the generators of $\vb*k_0$-little group on the Bilbao Crystalline Server (BCS). They are given in ``{\ttfamily mat2kp.in}" (the input file of {\ttfamily mat2kp}).

{\tiny 
  {\ttfamily
\#\#\#\#\#\#\#\# mat2kp.in - Bi2Se3 \#\#\#\#\#\#\#\#\\
Symmetry = \{\\
'C3z' : \{ \\
'rotation\_matrix': \\
Matrix([[-Rational(1,2), -sqrt(3)/2,0],[sqrt(3)/2, -Rational(1,2), 0],[0, 0, 1]]),\\
'repr\_matrix':\\
Matrix([[Rational(1,2)-I*sqrt(3)/2,0,0,0],[0,Rational(1,2)+I*sqrt(3)/2,0,0],\\
\ [0,0,Rational(1,2)-I*sqrt(3)/2,0],[0,0,0,Rational(1,2)+I*sqrt(3)/2]]),\\
'repr\_has\_cc': False\},\\
'C2x' : \{\\
'rotation\_matrix': Matrix([[1, 0, 0],[0, -1, 0],[0, 0, -1]]),\\
'repr\_matrix': \\
Matrix([[0,-Rational(1,2)-I*sqrt(3)/2,0,0],[Rational(1,2)-I*sqrt(3)/2,0,0,0],\\
\ [0,0,0,-Rational(1,2)-I*sqrt(3)/2],[0,0,Rational(1,2)-I*sqrt(3)/2,0]]),\\
'repr\_has\_cc': False\},\\
'P' : \{\\
'rotation\_matrix': Matrix([[-1,0,0],[0, -1, 0],[0, 0, -1]]),\\
'repr\_matrix':\\
Matrix([[1,0,0,0],[0,1,0,0],[0,0,-1,0],[0,0,0,-1]]),\\
'repr\_has\_cc': False\},\\
'T' : \{\\
'rotation\_matrix': eye(3),\# Identity Matrix\\
'repr\_matrix': Matrix([[0,1,0,0],[-1,0,0,0],[0,0,0,-1],[0,0,1,0]]),\\
'repr\_has\_cc': True\}\\
\}\\
\# optional parameters \\
vaspMAT = '../Bi2Se3/GMmat'  \# the path: to read eigenvalues, Pi, s, and R matrices in this folder. \\
order = 2       \qquad \ \# Order of the kp model : 2 (default) or 3.\\
print\_flag = 2  \# Where to output results: 1 (screen) or 2 (files, default).\\
kpmodel = 1    \quad\ \# Whether to compute Hkp: 0 or 1 (default).\\
gfactor = 1    \quad\ \# Whether to compute HZ : 0 or 1 (default).\\
log = 1 \qquad\quad\ \# Whether to output log files: 0 or 1 (default).\\
}

}

  \item Run {\ttfamily vasp2mat} to generate $\vb*\pi_{mn}$ ({\ttfamily vmat=11; vmat\_name='Pi'}), $\vb*s_{\alpha\beta}$  ({\ttfamily vmat=10; vmat\_name='sig'}), and  generators' [$D(R)^{\text{num}}_{\alpha\beta}$;  {\ttfamily vmat=12}] matrices directly from the VASP wavefunctions ({\ttfamily WAVECAR}) by the following settings in ``{\ttfamily INCAR.mat}" files, respectively. These numerical matrix representations are output in  {\ttfamily MAT\_Pi.m}, {\ttfamily MAT\_sig.m}, {\ttfamily MAT\_R.m} files. They are in the `vaspMAT' folder (given in {\ttfamily mat2kp.in}). The {\ttfamily vmat\_name} of the generators should be the same as those given in ``{\ttfamily mat2kp.in}".
  
{\tiny
{\ttfamily
\#\#\#\#\#\#\#\# INCAR.mat - Pi mat. \#\#\#\#\#\#\#\#\\
\&vmat\_para\\
    ! soc------------------------\\
    cfactor=1.0\\
    socfactor=1.0\\
    nosoc\_inH = .false.\\
    ! operator------------------\\
    vmat = 11\\
    vmat\_name = 'Pi'\\
    vmat\_k = 1\\
    bstart=1,  bend=400\\
    print\_only\_diagnal = .false.\\
    % ! rotation------------------\\
    % rot\_n(:) = 0  0  1\\
    % rot\_alpha = 0\\
    % rot\_det = 1\\
    % rot\_tau(:) = 0     0    0\\
    % rot\_spin2pi = .false.\\
    % time\_rev = .false.\\
/\\

\#\#\#\#\#\#\#\# INCAR.mat - sigma mat. \#\#\#\#\#\#\#\#\\
\&vmat\_para\\
    ! soc------------------------\\
    cfactor=1.0\\
    socfactor=1.0\\
    nosoc\_inH = .false.\\
    ! operator------------------\\
    vmat = 10\\
    vmat\_name = 'sig'\\
    vmat\_k = 1\\
    bstart=47,  bend=50\\
    print\_only\_diagnal = .false.\\
    % ! rotation------------------\\
    % rot\_n(:) = 0  0  1\\
    % rot\_alpha = 0\\
    % rot\_det = 1\\
    % rot\_tau(:) = 0     0    0\\
    % rot\_spin2pi = .false.\\
    % time\_rev = .false.\\
/\\

\#\#\#\#\#\#\#\# INCAR.mat - C3z / C2x / P / T  mat. - Bi2Se3 \#\#\#\#\#\#\#\#\\
\&vmat\_para\\
    ! soc------------------------\\
    cfactor=1.0\\
    socfactor=1.0\\
    nosoc\_inH = .false.\\
    ! operator------------------\\
    vmat = 12\\
    vmat\_name = 'C3z'\qquad/\qquad'C2x'\qquad/\qquad'P'\quad\qquad/\qquad'T'\\
    vmat\_k = 1\\
    bstart=47,  bend=50\\
    print\_only\_diagnal = .false.\\
    ! rotation------------------\\
    rot\_n(:) = 0  0  1~\qquad/\qquad1  0  0\qquad/\qquad0  0  1\qquad/\qquad0  0  1\\
    rot\_alpha = 120 \ \qquad/\qquad180  \ \qquad /\qquad0 \quad\ \qquad/\qquad0\\
    rot\_det =  1   \ \qquad\qquad/\qquad1\qquad\qquad\ /\qquad-1\qquad\ \quad/\qquad1\\
    rot\_tau(:) = 0     0    0\\
    rot\_spin2pi = .false.\\
    time\_rev = .false.\quad\ /\qquad.false.\quad/\qquad.false.\quad/\qquad.true.\\
/\\
}

}

% \boxed{text text math math 
% asdfadsf
% text text math math 
% asdfadsf
% text text math math 
% asdfadsf
% }

% \fcolorbox{green}{green!20}{\parabox{0.5\textwidth}{%
%         \color{red!70!black}%
% text text math math 
% asdfadsf
% text text math math 
% asdfadsf
%     }}

      \clearpage
		\item Run the post-processing python code {\ttfamily mat2kp} with the input files, to construct the standard $ k\cdot p $ Hamiltonian and Zeeman's coupling, and compute the values of $k\cdot p$ parameters and $g$-factors. The outputs are ``{\ttfamily kp-parameters.out}" and ``{\ttfamily g-factors.out}" files, as pasted below.

{\tiny
{\ttfamily
\#\#\#\#\#\#\#\# kp-parameters.out \#\#\#\#\#\#\#\#\\
kp Hamiltonian\\
==========  Result of kp Hamiltonian  ==========\\
Matrix([[a1 + a2 + c1*(kx**2 + ky**2) + c2*(kx**2 + ky**2) + c3*kz**2 + c4*kz**2, 0, -I*b2*kz, -b1*(kx*(sqrt(3) + 3*I) + ky*(3 - sqrt(3)*I))/3], [0, a1 + a2 + c1*(kx**2 + ky**2) + c2*(kx**2 + ky**2) + c3*kz**2 + c4*kz**2, b1*(kx*(sqrt(3) - 3*I) + ky*(3 + sqrt(3)*I))/3, I*b2*kz], [I*b2*kz, b1*(kx*(sqrt(3) + 3*I) + ky*(3 - sqrt(3)*I))/3, a1 - a2 + c1*(kx**2 + ky**2) - c2*(kx**2 + ky**2) + c3*kz**2 - c4*kz**2, 0], [-b1*(kx*(sqrt(3) - 3*I) + ky*(3 + sqrt(3)*I))/3, -I*b2*kz, 0, a1 - a2 + c1*(kx**2 + ky**2) - c2*(kx**2 + ky**2) + c3*kz**2 - c4*kz**2]])\\
Parameters:\\
a1 = 4.8898 ;\\
a2 = -0.2244 ;\\
b1 = -3.238 ;\\
b2 = 2.5562 ;\\
c1 = 19.5842 ;\\
c2 = 44.4746 ;\\
c3 = 1.8117 ;\\
c4 = 9.5034 ;\\
Error of the linear least square method: 3.93e-06\\
Sum of absolute values of numerical zero elements: 6.47e-02\\

\#\#\#\#\#\#\#\# g-factors.out \#\#\#\#\#\#\#\#\\
Zeeman's coupling\\
==========  Result of Zeeman's coupling  ==========\\
mu\_B/2*Matrix([[Bz*g3 + Bz*g4, g1*(Bx*(1 - sqrt(3)*I/3) + By*(-sqrt(3)/3 - I)) + g2*(Bx*(1 - sqrt(3)*I/3) + By*(-sqrt(3)/3 - I)), 0, 0], [g1*(Bx*(1 + sqrt(3)*I/3) + By*(-sqrt(3)/3 + I)) + g2*(Bx*(1 + sqrt(3)*I/3) + By*(-sqrt(3)/3 + I)), -Bz*g3 - Bz*g4, 0, 0], [0, 0, Bz*g3 - Bz*g4, g1*(Bx*(1 - sqrt(3)*I/3) + By*(-sqrt(3)/3 - I)) + g2*(Bx*(-1 + sqrt(3)*I/3) + By*(sqrt(3)/3 + I))], [0, 0, g1*(Bx*(1 + sqrt(3)*I/3) + By*(-sqrt(3)/3 + I)) + g2*(Bx*(-1 - sqrt(3)*I/3) + By*(sqrt(3)/3 - I)), -Bz*g3 + Bz*g4]])\\
Parameters:\\
g1 = -0.3244 ;\\
g2 = 5.761 ;\\
g3 = -7.8904 ;\\
g4 = -13.0138 ;\\
Error of the linear least square method: 6.11e-08\\
Sum of absolute values of numerical zero elements: 4.12e-03\\
}

}
	\end{enumerate}

        \begin{figure}[t!]
 		\includegraphics[width=0.94\linewidth]{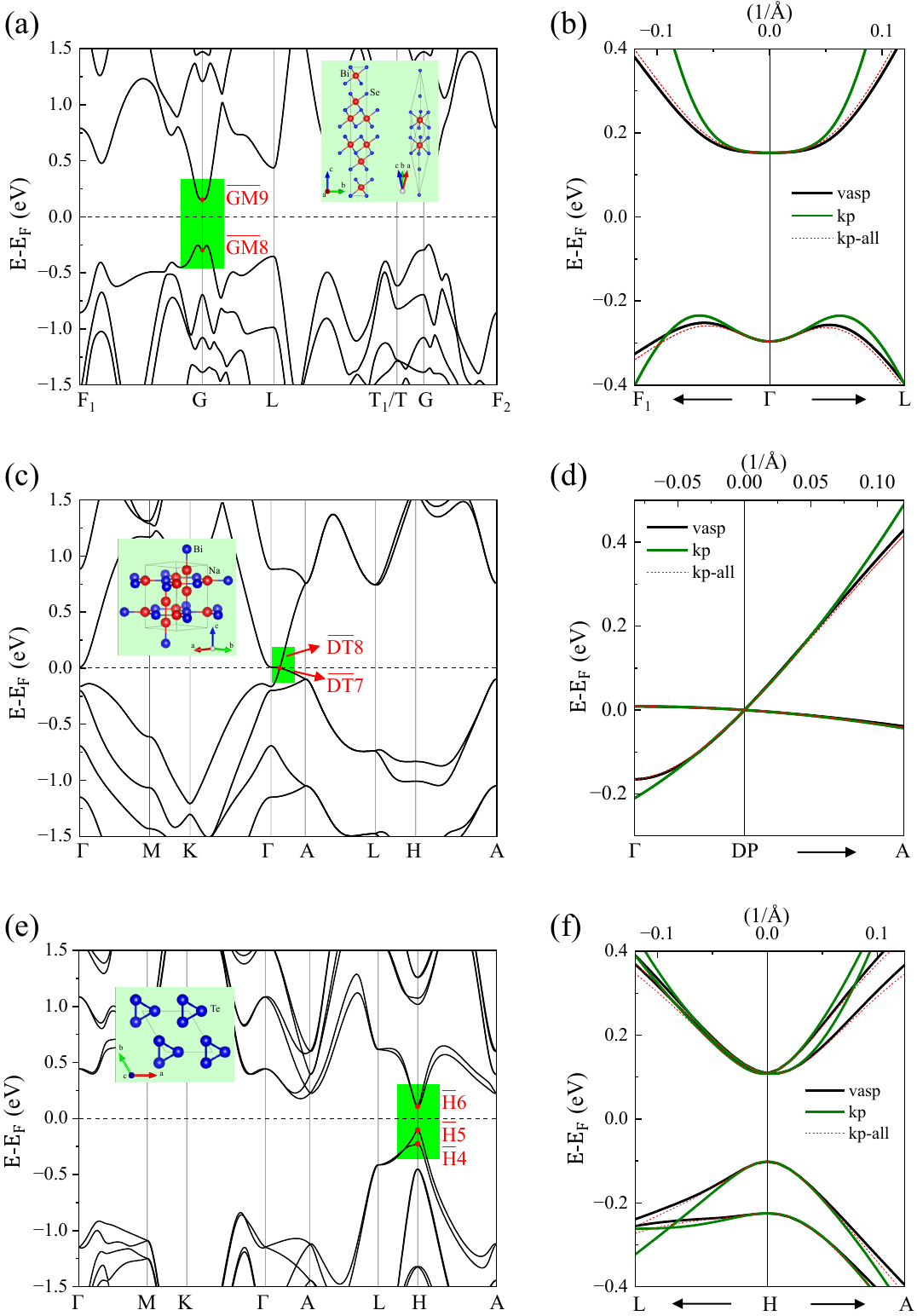}
            \caption{%Crystal structure, electronic structure, and model's band dispersions of Bi$_2$Se$_3$, Na$_3$Bi and Te. 
            Crystal structure and electronic band structure (VASP) of Bi$_2$Se$_3$ (a), Na$_3$Bi (c) and Te (e). Model dispersions of Bi$_2$Se$_3$ (b), Na$_3$Bi (d) and Te (f) [kp; Eq. (\ref{2kp-1}) and kp-all; Eq. (\ref{kp-initial})].
            }
 		\label{kp-fig}
 	    \end{figure}

	\section{Applications in materials}\label{exa}
	In this section, we apply this package to some typical materials, \ie Bi$_2$Se$_3$, Na$_3$Bi, Te, InAs, and 1H-TMD, to construct the effective models and to compute all the parameters.

	\subsection{Four-band model at \texorpdfstring{$\Gamma$}{} in \texorpdfstring{Bi$_2$Se$_3$}{}}
        As we know, the topological property of Bi$_2$Se$_3$ is due to the band inversion at $\Gamma$. By using {\ttfamily IRVSP}, the lowest conduction band belongs to $ \overline{\text{GM}}9 $ (twofold degenerate), while the highest valence band belongs to $ \overline{\text{GM}}8 $ (twofold degenerate), as depicted in Fig.~\ref{kp-fig}(a). Based on these states in the ascending order, the low-energy effective Hamiltonian at $\Gamma$ is constructed automatically. The generators of the $\Gamma$-little group are $ \{C_{3z}|0,0,0\} $, $ \{C_{2x}|0,0,0\} $, $ \{P|0,0,0\} $, and $ \{T|0,0,0\} $. Their standard matrix representations are given in TABLE \ref{Bi2Se3-rep}, which are needed by the code {\ttfamily mat2kp} to construct effective models.

        To the second order, the $k\cdot p$ Hamiltonian and Zeeman's coupling are given in Eq. (\ref{Bi2Se3-kp-Zeeman}), where $\mu_B$ is the Bohr magneton ($\sim$ 0.05788 meV/Tesla). 
        After their numerical matrix representations are computed by {\ttfamily vasp2mat} directly from the VASP wavefunctions, the parameters of $k\cdot p$ Hamiltonian and $g$-factors of Zeeman's coupling are computed, as shown in TABLE~\ref{Bi2Se3-kp-par}. The four-band model's dispersions are plotted in Fig. \ref{kp-fig}(b). They fit well with the VASP bands in the vicinity of $\Gamma$.  Moreover, the dispersions of the all-band $k\cdot p$ model without downfolding in Eq. (\ref{kp-initial}) are also plotted for comparison (labeled by `kp-all').
        
        \begin{equation}
            \begin{aligned}
                &H^{\text{eff}}(\vb*k,\vb*B)= H^{kp}+H^Z,\\
                &H^{kp}=\begin{pmatrix}
                D_1&0&-ib_2k_z&-\frac{3i+\sqrt{3}}{3}b_1k_-\\
                &D_1&\frac{\sqrt{3}-3i}{3}b_1k_+&ib_2k_z\\
                &&D_2&0\\\dagger&&&D_2
            \end{pmatrix},\\
            &H^Z=\frac{\mu_B}{2}\begin{pmatrix}
                h_{1}^{+}B_z&h_{2}^{+}B_+&0&0\\&-h_{1}^{+}B_z&0&0\\&&h_{1}^{-}B_z&h_{2}^{-}B_-\\\dagger&&&-h_{1}^{-}B_z
            \end{pmatrix},\\
                &D_1=a_1+a_2+(c_1+c_2)k_+k_-+(c_3+c_4)k_z^2,\\
                &D_2=a_1-a_2+(c_1-c_2)k_+k_-+(c_3-c_4)k_z^2,\\
                &h_{1}^{\pm}=g_3\pm g_4, h_{2}^{\pm}=\frac{3-\sqrt{3}i}{3}(g_1\pm g_2),\\
                &k_\pm=k_x\pm ik_y,\quad B_\pm=B_x\pm iB_y.\\
            \end{aligned}
            \label{Bi2Se3-kp-Zeeman}
        \end{equation}

\begin{table}[h]
 		\centering
 		\caption{The matrix representations of ${\rm\overline{GM}}8$ and $\rm{\overline{GM}}9$ irreps at $\Gamma$ are given on \href{https://www.cryst.ehu.es/cgi-bin/cryst/programs/corepresentations_out.pl?super=166.98&vecfinal=GM}{BCS server} for the generators.}
        \label{Bi2Se3-rep}
        \begin{ruledtabular}
 		\begin{tabular}{ccc}
 			&$ \overline{\text{GM}}8 $&$ \overline{\text{GM}}9 $\\
 			\hline
 			$ \{C_{3z}|0,0,0\} $&$\begin{pmatrix}
 				e^{-\frac{\pi i}{3}}&0\\0&e^{\frac{\pi i}{3}}
 			\end{pmatrix}$&$\begin{pmatrix}
 				e^{-\frac{\pi i}{3}}&0\\0&e^{\frac{\pi i}{3}}
 			\end{pmatrix}$\\
 			$ \{C_{2x}|0,0,0\} $&$\begin{pmatrix}
 				0&e^{-\frac{2\pi i}{3}}\\e^{-\frac{\pi i}{3}}&0
 			\end{pmatrix}$&$\begin{pmatrix}
 				0&e^{-\frac{2\pi i}{3}}\\e^{-\frac{\pi i}{3}}&0
 			\end{pmatrix}$\\
 			$ \{P|0,0,0\} $&$\begin{pmatrix}
 				1&0\\0&1
 			\end{pmatrix}$&$\begin{pmatrix}
 				-1&0\\0&-1
 			\end{pmatrix}$\\
 			$ \{T|0,0,0\} $&$\begin{pmatrix}
 				0&1\\-1&0
 			\end{pmatrix}\mathcal{K}$&$\begin{pmatrix}
 				0&-1\\1&0
 			\end{pmatrix}\mathcal{K}$\\
 		\end{tabular}
   \end{ruledtabular}
 	\end{table}
        \begin{table}[h!]
		\centering
            \caption{The computed values of parameters $ \{a_i, b_i, c_i, g_i\} $ for Bi$_2$Se$_3$ in Eq. (\ref{Bi2Se3-kp-Zeeman}), obtained from the VASP calculations directly.}
		\begin{tabular}{p{1.9cm}p{1.9cm}p{1.9cm}p{1.9cm}}
                 \hline\hline
                %\begin{align}
                $a_i$ (eV)& $b_i$ (eV$\cdot$\AA) & $c_i$ (eV$\cdot$\AA$^2$)&$g_i$ \\
			\hline
			$a_1=4.89$&$b_1=3.24$&$c_1=19.58$&$g_1=-0.32$\\
                $a_2=-0.22$&$b_2=-2.56$&$c_2=44.47$&$g_2=5.76$\\
                &&$c_3=1.81$&$g_3=-7.90$\\&&$c_4=9.50$&$g_4=-13.01$\\
                %\end{align}
			\hline\hline
		\end{tabular}
		\label{Bi2Se3-kp-par}
	\end{table}

   	\subsection{Four-band model at the Dirac point in \texorpdfstring{Na$_3$Bi}{}}
        The Dirac semimetal Na$_3$Bi has a Dirac point (DP: $\vb*k_D$) along $\Gamma$-A in Fig.~\ref{kp-fig}(c). It is formed by the crossing of the $\rm\overline{DT}7$-irrep and $\rm\overline{DT}8$-irrep bands. Thus, we construct the effective Hamiltonian at $\vb*k_D$. The standard matrix representations are presented in TABLE \ref{Na3Bi-rep1}. The $H^{kp}$ and $H^Z$ are obtained in Eq. (\ref{eq:nabi-kp-Zeeman}) with the computed parameters and $g$-factors in TABLE~\ref{Na3Bi-kp-par}. The four-band $k\cdot p$ model's dispersions are fitting well with the VASP bands, as shown in Fig.~\ref{kp-fig}(d).
        \begin{equation}
            \begin{aligned}
                H^{kp}=&\begin{pmatrix}
                    D_1&0&\xi_-c_2k_-^2&\Theta_{14}\\
                    &D_1&-i\xi_-k_+(b_1+c_4k_z)&\xi_+c_2k_+^2\\
                    &&D_2&0\\ \dagger&&&D_2
                \end{pmatrix},\\
                &D_1=a_1+a_2+(b_2+b_3)k_z+d_1^{+}k_+k_-+d_2^+k_z^2,\\
                &D_2=a_1-a_2+(b_2-b_3)k_z+d_1^-k_+k_-+d_2^-k_z^2,\\
                &d_1^\pm=c_1\pm c_3,\quad d_2^\pm=c_5\pm c_6,\\                &\xi_\pm=1\pm\sqrt{3}i,\quad\Theta_{14}=-i\zeta_+k_-(b_1+c_4k_z),\\
                H^Z=&\frac{\mu_B}{2}\begin{pmatrix}
                h_{1}^{+}B_z&0&0&\frac{3-\sqrt{3}i}{3}g_2B_-\\
                &-h_{1}^{+}B_z&\frac{3+\sqrt{3}i}{3}g_2B_-&0\\
                &&h_{1}^{-}B_z&\frac{6+2\sqrt{3}i}{3}g_1B_+\\ \dagger&&&-h_{1}^{-}B_z
            \end{pmatrix},\\
            &h_{1}^{\pm}=g_3\pm g_4
            \end{aligned}
            \label{eq:nabi-kp-Zeeman}
        \end{equation}

 	\subsection{Four-band model at H in Te}	
    Element tellurium is a narrow-gap semiconductor. The direct gap is at H. The low-energy bands at H are the $\rm\overline{H}4$-irrep and $\rm\overline{H}5$-irrep valence bands and $\rm\overline{H}6$-irrep (doubly degenerate) conduction bands. The four-band effective model is constructed accordingly. The standard matrix representations of the generators of the H-little group are given in TABLE~\ref{Te-rep}. 
    The $k\cdot p$ model at H is expressed as
{\tiny
	\begin{equation}
	\begin{aligned}
		H^{kp}_{11}=&a_{1} + 2 a_{2} + a_{3} + (c_{1}+2c_3+c_8)k_-k_+ \\
        &+ (c_{14} +2 c_{15}+ c_{16})k_{z}^{2},\\
		H^{kp}_{12}=&2 (b_{6} - i b_{7}) k_{z},\\
		H^{kp}_{13}=&-\frac{\sqrt{3}}{6}\left[\left(2i\xi_+b_2+\sqrt{3}i\xi_+ b_3+\sqrt{3}\xi_+ b_4+4 ib_5\right)k_+\right.\\
		&+\left(2i\xi_+ c_{10}+\sqrt{3}i\xi_+c_{11}-\sqrt{3}\xi_+c_{12}-4 ic_{13}\right)k_+k_z\\
        &+\left.\left(2i\xi_+ c_4+\sqrt{3}i\xi_+c_5+\sqrt{3}\xi_+c_6+4 ic_7\right)k_-^2\right],\\
		H^{kp}_{14}=&\frac{\sqrt{3}}{3}\left[\left(2 b_2+\sqrt{3} b_3-\sqrt{3} i b_4+\xi_- b_5\right)k_-\right.\\
		&-\left(2 c_{10}-\sqrt{3}c_{11}+\sqrt{3}i c_{12}-\xi_-c_{13}\right)k_-k_z\\
        &\left.+\left(2c_{4}+\sqrt{3}c_{5}-\sqrt{3}ic_{6}+\xi_-c_{7}\right)k_+^2\right],\\
		H^{kp}_{22}=&a_{1} - 2 a_{2} + a_{3} + (c_{1}-2c_3+c_8)k_-k_+ \\
        &+ (c_{14}- 2 c_{15}+ c_{16})k_{z}^{2},\\
		H^{kp}_{23}=&\frac{\sqrt{3}}{3}\left[\left(-2b_{2}+\sqrt{3}b_{3}+\sqrt{3}i b_{4}-\xi_+b_{5}\right)k_+\right.\\
		&+\left(2c_{4}-\sqrt{3}c_{5}-\sqrt{3}ic_{6}+\xi_+c_{7}\right)k_-^2\\
        &\left.+\left(2 c_{10}+\sqrt{3}c_{11}+\sqrt{3}ic_{12}-\xi_+c_{13}\right)k_+k_z\right],\\
		H^{kp}_{24}=&\frac{\sqrt{3}}{6}\left[\left(2i\xi_-b_{2}-\sqrt{3} i\xi_-b_{3}+\sqrt{3}\xi_-b_{4}+4ib_{5}\right)k_-\right.\\
		&+\left(2i\xi_-c_{10}+\sqrt{3} i\xi_-c_{11}-\sqrt{3}\xi_-c_{12}-4ic_{13}\right)k_-k_z\\
        &\left.+\left(2i\xi_-c_{4}-\sqrt{3}i\zeta_-c_{5}-\sqrt{3}\xi_-c_{6}+4ic_{7}\right)k_+^2\right],\\
		H^{kp}_{33}=&H^{kp}_{44}=a_1-a_3+2 b_8k_z+(c_1-c_8)k_-k_+\\
        &+(c_{14}-c_{16})k_z^2,\\
		H^{kp}_{34}=&\frac{2\sqrt{3}}{3}i\xi_-b_{1}k_++\frac{2\sqrt{3}}{3} i\xi_-c_2k_-^2+2\xi_-c_{9}k_+k_z\\
        &\text{with} \quad\xi_\pm=1\pm\sqrt{3}i
	\end{aligned}
        \label{Te-kp}
        \end{equation}}
	Zeeman's coupling is expressed by
        \begin{equation}
            \begin{aligned}
            H^Z=&\frac{\mu_B}{2}\begin{pmatrix}
                0&(g_6-ig_7)B_z&h_1B_+&h_2B_-\\
                &0&h_3B_+&h_4B_+\\
                &&g_8B_z&\frac{6+2\sqrt{3}i}{3}g_1B_+\\
                \dagger&&&-g_8B_z
            \end{pmatrix}, \\
                h_1=&\frac{3-\sqrt{3}i}{3}g_{2}+\frac{\sqrt{3}-i}{2}g_{3}-\frac{\sqrt{3}i+1}{2}g_{4}-\frac{2\sqrt{3}}{3}i g_{5},\\
                h_2=&\frac{2\sqrt{3}}{3}g_2+g_3-ig_4+\frac{\sqrt{3}-3i}{3}g_5,\\
                h_3=&-\frac{2\sqrt{3}}{3}g_2+g_3+ig_4-\frac{\sqrt{3}+3i}{3}g_5,\\
                h_4=&\frac{3+\sqrt{3}i}{3}g_2-\frac{\sqrt{3}+i}{2}g_3-\frac{\sqrt{3}i-1}{2}g_4+\frac{2\sqrt{3}}{3}ig_5.
            \end{aligned}
            \label{Te-Zeeman}
        \end{equation}
	The computed parameters $ \{a_i, b_i, c_i, g_i\} $ are presented in TABLE \ref{Te-kp-par}. The four-band $k\cdot p$ model's dispersions agree well with the VASP bands in Fig. \ref{kp-fig}(f).
        \begin{table}[h]
        \centering
            \caption{The computed values of parameters $ \{a_i, b_i, c_i, g_i\} $ in Eq. (\ref{eq:nabi-kp-Zeeman}) for Na$_3$Bi, obtained from the VASP calculations directly.}
        \begin{tabular}{p{1.9cm}p{1.9cm}p{1.9cm}p{1.9cm}}
            \hline\hline
            $a_i$ (eV)& $b_i$ (eV$\cdot$\AA) & $c_i$ (eV$\cdot$\AA$^2$)&$g_i$ \\
            \hline
            $a_1=2.22$&$b_1=0.99$&$c_1=7.39$&$g_1=-3.78$\\$a_2=0.00$&$b_2=1.48$&$c_2=-4.02$&$g_2=-5.24$\\&$b_3=-1.69$&$c_3=-0.76$&$g_3=2.74$\\&&$c_4=7.24$&$g_4=8.58$\\&&$c_5=3.17$&\\&&$c_6=-4.43$&\\
            \hline\hline
        \end{tabular}
        \label{Na3Bi-kp-par}
    \end{table}
    
	\begin{table}[h!]
		\centering
              \caption{The computed values of parameters in Eqs. (\ref{Te-kp})-(\ref{Te-Zeeman}) for Te, obtained from the VASP calculations directly.}
		\begin{tabular}{p{1.9cm}p{1.9cm}p{1.9cm}p{1.9cm}}
			\hline\hline
			$a$ (eV)& $b$ (eV$\cdot$\AA) & $c$ (eV$\cdot$\AA$^2$)&$g$ \\
			\hline
			$a_1=5.72$&$b_1=-0.16$&$c_1=6.13$&$g_1=1.52$\\
                $a_2=-0.03$&$b_2=0.84$&$c_2=-0.03$&$g_2=-3.12$\\
                $a_3=-0.13$&$b_3=-1.32$&$c_3=2.51$&$g_3=-2.64$\\
                &$b_4=-1.30$&$c_4=4.16$&$g_4=0.18$\\
                &$b_5=-0.55$&$c_5=1.63$&$g_5=4.08$\\
                &$b_6=0.65$&$c_6=-7.48$&$g_6=-0.08$\\
                &$b_7=-0.98$&$c_7=-1.72$&$g_7=-8.92$\\
                &$b_8=-0.29$&$c_8=-5.96$&$g_8=-10.84$\\
                &&$c_9=1.41$&\\
                &&$c_{10}=-2.56$&\\
                &&$c_{11}=5.52$&\\
                &&$c_{12}=-6.40$&\\
                &&$c_{13}=-1.04$&\\
                &&$c_{14}=8.47$&\\
                &&$c_{15}=-0.09$&\\
                &&$c_{16}=-46.46$&\\
			\hline\hline
		\end{tabular}
		\label{Te-kp-par}
	\end{table}

        \subsection{Two-band model at \texorpdfstring{$\Gamma$}{} in InAs}
\begin{figure}[t!]
 		\includegraphics[width=0.99\linewidth]{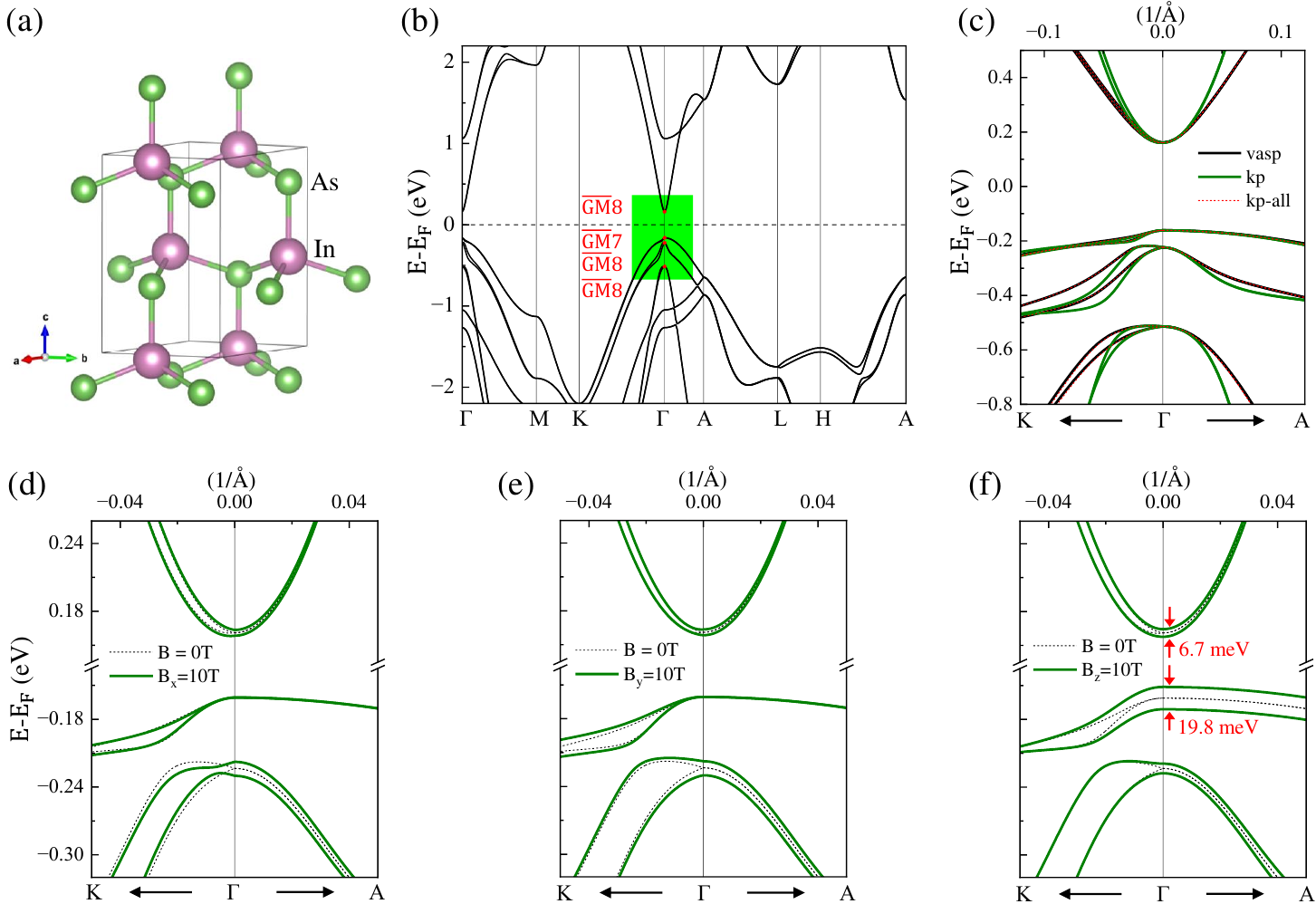}
            \caption{Crystal structure, electronic structure, and $k\cdot p$ band dispersions of InAs. (a) Crystal structure, (b) Electronic band structure (VASP), (c) Model dispersions (kp; Eq. (\ref{2kp-1}) and kp-all; Eq. (\ref{kp-initial})), and (d-f) the Zeeman effect with magnetic field along $\hat{x}$, $\hat{y}$ and $\hat{z}$ respectively. The splitting due to the magnetic field is depicted explicitly. 
            }
 		\label{InAs-fig}
 	\end{figure}

        Here, we compute the band structure of the wurtzite (WZ) InAs semiconductor. Since InAs is usually of $n$-type, we consider the $\overline{\rm GM}8$ conduction bands (doubly degenerate) for the electron doped samples. The matrix representations of generators are presented in TABLE \ref{InAs-rep}. The two-conduction-band effective model is constructed in Eq. (\ref{InAs-conductance}), and the $k\cdot p$ parameters and $g$-factors are computed as listed in TABLE \ref{InAs-kp-c-par}. 
        In Fig.~\ref{InAs-fig}(f), the splitting of the conduction bands at field 10T is 6.7 meV, indicating an effective $g$-factor of $11.6$. 
        It is consistent with the experimental value in the bulk material~\cite{InAs-nature17162, PhysRevLett.119.037701}. 
                \begin{equation}
            \begin{aligned}
                H^{kp}=&a_1+\begin{pmatrix}
                c_1k_+k_-+c_2k_z^2&(1+\sqrt{3}i)b_1k_-\\
                \dagger&c_1k_+k_-+c_2k_z^2
            \end{pmatrix}\\
                H^{Z}=&\frac{\mu_B}{2}\begin{pmatrix}
                g_2B_z&\frac{3-\sqrt{3}i}{3}g_1B_-\\
                \dagger&-g_2B_z
            \end{pmatrix}
            \label{InAs-conductance}
            \end{aligned}
        \end{equation}
        
        In addition, another effective model for the six valence bands is constructed in Appendix~\ref{InAssixband}. Surprisingly, the highest valence bands show a splitting of 19.8 meV under 10T magnetic field in Fig.~\ref{InAs-fig}(f), indicating a remarkable effective $g$-factor, being three times of that of the conduction bands.
 
         \begin{table}[H]
		\centering
		\caption{The computed values of parameters $ \{a_i, b_i, c_i, g_i\} $ in Eq. (\ref{InAs-conductance}) in wurtzite InAs are obtained from the VASP calculations directly.}
		\begin{tabular}{p{1.9cm}p{1.9cm}p{1.9cm}p{1.9cm}}
			\hline\hline
			$a_i$ (eV)& $b_i$ (eV$\cdot$\AA) & $c_i$ (eV$\cdot$\AA$^2$)&$g_i$ \\
			\hline
			$a_1=4.37$&$b_1=-0.20$&$c_1=124.41$&$g_1=-7.66$\\
            &&$c_2=123.40$&$g_2=-11.60$\\
			\hline\hline
		\end{tabular}
		\label{InAs-kp-c-par}
	\end{table}

    \subsection{Two-band model in 1H-TMD monolayers}
\begin{figure}[t!]
 		\includegraphics[width=0.98\linewidth]{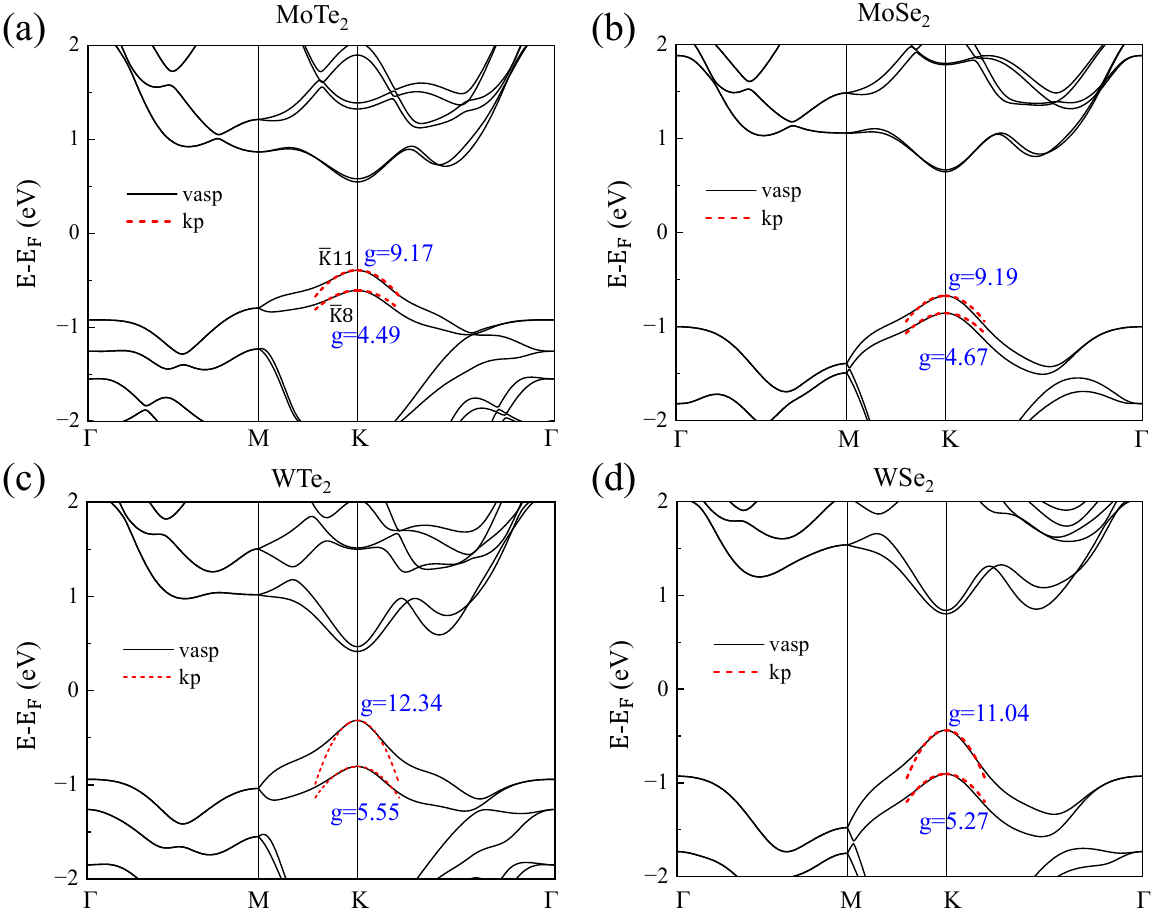}
            \caption{Electronic structures and $k\cdot p$ band dispersions of 1H-TMD monolayers. (a) MoTe$_2$, (b) MoSe$_2$, (c) WTe$_2$, (d) WSe$_2$. 
            }
 		\label{TMD-fig}
 	\end{figure}

    In 1H-phase transition metal chalcogenide (TMD) monolayers, their direct gaps are at K. The two valence bands at K belong to the $\rm\overline{K}8$-irrep  and $\rm\overline{K}11$-irrep, respectively.
    The standard matrix representations of the generators of the K-little group are given in TABLE~\ref{TMD-rep}. The two-band effective models are constructed in Eq. (\ref{TMD-kp-Zeeman}) (to the second order), which are plotted in Fig. \ref{TMD-fig}. The computed $k\cdot p$ parameters and $g$-factors are listed in TABLE~\ref{TMD-kp-par}. 
    \begin{equation}
        \begin{aligned}
        H^{kp}=&a_1+\begin{pmatrix}
		(c_1+c_2)k_-k_+ +a_2&0\\
		0&(c_1-c_2)k_-k_+-a_2
	\end{pmatrix},\\
        H^Z=&\frac{\mu_B}{2}B_z\begin{pmatrix}
            g_1+g_2&0\\
            0&g_1-g_2
        \end{pmatrix}.
        \end{aligned}
        \label{TMD-kp-Zeeman}
    \end{equation}

 \begin{table}[htb]
		\centering
            \caption{The computed values of parameters $ \{a_i, b_i, c_i, g_i\} $ in Eq.~(\ref{TMD-kp-Zeeman}) for 1H-TMD monolayers, obtained from the VASP calculations.}
            \begin{ruledtabular}
		\begin{tabular}{lcccc}
			%\hline\hline
			&MoTe$_2$&MoSe$_2$&WTe$_2$&WSe$_2$\\
			\hline
			$a_1$ (eV)&$-1.92$ &$-2.93$&$-1.79$&$-2.75$\\
            $a_2$ (eV)&$-0.11$ &$-0.09$&$-0.24$&$-0.23$\\
            $c_1$ (eV$\cdot$\AA$^2$)&$-3.83$&$-3.89$&$-8.06$&$-6.48$\\
            $c_2$ (eV$\cdot$\AA$^2$)&$0.60$&$0.47$&$2.75$&$1.73$\\
            $g_1$ &$6.82$&$6.92$&$8.94$&$8.16$\\
            $g_2$ &$-2.34$&$-2.26$&$-3.40$&$-2.88$\\
			%\hline\hline
		\end{tabular}
            \end{ruledtabular}
		\label{TMD-kp-par}
	\end{table}
	%%%%%%%%%%%%%%%%%%%%%%%%%%%%%%%%%%%%%%%%%%%%%

     \begin{table}[t!]
		\centering
            \caption{The computed effective $g$-factors for the two conduction/valence bands only in Bi$_2$Se$_3$, Te, InAs, and 1H-TMD.  The conduction bands of Te have the larger $g$-factor than the valence bands.  The 1st valence (v$_1$) band has the larger $g$-factor than the 2nd valence band (v$_2$) in 1H-TMD.}
            \begin{ruledtabular}
		\begin{tabular}{lcccc}
			%\hline\hline
			&band&$|g_z|$&$|g_\perp|$&experiment\\
			\hline
        Bi$_2$Se$_3$&c&$20.52$&$14.9$&$32$,$23$~\cite{https://doi.org/10.1002/pssb.2220670229}\\
            Te&c&$6.40$&$3.04$&\\
            WZ-InAs&c&$11.60$&$7.66$&$13$~\cite{WZ-InAs-exp}\\
            MoTe$_2$&v$_1$/v$_2$&$9.17$/$4.49$&&\\
            MoSe$_2$&v$_1$/v$_2$&$9.19$/$4.67$&&\\
            WTe$_2$&v$_1$/v$_2$&$12.34$/$5.55$&&\\
            WSe$_2$&v$_1$/v$_2$&$11.04$/$5.27$&&$12.2$~\cite{WSe2-exp}\\
            WSe$_2$&c$_1$/c$_2$&$1.36$/$6.95$&&$1.72$/$7.68$~\cite{WSe2-exp}\\
			%\hline\hline
		\end{tabular}
        \label{expg}
        \end{ruledtabular}
	\end{table}

 \section{Discussion}\label{dis}
 
	In this work, we develop an \emph{open-source} package {\ttfamily VASP2KP} to construct the $ k\cdot p $ Hamiltonian and Zeeman's coupling, and to compute the $ k\cdot p $ parameters and Land\'e $ g $-factors from the VASP calculations  directly. By applying this package in many typical materials, we get $ k\cdot p $ effective models, whose band dispersions are in good agreement with the VASP data around the specific wave vector in the Brillouin zone. 
    As the orbital contribution between the low-energy bands is usually remarkable (due to the small energy difference), the computed $g$-factor is not accurate in the multiple-band model. Thus, we recompute the effective $g$-factor for the two conduction bands only in Bi$_2$Se$_3$ and Te. The recomputed values are mainly consistent with the previous experimental data listed in TABLE~\ref{expg}. 
    We reveal that the conduction bands of Te have the larger $g$-factor than the valence bands, and the 1st valence band has the larger $g$-factor than the 2nd valence band in 1H-TMD monolayers. Our results indicate that {\ttfamily VASP2KP} has the capabilities to handle more materials.

    The obtained models provide the effective masses and $g$-factors, which are important physical quantities of the materials.
    The minor discrepancy from the experimental data is due to the non-precise band gap in the DFT calculations. To improve these parameters, one can use our code in the hybrid functionals or $GW$ calculations. 
    For the new synthesized or predicted materials, for which there is no experimental data available, our code can be used to predict reliable parameters. In conclusion, {\ttfamily VASP2KP}~\cite{vasp2kp.com} would be widely used in the Materials Science.

	%%%%%%%%%%%%%%%%%%%%%%%%%%%%%%%%%%%%%%%%%%%%%
	\section*{Acknowledgements}
 This work was supported by the National Key R\&D Program of Chain (Grant No. 2022YFA1403800),  National Natural Science Foundation of China (Grants No. 11974395, No. 12188101, No. 11925408, No. 12274436, and No. 11921004), the Strategic Priority Research Program of Chinese Academy of Sciences (Grant No. XDB33000000), and the Center for Materials Genome.
Zhi-Da Song was supported by the Innovation Program for Quantum Science and Technology (No. 2021ZD0302403), National Natural Science Foundation of China (General Program No. 12274005), and the National Key Research and Development Program of China (No. 2021YFA1401900).
Hongming Weng and Quansheng Wu were also supported by the Informatization Plan of the Chinese Academy of Sciences (Grant No. CASWX2021SF-0102).

	%%%%%%%%%%%%%%%%%%%%%%%%%%%%%%%%%%%%%%%%%%%%%

 % \paragraph*{Note added.}
 % In the preparation of this work, we notice a similar work to generate the $k\cdot p$ parameters automatically. The python package {\ttfamily DFT2kp}, which was developed by Cassiano \ea\ recently~\cite{cassiano2023dft2kp} at \url{https://gitlab.com/dft2kp/dft2kp}, provides an interface to use the DFT calculation data obtained by Quantum Espresso to obtain the values of the parameters of the $ k\cdot p $ Hamiltonian automatically, but it is not able to generate the Zeeman's coupling and calculate the Land\'e $ g $-factors, which are quite significant when considering magnetic fields.

	\onecolumngrid
	
	\appendix
	\setcounter{equation}{0}
	\renewcommand{\theequation}{A.\arabic{equation}}
    \renewcommand{\theHequation}{A.\arabic{equation}}
    \setcounter{figure}{0}
	\renewcommand{\thefigure}{A\arabic{figure}}
    \renewcommand{\theHfigure}{A\arabic{figure}}
	\section{Löwdin partitioning theory to obtain \texorpdfstring{$ k\cdot p $}{} Hamiltonian}\label{lowdin-app}

\begin{figure}[htbp]
		\centering
		\includegraphics[width=0.35\columnwidth]{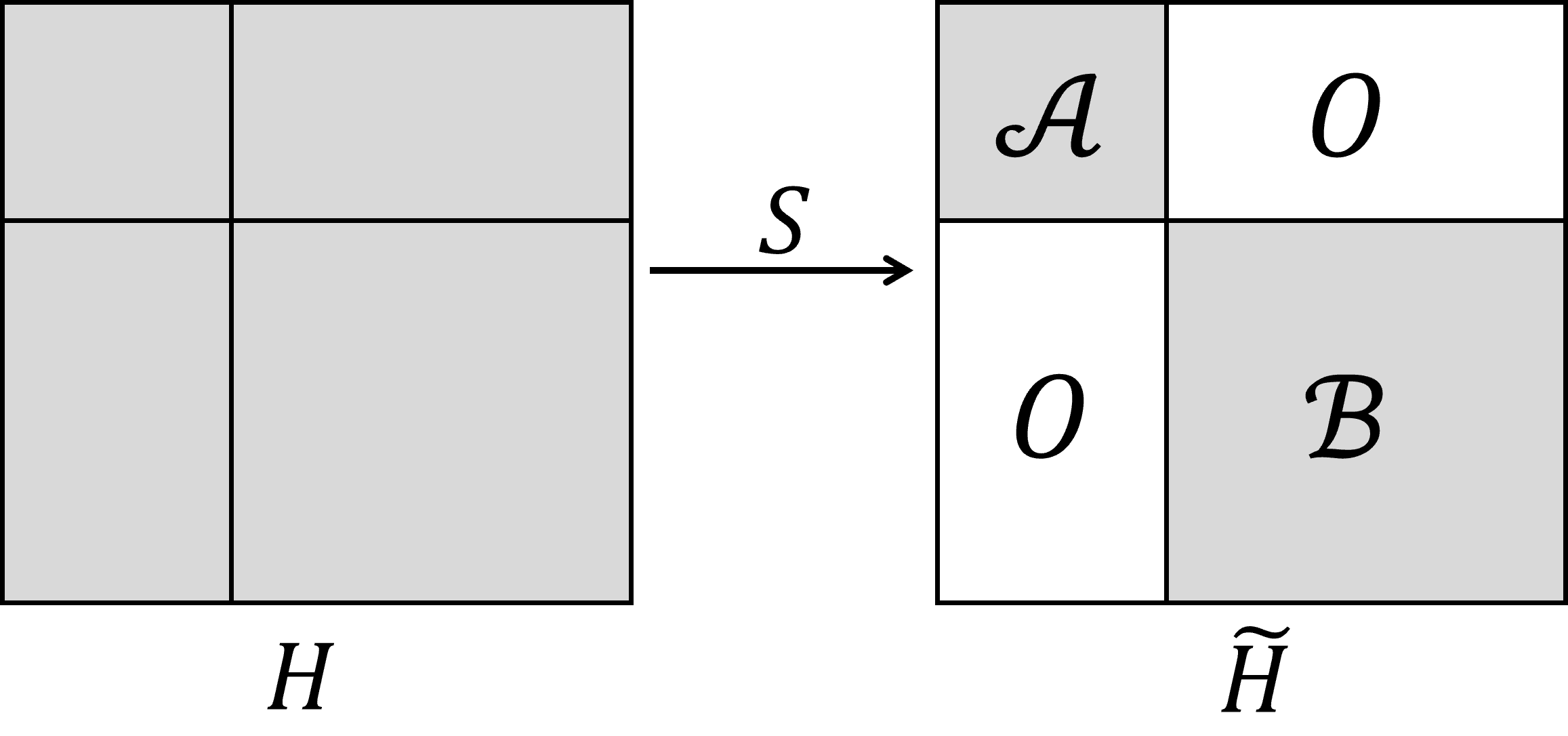}
		\caption{Transform the original Hamiltonian $ H $ to the optimal Hamiltonian $ \widetilde{H} $. The gray parts represent nontrivial matrix elements while the white parts represent trivial matrix elements (zero elements).}
		\label{figB1}
	\end{figure}

	The Löwdin partitioning theory, also called as quasi-degenerate perturbation, is really a useful and important method to make an approximation to simplify the Hamiltonian by reducing the dimension. The main idea of it is to introduce an anti-Hermitian matrix $ S $, which can transform the origin Hamiltonian matrix $ H $ (such as $ k\cdot p $ Hamiltonian $ H^{kp} $ defined by Eq. (\ref{kp-initial})) to the optimal Hamiltonian matrix $ \widetilde{H} $, which is expressed by
	\begin{equation}
		\widetilde{H}=e^{-S}He^S
		\label{eq-b1}
	\end{equation}
	It is obvious that $ \widetilde{H} $ has the same eigenvalues as $ H $, thus making the corresponding bands all the same. Moreover, $ \widetilde{H} $ is expected to be block diagonal, as shown in Fig. \ref{figB1}, where $ \mathcal{A} $ corresponds to the subspace of the bands of interest and $ \mathcal{B} $ corresponds to the subspace of other bands. After this transformation, we can directly use the matrix block corresponding to $ \mathcal{A} $ to replace the original Hamiltonian matrix. However, it is difficult to get the analytic or accurate matrix $ S $, so we must use the perturbation expansion method to find the series solution.

\begin{figure}[htbp]
		\centering
		\includegraphics[width=0.6\columnwidth]{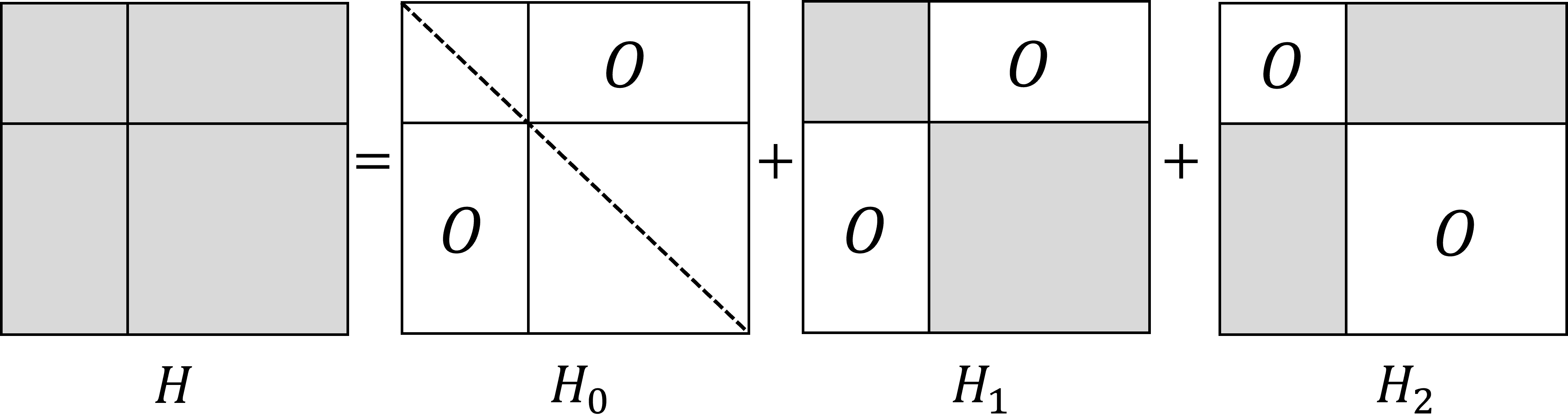}
		\caption{The schematic of $ H $, $ H_0 $, $ H_1 $ and $ H_2 $. The gray parts represent nontrivial matrix elements while the white parts represent trivial matrix elements.}
		\label{figB2}
	\end{figure}

	First, suppose that the Hamiltonian $ H=H_0+H' $, where $ H_0 $ is a diagonal matrix, which is the main part of the Hamiltonian while $ H' $ can be treated as a perturbation. For instance, the $ k\cdot p $ Hamiltonian $ H^{kp} $ can be the sum of the diagonal matrix whose diagonal elements are the eigenvalues $ \epsilon_n({\vb*k}_0) $
	\begin{equation}
		(H_0^{kp})_{mn}=\left(\epsilon_n({\vb*k}_0)+\frac{\hbar^2k^2}{2m}\right)\delta_{mn}
		\label{kp-h0}
	\end{equation}
	and the perturbation terms 
	\begin{equation}
		H'^{kp}_{mn}=\frac{\hbar}{m}{\vb*\pi}_{mn}\cdot{\vb*k}  \ , \ (m \neq n)
		\label{kp-h'}
	\end{equation}
	with small $ {\vb*k} $. Furthermore, $ H' $ can be separated as the sum of $ H_1 $ and $ H_2 $ which only have nonzero elements in and between the subspaces $ \mathcal{A} $ and $ \mathcal{B} $, respectively, as shown in Fig. \ref{figB2}. Therefore, we can rewrite the origin Hamiltonian as
	\begin{equation}
		H=H_0+H_1+H_2
	\end{equation}


	We suppose that the matrix $ S $ is anti-Hermitian and in the same shape as $ H_2 $ (only has nonzero matrix elements between subspaces $ \mathcal{A} $ and $ \mathcal{B} $), making $ e^S $ a unitary matrix. It is simple to find that $ [H_1,S] $ is in the same shape as $ H_2 $ and $ [H_2,S] $ is in the same shape as $ H_1 $, no matter what $ H_1 $ and $ H_2 $ are. According to the Baker–Campbell–Hausdorff formula, Eq. (\ref{eq-b1}) can be rewritten as
	\begin{equation}
		\begin{aligned}
			&\widetilde{H}=\sum_{i=0}^{+\infty}\frac{1}{i!}[H,S]^{(i)}\\
			=&\sum_{i=0}^{+\infty}\frac{1}{(2i)!}[H_0+H_1,S]^{(2i)}+\sum_{i=0}^{+\infty}\frac{1}{(2i+1)!}[H_2,S]^{(2i+1)}+\sum_{i=0}^{+\infty}\frac{1}{(2i+1)!}[H_0+H_1,S]^{(2i+1)}+\sum_{i=0}^{+\infty}\frac{1}{(2i)!}[H_2,S]^{(2i)}
		\end{aligned}
		\label{H-expand}
	\end{equation}
	The sum of first two terms of Eq. (\ref{H-expand}) is block diagonal in the same shape as $ H_1 $, which is denoted by 
	\begin{equation}
		\widetilde{H}_D=\sum_{i=0}^{+\infty}\frac{1}{(2i)!}[H_0+H_1,S]^{(2i)}+\sum_{i=0}^{+\infty}\frac{1}{(2i+1)!}[H_2,S]^{(2i+1)}
		\label{HD}
	\end{equation}
	and the sum of last two terms of Eq. (\ref{H-expand}) is non-block diagonal in the same shape as $ H_2 $, which is denoted by 
	\begin{equation}
		\widetilde{H}_N=\sum_{i=0}^{+\infty}\frac{1}{(2i+1)!}[H_0+H_1,S]^{(2i+1)}+\sum_{i=0}^{+\infty}\frac{1}{(2i)!}[H_2,S]^{(2i)}
	\end{equation}
	To make $ \widetilde{H} $ block diagonal, we have $ \widetilde{H}_{N}=0 $. Use the ansatz that $ S $ can be expanded as 
	\begin{equation}
		S=S^{(1)}+S^{(2)}+S^{(3)}+\cdots
	\end{equation}
	
	Extract the small quantities of each order in $ H_N $ and let them be 0:
	
	1st order
	\begin{equation}
		[H_0,S^{(1)}]+H_2=0
		\label{1-order}
	\end{equation}
	
	2nd order
	\begin{equation}
		[H_0,S^{(2)}]+[H_1,S^{(1)}]=0
		\label{2-order}
	\end{equation}
	
	3rd order
	\begin{equation}
		[H_0,S^{(3)}]+\frac{1}{6}[H_0,S^{(1)}]^{(3)}+[H_1,S^{(2)}]+\frac{1}{2}[H_2,S^{(1)}]^{(2)}=0
		\label{3-order}
	\end{equation}
	
	$ \cdots $
	
	By solving Eqs. (\ref{1-order})-(\ref{3-order}), the matrix elements of $ S^{(i)} $ can be written as
	\begin{equation}
		\left\{
		\begin{aligned}
			&S^{(1)}_{\alpha l}=-\frac{H'_{\alpha l}}{E_\alpha-E_l}\\
			&S_{\alpha l}^{(2)}= \frac{1}{E_\alpha-E_l}\left[\sum_{\alpha^{\prime}\in\mathcal{A}} \frac{H'_{\alpha \alpha^{\prime}} H'_{\alpha^{\prime} l}}{E_{\alpha^{\prime}}-E_l}-\sum_{l^{\prime}\in\mathcal{B}} \frac{H'_{\alpha l^{\prime}} H'_{l^{\prime} l}}{E_\alpha-E_{l^{\prime}}}\right]\\
			&\begin{aligned}
				S_{\alpha l}^{(3)}= & \frac{1}{E_\alpha-E_l} \\
				\times & \left[-\sum_{\alpha^{\prime}, \alpha^{\prime \prime}\in\mathcal{A}} \frac{H'_{\alpha \alpha^{\prime \prime}} H'_{\alpha^{\prime \prime} \alpha^{\prime}} H'_{\alpha^{\prime} l}}{\left(E_{\alpha^{\prime \prime}}-E_l\right)\left(E_{\alpha^{\prime}}-E_l\right)}-\sum_{l^{\prime}, l^{\prime \prime}\in\mathcal{B}} \frac{H'_{\alpha l^{\prime}} H'_{l^{\prime} l^{\prime \prime}} H'_{l^{\prime \prime} l}}{\left(E_\alpha-E_{l^{\prime \prime}}\right)\left(E_\alpha-E_{l^{\prime}}\right)}\right. \\
				& +\sum_{l^{\prime}\in\mathcal{B}, \alpha^{\prime}\in\mathcal{A}} \frac{H'_{\alpha \alpha^{\prime}} H'_{\alpha^{\prime} l^{\prime}} H'_{l^{\prime} l}}{\left(E_{\alpha^{\prime}}-E_l\right)\left(E_{\alpha^{\prime}}-E_{l^{\prime}}\right)}+\sum_{l^{\prime}\in\mathcal{B}, \alpha^{\prime}\in\mathcal{A}} \frac{H'_{\alpha \alpha^{\prime}} H'_{\alpha^{\prime} l^{\prime}} H'_{l^{\prime} l}}{\left(E_m-E_{l^{\prime}}\right)\left(E_{\alpha^{\prime}}-E_{l^{\prime}}\right)} \\
				& +\frac{1}{3} \sum_{l^{\prime}\in\mathcal{B}, \alpha^{\prime}\in\mathcal{A}} \frac{H'_{\alpha l^{\prime}} H'_{l^{\prime} \alpha^{\prime}} H'_{\alpha^{\prime} l}}{\left(E_{\alpha^{\prime}}-E_{l^{\prime}}\right)\left(E_{\alpha^{\prime}}-E_l\right)}+\frac{1}{3} \sum_{l^{\prime}\in\mathcal{B}, \alpha^{\prime}\in\mathcal{A}} \frac{H'_{\alpha l^{\prime}} H'_{l^{\prime} \alpha^{\prime}} H'_{\alpha^{\prime} l}}{\left(E_\alpha-E_{l^{\prime}}\right)\left(E_{\alpha^{\prime}}-E_{l^{\prime}}\right)}\left.+\frac{2}{3} \sum_{l^{\prime}\in\mathcal{B}, \alpha^{\prime}\in\mathcal{A}} \frac{H'_{\alpha l^{\prime}} H'_{l^{\prime} \alpha^{\prime}} H'_{\alpha^{\prime} l}}{\left(E_\alpha-E_{l^{\prime}}\right)\left(E_{\alpha^{\prime}}-E_l\right)}\right]
			\end{aligned}\\
			&\cdots=\cdots
		\end{aligned}
		\right.
	\end{equation}
	where $ \alpha\in\mathcal{A} $ and $ l\in\mathcal{B} $. This matrix $ S $ makes $ \widetilde{H}_N=0 $ so that $ \widetilde{H}=\widetilde{H}_D $. After obtaining the matrix $ S $, we can directly obtain the optimal Hamiltonian by Eq. (\ref{HD}), which is expressed by
	\begin{equation}
		\widetilde{H}=\widetilde{H}^{(0)}+\widetilde{H}^{(1)}+\widetilde{H}^{(2)}+\widetilde{H}^{(3)}+\cdots
	\end{equation}
	where
	\begin{equation}
		\left\{
		\begin{aligned}
			H_{\alpha \alpha^{\prime}}^{(0)}= & H_{\alpha \alpha^{\prime}}^0 \\
			H_{\alpha \alpha^{\prime}}^{(1)}= & H_{\alpha \alpha^{\prime}}^{\prime} \\
			H_{\alpha \alpha^{\prime}}^{(2)}= & \frac{1}{2} \sum_{l\in\mathcal{B}} H_{\alpha l}^{\prime} H_{l \alpha^{\prime}}^{\prime}\left[\frac{1}{E_\alpha-E_l}+\frac{1}{E_{\alpha^{\prime}}-E_l}\right] \\
			H_{\alpha \alpha^{\prime}}^{(3)}= & -\frac{1}{2} \sum_{l\in\mathcal{B}, \alpha^{\prime \prime}\in\mathcal{A}}\left[\frac{H_{\alpha l}^{\prime} H_{l \alpha^{\prime \prime}}^{\prime} H_{\alpha^{\prime \prime} \alpha^{\prime}}^{\prime}}{\left(E_{\alpha^{\prime}}-E_l\right)\left(E_{\alpha^{\prime \prime}}-E_l\right)}+\frac{H_{\alpha \alpha^{\prime \prime}}^{\prime} H_{\alpha^{\prime \prime} l}^{\prime} H_{l \alpha^{\prime}}^{\prime}}{\left(E_\alpha-E_l\right)\left(E_{\alpha^{\prime \prime}}-E_l\right)}\right] \\
			& +\frac{1}{2} \sum_{l, l^{\prime}\in\mathcal{B}} H_{\alpha l}^{\prime} H_{l l^{\prime}}^{\prime} H_{l^{\prime} \alpha^{\prime}}^{\prime}\left[\frac{1}{\left(E_\alpha-E_l\right)\left(E_\alpha-E_{l^{\prime}}\right)}+\frac{1}{\left(E_{\alpha^{\prime}}-E_l\right)\left(E_{\alpha^{\prime}}-E_{l^{\prime}}\right)}\right]\\
			\cdots=\cdots
		\end{aligned}\right.
		\label{lowdin-origin}
	\end{equation}
	
	Up to now, we have already obtained the optimal Hamiltonian. In the case of $ k\cdot p $ Hamiltonian, by substituting Eq. (\ref{kp-h0}) and Eq. (\ref{kp-h'}) into Eq. (\ref{lowdin-origin}), we can obtain
	\begin{equation}
		\left\{
		\begin{aligned}
			&H^{kp(0)}_{\alpha\beta}=\left(\epsilon_\alpha({\vb*k}_0)+\frac{\hbar^2k^2}{2m}\right)\delta_{\alpha\beta}\\
			&H^{kp(1)}_{\alpha\beta}=\frac{\hbar}{m}{\vb*\pi}_{\alpha\beta}\cdot{\vb*k}\\
			&H^{kp(2)}_{\alpha\beta}=\frac{\hbar^2}{2m^2}\sum_{l\in \mathcal{B}}\sum_{ij}\left[\frac{1}{\epsilon_\alpha({\vb* k}_0)-\epsilon_l({\vb* k}_0)}+\frac{1}{\epsilon_\beta({\vb* k}_0)-\epsilon_l({\vb* k}_0)}\right]\pi^{i}_{\alpha l}\pi^{j}_{l\beta}k^ik^j\\
			&\begin{aligned}
				H^{kp(3)}_{\alpha\beta}=-\frac{\hbar^3}{2m^3}\sum_{ijq}\left\{\sum_{l\in\mathcal{B},\gamma\in\mathcal{A}}\left[\frac{\pi^i_{\alpha l}\pi^j_{l\gamma}\pi^q_{\gamma\beta}}{(\epsilon_\beta({\vb* k}_0)-\epsilon_l({\vb* k}_0))(\epsilon_\gamma({\vb* k}_0)-\epsilon_l({\vb* k}_0))}+\frac{\pi^i_{\alpha \gamma}\pi^j_{\gamma l}\pi^q_{l\beta}}{(\epsilon_\alpha({\vb* k}_0)-\epsilon_l({\vb* k}_0))(\epsilon_\gamma({\vb* k}_0)-\epsilon_l({\vb* k}_0))}\right]\right.\\
				\left.-\sum_{l,l'\in\mathcal{B}}\left[\frac{1}{(\epsilon_\alpha({\vb* k}_0)-\epsilon_l({\vb* k}_0))(\epsilon_\alpha({\vb* k}_0)-\epsilon_{l'}({\vb* k}_0))}+\frac{1}{(\epsilon_\beta({\vb* k}_0)-\epsilon_l({\vb* k}_0))(\epsilon_\beta({\vb* k}_0)-\epsilon_{l'}({\vb* k}_0))}\right]\pi^i_{\alpha l}\pi^{j}_{ll'}\pi^{q}_{l'\beta}\right\}k^ik^jk^q
			\end{aligned}
		\end{aligned}\right.
	\end{equation}
	Therefore, the $ k\cdot p $ Hamiltonian of order 2 is the sum of $ H^{kp(0)} $, $ H^{kp(1)} $ and $ H^{kp(2)} $, which is equal to Eq. (\ref{eq-kp-origin}). Moreover, the $ k\cdot p $ Hamiltonian of order 3 can be expressed by
	\begin{equation}
		\begin{aligned}
			&H^{kp(\leq3)}_{\alpha\beta}\\
			=&\left(\epsilon_\alpha({\vb*k}_0)+\frac{\hbar^2k^2}{2m}\right)\delta_{\alpha\beta}+\frac{\hbar}{m}{\vb*\pi}_{\alpha\beta}\cdot{\vb*k}+\frac{\hbar^2}{2m^2}\sum_{l\in \mathcal{B}}\sum_{ij}\left[\frac{1}{\epsilon_\alpha({\vb* k}_0)-\epsilon_l({\vb* k}_0)}+\frac{1}{\epsilon_\beta({\vb* k}_0)-\epsilon_l({\vb* k}_0)}\right]\pi^{i}_{\alpha l}\pi^{j}_{l\beta}k^ik^j\\
			&-\frac{\hbar^3}{2m^3}\sum_{ijq}\left\{\sum_{l\in\mathcal{B},\gamma\in\mathcal{A}}\left[\frac{\pi^i_{\alpha l}\pi^j_{l\gamma}\pi^q_{\gamma\beta}}{(\epsilon_\beta({\vb* k}_0)-\epsilon_l({\vb* k}_0))(\epsilon_\gamma({\vb* k}_0)-\epsilon_l({\vb* k}_0))}+\frac{\pi^i_{\alpha \gamma}\pi^j_{\gamma l}\pi^q_{l\beta}}{(\epsilon_\alpha({\vb* k}_0)-\epsilon_l({\vb* k}_0))(\epsilon_\gamma({\vb* k}_0)-\epsilon_l({\vb* k}_0))}\right]\right.\\
			&\qquad\quad\left.-\sum_{l,l'\in\mathcal{B}}\left[\frac{1}{(\epsilon_\alpha({\vb* k}_0)-\epsilon_l({\vb* k}_0))(\epsilon_\alpha({\vb* k}_0)-\epsilon_{l'}({\vb* k}_0))}+\frac{1}{(\epsilon_\beta({\vb* k}_0)-\epsilon_l({\vb* k}_0))(\epsilon_\beta({\vb* k}_0)-\epsilon_{l'}({\vb* k}_0))}\right]\pi^i_{\alpha l}\pi^{j}_{ll'}\pi^{q}_{l'\beta}\right\}k^ik^jk^q
		\end{aligned}
	\end{equation}
	
	\setcounter{equation}{0}
	\renewcommand{\theequation}{B.\arabic{equation}}
    \renewcommand{\theHequation}{B.\arabic{equation}}
    \section{Derivation of Zeeman's coupling}\label{Zeeman-app}
	It is easy to compute the commutator $[\partial^i,A^j]$ as
	\begin{equation}
		[\partial^i,A^j]\phi=\partial^i(A^j\phi)-A^j\partial^i\phi=(\partial^iA^j)\phi+A^j\partial^i\phi-A^j\partial^i\phi=(\partial^iA^j)\phi
	\end{equation}
	or in a simple form
	\begin{equation}
		[\partial^i,A^j]=\partial^iA^j.
	\end{equation}
	In addition, the components of the magnetic field can be expressed by
	\begin{equation}
		B_k=(\nabla\times \vb*A)\cdot {\vb*e}^k=\sum_{lmn}\epsilon_{lmn}\partial^mA^n{\vb*e^l}\cdot {\vb*e}^k=\sum_{lmn}\epsilon_{lmn}\partial^mA^n\delta_{lk}=\sum_{mn}\epsilon_{kmn}\partial^mA^n
	\end{equation}
	where $\epsilon_{lmn}$ is the Levi-Civita symbol and $A$ is the magnetic vector potential. Therefore, we can establish the relation that
	\begin{equation}
		\begin{aligned}
			\sum_{k}\epsilon^{ijk}B_k&=\sum_{mnk}\epsilon^{ijk}\epsilon_{kmn}\partial^mA^n=\sum_{mn}\left(\sum_{k}\epsilon^{ijk}\epsilon_{kmn}\right)\partial^mA^n=\sum_{mn}(\delta_{im}\delta_{jn}-\delta_{in}\delta_{jm})\partial^mA^n=\partial^iA^j-\partial^jA^i
		\end{aligned}
	\end{equation}
	Therefore, we can obtain the relation in Sec. \ref{th.Zeeman} that
	\begin{equation}
		\begin{aligned}
			\left[-i\hbar\partial^i+eA^i,-i\hbar\partial^j+eA^j\right]=-ie\hbar(\left[\partial^i,A^j\right]-\left[\partial^j,A^i\right])=-i\hbar e(\partial^iA^j-\partial^jA^i)=-i\hbar e\sum_{k}\epsilon^{ijk}B_k
		\end{aligned}
	\end{equation}
	Furthermore, after replacing $k^i$ by  $-i\hbar\partial^i+eA^i$ (Peierls substitution) in Eq. (\ref{eq-kp-origin}), the last summation can be transformed as
	\begin{equation}
		\begin{aligned}
			&\sum_{ij}\pi^i_{\alpha l}\pi^j_{l\beta}(-i\hbar\partial^i+eA^i)(-i\hbar\partial^j+eA^j)\\
			=&\sum_{ij}\pi^i_{\alpha l}\pi^j_{l\beta}\left(\frac{1}{2}\left[-i\hbar\partial^i+eA^i,-i\hbar\partial^j+eA^j\right]+\frac{1}{2}\left\{-i\hbar\partial^i+eA^i,-i\hbar\partial^j+eA^j\right\}\right)\\
			=&-\frac{i\hbar e}{2}\sum_{ijk}\pi^i_{\alpha l}\pi^j_{l\beta}\epsilon^{ijk}B_k+\frac{\hbar^2}{2}\sum_{ij}\pi^i_{\alpha l}\pi^j_{l\beta}\left(-i\partial^i+\frac{e}{\hbar}A^i\right)\left(-i\partial^j+\frac{e}{\hbar}A^j\right)+\frac{\hbar^2}{2}\sum_{ij}\pi^i_{\alpha l}\pi^j_{l\beta}\left(-i\partial^j+\frac{e}{\hbar}A^j\right)\left(-i\partial^i+\frac{e}{\hbar}A^i\right)\\
			=&-\frac{i\hbar e}{2}\sum_{ijk}\pi^i_{\alpha l}\pi^j_{l\beta}\epsilon^{ijk}B_k+\frac{\hbar^2}{2}\sum_{ij}\pi^i_{\alpha l}\pi^j_{l\beta}\left(-i\partial^i+\frac{e}{\hbar}A^i\right)\left(-i\partial^j+\frac{e}{\hbar}A^j\right)+\frac{\hbar^2}{2}\sum_{ij}\pi^j_{\alpha l}\pi^i_{l\beta}\left(-i\partial^i+\frac{e}{\hbar}A^i\right)\left(-i\partial^j+\frac{e}{\hbar}A^j\right)\\
			=&-\frac{i\hbar e}{2}\sum_{ijk}\pi^i_{\alpha l}\pi^j_{l\beta}\epsilon^{ijk}B_k+\hbar^2\sum_{ij}\frac{\pi^i_{\alpha l}\pi^j_{l\beta}+\pi^j_{\alpha l}\pi^i_{l\beta}}{2}\left(-i\partial^i+\frac{e}{\hbar}A^i\right)\left(-i\partial^j+\frac{e}{\hbar}A^j\right)
		\end{aligned}
	\end{equation}
	The Hamiltonian of Zeeman's coupling is the gauge independent part in Eq. (\ref{eq-kp-origin}) (after Peierls substitution), which is expressed as
	\begin{equation}
		H^{Z}_{\alpha\beta}=\frac{\mu_B}{\hbar}\left({\vb *L}_{\alpha\beta}+2{\vb *s}_{\alpha\beta}\right)\cdot{\vb *B}
		\label{Zeeman-ap}
	\end{equation}
	where
	\begin{equation}
			L_{\alpha\beta}^{k}=- \frac{i\hbar}{2m}\sum_{l\in\mathcal{B}}\sum_{ij}\epsilon^{ijk}\pi^{i}_{\alpha l}\pi^{j}_{l\beta}\left(\frac{1}{\epsilon_\alpha({\vb *k}_0)-\epsilon_l({\vb *k}_0)}+\frac{1}{\epsilon_\beta({\vb *k}_0)-\epsilon_l({\vb *k}_0)}\right)
	\end{equation}
	and $\frac{2}{\hbar}\mu_B{\vb*s}\cdot{\vb*B}$ is the Zeemans's coupling of the bare electron.
	In addition, the gauge dependent part in Eq. (\ref{eq-kp-origin}) (after Peierls substitution) is expressed by
	\begin{equation}
		\begin{aligned}
		H^{kp}_{\alpha\beta}=\epsilon_\alpha({\vb* k}_0)\delta_{\alpha\beta}+\frac{\hbar}{m}{\vb*\pi}_{\alpha\beta}\cdot\left(-i\nabla+\frac{e}{\hbar}{\vb* A}\right)+\sum_{ij}M_{\alpha\beta}^{ij}\left(-i\partial^i+\frac{e}{\hbar}A^i\right)\left(-i\partial^j+\frac{e}{\hbar}A^j\right)
        \end{aligned}
	\end{equation}
	where
	\begin{equation}
		\begin{aligned}
	M_{\alpha\beta}^{ij}=\frac{\hbar^2}{2m}\delta_{\alpha\beta}\delta_{ij}+\frac{\hbar^2}{4m^2}\sum_{l\in\mathcal{B}}\left(\pi^{i}_{\alpha l}\pi^{j}_{l\beta}+\pi^{j}_{\alpha l}\pi^{i}_{l\beta}\right)
     \left(\frac{1}{\epsilon_\alpha({\vb *k}_0)-\epsilon_l({\vb *k}_0)}+\frac{1}{\epsilon_\beta({\vb *k}_0)-\epsilon_l({\vb *k}_0)}\right)
		\end{aligned}.
		\label{Zeeman2-ap}
	\end{equation}
	The Eqs. (\ref{Zeeman-ap})-(\ref{Zeeman2-ap}) are the same as Eqs. (\ref{2kp-1})-(\ref{Zeeman-L}).

    \setcounter{equation}{0}
	\renewcommand{\theequation}{C.\arabic{equation}}
    \renewcommand{\theHequation}{C.\arabic{equation}}
	\section{Construction of the coefficient matrix \texorpdfstring{$ Q $}{} for finding the unitary transformation \texorpdfstring{$ U $}{}}\label{Q-app}
	Only the generators of the group $L$ should be taken into account when finding the unitary transformation matrix $U$. $T$ is an anti-unitary generator, while $S$ are the unitary generators.
   The Eq. (\ref{U-all}) can be rewritten as 
	\begin{equation}
		D^{\text{num}}(S)U-UD^{\text{std}}(S)=\mathcal{O}
		\label{U-unitary}
	\end{equation}
	
	\begin{equation}
		D^{\text{num}}(T)U^*-UD^{\text{std}}(T)=\mathcal{O}
		\label{U-anti-unitary}
	\end{equation}
	where $ \mathcal{O} $ is a zero matrix. The matrices $ U $, $ D^{\text{std}}(R) $ and $ D^{\text{num}}(R) $ are complex, so we can consider the real parts and the imaginary parts separately, thus transforming Eqs. (\ref{U-unitary})-(\ref{U-anti-unitary}) into
	\begin{equation}
		\left\{
		\begin{aligned}
			D^{\text{num}}_r(S)U_r-U_rD^{\text{std}}_r(S)-D^{\text{num}}_i(S)U_i+U_iD^{\text{std}}_i(S)=\mathcal{O}\\
			D^{\text{num}}_i(S)U_r-U_rD^{\text{std}}_i(S)+D^{\text{num}}_r(S)U_i-U_iD^{\text{std}}_r(S)=\mathcal{O}
		\end{aligned}\right.
		\label{U-unitary-1}
	\end{equation}
	and
	\begin{equation}
		\left\{
		\begin{aligned}
			D^{\text{num}}_r(T)U_r-U_rD^{\text{std}}_r(T)+D^{\text{num}}_i(T)U_i+U_iD^{\text{std}}_i(T)=\mathcal{O}\\
			-D^{\text{num}}_i(T)U_r+U_rD^{\text{std}}_i(T)+D^{\text{num}}_r(T)U_i+U_iD^{\text{std}}_r(S)=\mathcal{O}
		\end{aligned}\right.
		\label{U-anti-unitary-1}
	\end{equation}
	where the subscripts $ r $ represent the real parts of $ U $, $ D^{\text{std}}(R) $ or $ D^{\text{num}}(R) $ and the subscripts $ i $ represent the imaginary parts. Consider the real part and the imaginary part of each elements of $ U $ as independent variables, which are denoted as $ U_{r11}, U_{r12}, \cdots U_{rnn} $ and $ U_{i11}, U_{i12}, \cdots U_{inn} $, respectively. From Eqs. (\ref{U-unitary-1})-(\ref{U-anti-unitary-1}), it is clear to find that the matrix equations are all linear equations and all the parameters and variables are real. Introduced a column vector $ {\vb* u}=(U_{r11}, U_{r12},\cdots,U_{rnn},U_{i11},U_{i12},\cdots,U_{inn})^T $, which is comprised of all the independent variables to be solved, Eqs. (\ref{U-unitary-1})-(\ref{U-anti-unitary-1}) can be rewritten as
	\begin{equation}
		A(S){\vb* u}=\vb*0
		\label{m-unitary}
	\end{equation}
	where
	\begin{equation}
		\begin{aligned}
			A(S)=
			\left(\begin{array}{cc}
				D_r^{\text{num}}(S)\otimes I-I\otimes D_r^{\text{std}T}(S)&-D_i^{\text{num}}(S)\otimes I+I\otimes D_i^{\text{std}T}(S)\\
				D_i^{\text{num}}(S)\otimes I-I\otimes D_i^{\text{std}T}(S)&D_r^{\text{num}}(S)\otimes I-I\otimes D_r^{\text{std}T}(S)
			\end{array}\right)
		\end{aligned}
		\label{m-anti-unitary}
	\end{equation}
	and
	\begin{equation}
		B(T){\vb* u}=\vb*0
	\end{equation}
	where
	\begin{equation}
		\begin{aligned}
			B(T)=\left(\begin{array}{cc}
				D_r^{\text{num}}(T)\otimes I-I\otimes D_r^{\text{std}T}(T)&D_i^{\text{num}}(T)\otimes I+I\otimes D_i^{\text{std}T}(T)\\
				-D_i^{\text{num}}(T)\otimes I+I\otimes D_i^{\text{std}T}(T)&D_r^{\text{num}}(T)\otimes I+I\otimes D_r^{\text{std}T}(T)
			\end{array}\right).
		\end{aligned}
	\end{equation}
	
	The Eq. (\ref{m-unitary}) holds for all unitary generators $ S_1,S_2,\cdots,S_n $. Combined Eq. (\ref{m-unitary}) with Eq. (\ref{m-anti-unitary}), we can construct the large parameter matrix $ Q $ so that the vector $ {\vb*u} $ corresponds to the transformation matrix $ U $ satisfies $ Q{\vb*u}=\vb*0 $, where 
	\begin{equation}
		Q=(A^T(S_1),A^T(S_2),\cdots,A^T(S_n),B^T(T))^T.
		\label{Q-define}
	\end{equation}
	
	Therefore, all vectors in the null space of the matrix $ Q $ is the solution to Eq. (\ref{eq-transform}).
	
	\setcounter{table}{0}
	\renewcommand{\thetable}{D\arabic{table}}
    \renewcommand{\theHtable}{D\arabic{table}}
	\section{Functions of {\ttfamily vasp2mat}}\label{vasp_kp_all}
	Except for the calculation of matrices of generalized momentum $ \hat{\vb* \pi}=\hat{\vb* p}+\frac{1}{2mc^2}\left(\hat{{\vb* s}}\times\nabla V({\vb* r})\right) $, spin $ \hat{\vb*s} $, time reversal operator $ \hat{T} $ and crystalline symmetry operators $\hat{R}$, the patch {\ttfamily vasp2mat} can also do other calculations by setting the parameter {\ttfamily vmat} in {\ttfamily INCAR.mat}. All functions of {\ttfamily vasp2mat} are shown in TABLE \ref{vmattable}, where $ \hat{\vb*\sigma} $ is Pauli operator. The matrix of time reversal operator $ \hat{T} $ can be calculated with {\ttfamily vmat=12; time\_rev=.true.}.
	
	\begin{table}[H]
        \centering
        \caption{The function of different {\ttfamily vmat} of {\ttfamily vasp2mat}.}
		\begin{tabular}{|c|c|}
			\hline
			vmat&Functions\\
			\hline
			1&Calculate overlap matrix $ \braket{m(\vb*K)}{n(\vb*K)} $\\
			\hline
			2&Calculate soft local potential matrix $ \mel{\widetilde{m}(\vb*K)}{\hat{V}_{\text{eff}}}{\widetilde{n}(\vb*K)} $\\
			\hline
			3&Calculate kinetic energy matrix of pseudo wavefunctions $ \mel{\widetilde{m}(\vb*K)}{\hat{T}_k}{\widetilde{n}(\vb*K)} $\\
			\hline
			4&Calculate nonlocal potential matrix $ \mel{\widetilde{m}(\vb*K)}{\hat{V}_{NL}}{\widetilde{n}(\vb*K)} $\\
			\hline
			5&Calculate Hamiltonian matrix $ \mel{m(\vb*K)}{\hat{H}}{n(\vb*K)} $\\
			\hline
			7&Calculate momentum matrices $ \mel{m(\vb*K)}{\hat{{\vb* p}}}{n(\vb*K)} $\\
			\hline
			8&Calculate SOC Hamiltonian matrix $ \mel{m(\vb*K)}{\hat{H}_{SOC}}{n(\vb*K)} $\\
			\hline
			10&Calculate spin matrices $ \mel{m(\vb*K)}{\hat{{\vb* \sigma}}}{n(\vb*K)} $\\
			\hline
			11&Calculate generalized momentum matrices $ \mel{m(\vb*K)}{\hat{{\vb* \pi}}}{n(\vb*K)} $\\
			\hline
			12&Calculate matrix representation of a symmetry operator $ \mel{m(\vb*K)}{\hat{R}}{n(\vb*K)} $\\
			\hline
			13&Calculate Berry curvature, anomalous Hall conductance and spin Hall conductance\\
			\hline
			14&Calculate Wilson loops to obtain Berry phases\\
			\hline
		\end{tabular}
        \label{vmattable}
	\end{table}

\clearpage
        \setcounter{table}{0}
	\renewcommand{\thetable}{E\arabic{table}}
    \renewcommand{\theHtable}{E\arabic{table}}
        \setcounter{equation}{0}
	\renewcommand{\theequation}{E.\arabic{equation}}
    \renewcommand{\theHequation}{E.\arabic{equation}}
        \section{The standard matrix representations}

    \begin{table}[h!]
         \centering
         \caption{In Na$_3$Bi, the matrix representations of ${\rm\overline{DT}}7$ and $\rm{\overline{DT}}8$ irreps at $\vb*k_D$ (0 0 $w$) are given on BCS server, \href{https://www.cryst.ehu.es/cgi-bin/cryst/programs/corepresentations_out.pl?super=194.264&vecfinal=DT}{https://www.cryst.ehu.es/cgi-bin/cryst/programs/corepresentations\_out.pl?super=194.264\&vecfinal=DT}.}
         \label{Na3Bi-rep1}
         \begin{tabular}{|c|c|c|}
             \hline
             \diagbox{$ R $}{$ D^{\text{std}}(R) $}{irrep}&$ \overline{\text{DT}}7 $&$ \overline{\text{DT}}8 $\\
             \hline
             $ \{C_{3z}|0,0,0\} $&$\begin{pmatrix}
                 -1&0\\0&-1
             \end{pmatrix}$&$\begin{pmatrix}
                 e^{-\frac{\pi i}{3}}&0\\0&e^{\frac{\pi i}{3}}
             \end{pmatrix}$\\
             \hline
             $ \{C_{2z}|0,0,\frac{1}{2}\} $&$\begin{pmatrix}
                 -ie^{i\pi w}&0\\0&ie^{i\pi w}
             \end{pmatrix}$&$\begin{pmatrix}
                 -ie^{i\pi w}&0\\0&ie^{i\pi w}
             \end{pmatrix}$\\
             \hline
             $ \{M_x|0,0,0\} $&$\begin{pmatrix}
                 0&1\\-1&0
             \end{pmatrix}$&$\begin{pmatrix}
                 0&e^{-\frac{2\pi i}{3}}\\e^{-\frac{\pi i}{3}}&0
             \end{pmatrix}$\\
             \hline
             $ \{TP|0,0,0\} $&$\begin{pmatrix}
                 0&1\\-1&0
             \end{pmatrix}\mathcal{K}$&$\begin{pmatrix}
                 0&1\\-1&0
             \end{pmatrix}\mathcal{K}$\\
             \hline
         \end{tabular}
     \end{table}

    % \begin{table}[h!]
    %    \centering
    %    \caption{In Na$_3$Bi, the matrix representations of $\overline{\text{GM}}8$ and $\overline{\text{GM}}10$ irreps at $\Gamma$ are given on BCS server, \href{https://www.cryst.ehu.es/cgi-bin/cryst/programs/corepresentations_out.pl?super=194.264&vecfinal=GM}{https://www.cryst.ehu.es/cgi-bin/cryst/programs/corepresentations\_out.pl?super=194.264\&vecfinal=GM}.}
    %    \begin{tabular}{|c|c|c|}
    %     \hline
    %     \diagbox{$ R $}{$ D^{\text{std}}(R) $}{irrep}&$ \overline{\text{GM}}8 $&$ \overline{\text{GM}}10 $\\
    %     \hline
    %     $ \{C_{3z}|0,0,0\} $&$\begin{pmatrix}
    %      e^{-\frac{\pi i}{3}}&0\\0&e^{\frac{\pi i}{3}}
    %     \end{pmatrix}$&$\begin{pmatrix}
    %      -1&0\\0&-1
    %     \end{pmatrix}$\\
    %     \hline
    %     $ \{C_{2z}|0,0,\frac{1}{2}\} $&$\begin{pmatrix}
    %      -i&0\\0&i
    %     \end{pmatrix}$&$\begin{pmatrix}
    %      -i&0\\0&i
    %     \end{pmatrix}$\\
    %     \hline
    %     $ \{C_{2x}|0,0,0\} $&$\begin{pmatrix}
    %      0&e^{-\frac{2\pi i}{3}}\\e^{-\frac{\pi i}{3}}&0
    %     \end{pmatrix}$&$\begin{pmatrix}
    %      0&1\\-1&0
    %     \end{pmatrix}$\\
    %     \hline
    %                 $ \{P|0,0,0\} $&$\begin{pmatrix}
    %      1&0\\0&1
    %     \end{pmatrix}$&$\begin{pmatrix}
    %      -1&0\\0&-1
    %     \end{pmatrix}$\\
    %                 \hline
    %                 $ \{T|0,0,0\} $&$\begin{pmatrix}
    %      0&1\\-1&0
    %     \end{pmatrix}\mathcal{K}$&$\begin{pmatrix}
    %      0&-1\\1&0
    %     \end{pmatrix}\mathcal{K}$\\
    %                 \hline
    %    \end{tabular}
    %    \label{Na3Bi-gamma-rep}
    %   \end{table}
          
        \begin{table}[h!]
 		\centering
 		\caption{In Te, the matrix representations of ${\rm\overline{H}}4,5$ and $\rm{\overline{H}}6$ irreps at H are given on BCS server, \href{https://www.cryst.ehu.es/cgi-bin/cryst/programs/corepresentations_out.pl?super=152.34&vecfinal=H}{https://www.cryst.ehu.es/cgi-bin/cryst/programs/corepresentations\_out.pl?super=152.34\&vecfinal=H}.}
 		\begin{tabular}{|c|c|c|c|}
 			\hline
 			\diagbox{$ R $}{$ D^{\text{std}}(R) $}{irrep}&$ \overline{\text{H}}4 $&$ \overline{\text{H}}5 $&$\overline{\text{H}}6 $\\
 			\hline
 			$ \{C_{3z}|0,0,\frac{1}{3}\} $&$1$&$1$&$\begin{pmatrix}
 				e^{-\frac{2\pi i}{3}}&0\\0&e^{\frac{2\pi i}{3}}
 			\end{pmatrix}$\\
 			\hline
 			$ \{C_{2x}|0,0,\frac{2}{3}\} $&$i$&$-i$&$\begin{pmatrix}
 				0&e^{\frac{2\pi i}{3}}\\e^{\frac{\pi i}{3}}&0
 			\end{pmatrix}$\\
 			\hline
 		\end{tabular}
 		\label{Te-rep}
 	\end{table}

        \begin{table}[tbh!]
 		\centering
 		\caption{In WZ InAs, the matrix representations of $ \overline{\text{GM}}7 $ and $ \overline{\text{GM}}8$ irreps at $\Gamma$ are given on BCS server, \href{https://www.cryst.ehu.es/cgi-bin/cryst/programs/corepresentations_out.pl?super=186.204&vecfinal=GM}{https://www.cryst.ehu.es/cgi-bin/cryst/programs/corepresentations\_out.pl?super=186.204\&vecfinal=GM}.}
 		\begin{tabular}{|c|c|c|}
 			\hline
 			\diagbox{$ R $}{$ D^{\text{std}}(R) $}{irrep}&$ \overline{\text{GM}}7 $&$\overline{\text{GM}}8 $\\
 			\hline
 			$ \{C_{3z}|0,0,0\} $&$\begin{pmatrix}
 				-1&0\\0&-1
 			\end{pmatrix}$&$\begin{pmatrix}
 				e^{-\frac{\pi i}{3}}&0\\0&e^{\frac{\pi i}{3}}
 			\end{pmatrix}$\\
 			\hline
 			$ \{C_{2z}|0,0,\frac{1}{2}\} $&$\begin{pmatrix}
 				-i&0\\0&i
 			\end{pmatrix}$&$\begin{pmatrix}
 				-i&0\\0&i
 			\end{pmatrix}$\\
 			\hline
                $ \{M_{x}|0,0,0\} $&$\begin{pmatrix}
 				0&1\\-1&0
 			\end{pmatrix}$&$\begin{pmatrix}
 				0&e^{-\frac{2\pi i}{3}}\\e^{-\frac{\pi i}{3}}&0
 			\end{pmatrix}$\\
                \hline
                $ \{T|0,0,0\} $&$\begin{pmatrix}
 				0&1\\-1&0
 			\end{pmatrix}\mathcal{K}$&$\begin{pmatrix}
 				0&1\\-1&0
 			\end{pmatrix}\mathcal{K}$\\
 			\hline
 		\end{tabular}
 		\label{InAs-rep}
 	\end{table}
  
    \begin{table}[tbh!]
 		\centering
 		\caption{In 1H-TMD monolayers, the matrix representations of $ \overline{\text{K}}8 $ and $ \overline{\text{K}}11$ irreps at K are given on BCS server, \href{https://www.cryst.ehu.es/cgi-bin/cryst/programs/corepresentations_out.pl?super=187.210&vecfinal=K}{https://www.cryst.ehu.es/cgi-bin/cryst/programs/corepresentations\_out.pl?super=187.210\&vecfinal=K}.}
 		\begin{tabular}{|c|c|c|}
 			\hline
 			\diagbox{$ R $}{$ D^{\text{std}}(R) $}{irrep}&$ \overline{\text{K}}8 $&$\overline{\text{K}}11 $\\
 			\hline
 			$ \{C_{3z}|0,0,0\} $&$-1$&$e^{\frac{\pi i}{3}}$\\
 			\hline
 			$ \{M_{z}|0,0,0\} $&$i$&$-i$\\
 			\hline
            $ \{M_{x}T|0,0,0\} $&$-1\cdot\mathcal{K}$&$e^{-\frac{\pi i}{3}}\cdot \mathcal{K}$\\
 			\hline
 		\end{tabular}
 		\label{TMD-rep}
 	\end{table}
  %       \begin{table}[H]
 	% 	\centering
 	% 	\caption{In Cd$_3$As$_2$, the matrix representations of $\overline{\text{GM}}8$ and $\overline{\text{GM}}9$ irreps at $\Gamma$ are given on BCS server, \href{https://www.cryst.ehu.es/cgi-bin/cryst/programs/corepresentations_out.pl?super=142.562&vecfinal=GM}{https://www.cryst.ehu.es/cgi-bin/cryst/programs/corepresentations\_out.pl?super=142.562\&vecfinal=GM}.}
 	% 	\begin{tabular}{|c|c|c|}
 	% 		\hline
 	% 		\diagbox{$ R $}{$ D^{\text{std}}(R) $}{irrep}&$ \overline{\text{GM}}8 $&$ \overline{\text{GM}}9 $\\
 	% 		\hline
 	% 		$ \{C_{2z}|0,\frac{1}{2},0\} $&$\begin{pmatrix}
 	% 			-i&0\\0&i
 	% 		\end{pmatrix}$&$\begin{pmatrix}
 	% 			-i&0\\0&i
 	% 		\end{pmatrix}$\\
 	% 		\hline
 	% 		$ \{C_{4z}|\frac{1}{4},\frac{3}{4},\frac{1}{4}\} $&$\begin{pmatrix}
 	% 			e^{\frac{3\pi i}{4}}&0\\0&e^{-\frac{3\pi i}{4}}
 	% 		\end{pmatrix}$&$\begin{pmatrix}
 	% 			e^{-\frac{\pi i}{4}}&0\\0&e^{\frac{\pi i}{4}}
 	% 		\end{pmatrix}$\\
 	% 		\hline
 	% 		$ \{C_{2y}|\frac{1}{2},0,0\} $&$\begin{pmatrix}
 	% 			0&e^{-\frac{\pi i}{4}}\\e^{-\frac{3\pi i}{4}}&0
 	% 		\end{pmatrix}$&$\begin{pmatrix}
 	% 			0&e^{\frac{3\pi i}{4}}\\e^{\frac{\pi i}{4}}&0
 	% 		\end{pmatrix}$\\
 	% 		\hline
 	% 		$ \{P|0,0,0\} $&$\begin{pmatrix}
 	% 			-1&0\\0&-1
 	% 		\end{pmatrix}$&$\begin{pmatrix}
 	% 			-1&0\\0&-1
 	% 		\end{pmatrix}$\\
 	% 		\hline
 	% 		$ \{T|0,0,0\} $&$\begin{pmatrix}
 	% 			0&-1\\1&0
 	% 		\end{pmatrix}\mathcal{K}$&$\begin{pmatrix}
 	% 			0&-1\\1&0
 	% 		\end{pmatrix}\mathcal{K}$\\
 	% 		\hline
 	% 	\end{tabular}
 	% 	\label{Cd3As2-rep}
 	% \end{table}
  \clearpage
         \setcounter{table}{0}
	\renewcommand{\thetable}{F\arabic{table}}
    \renewcommand{\theHtable}{F\arabic{table}}
        \setcounter{equation}{0}
	\renewcommand{\theequation}{F.\arabic{equation}}
    \renewcommand{\theHequation}{F.\arabic{equation}}
        \section{Six-band model at \texorpdfstring{$\Gamma$}{} in wurtzite InAs }
        \label{InAssixband}
        At $ \Gamma $ in the Brillouin zone of InAs, we consider six valence bands for the hole doping sample. The band representations are $  \overline{\text{GM}}8 $, $  \overline{\text{GM}}8 $ and $ \overline{\text{GM}}7 $ in the ascending order. The representation matrices of generators are presented in TABLE \ref{InAs-rep}. The second order $ k\cdot p $ Hamiltonian of six valence bands of InAs at $ \Gamma $ is expressed by
        \begin{equation}
            \left\{\begin{aligned}
                H^{kp}_{11}=&H^{kp}_{22}=a_1+\frac{4\sqrt{3}}{3}a_4+d_1k_+k_-+d_2k_z^2\\
                &d_1=c_1+\frac{4\sqrt{3}}{3}c_5,d_2=c_{10}+\frac{4\sqrt{3}}{3}c_{13}k_z^2\\
                H^{kp}_{12}=&\xi_+(2b_1+4b_2-3b_3-2b_4)k_-\\
                H^{kp}_{13}=&H^{kp}_{24}=2a_3-2ib_6k_z+d_5k_+k_-+c_{12}k_z^2\\
                &d_5=2c_3+\frac{2\sqrt{3}}{3}c_4\\
                H^{kp}_{14}=&2\xi_+b_2k_-+\frac{2\sqrt{3}}{3}i\xi_+c_8k_-k_z\\
                H^{kp}_{15}=&H^{kp*}_{26}=\left(-2c_3-\frac{2\sqrt{3}}{3}c_4+2c_6\right)\xi_-k_-^2\\
                H^{kp}_{16}=&2\xi_-(b_2-b_4)k_+-\frac{\sqrt{3}}{3}i\xi_-(2c_7+2c_8+3c_9)k_+k_z\\
                H^{kp}_{23}=&2\xi_-b_2k_++\frac{2\sqrt{3}}{3}i\xi_-c_8k_+k_z\\
                H^{kp}_{25}=&2\xi_+(b_2-b_4)k_--\frac{\sqrt{3}}{3}i\xi_+(2c_7+2c_8+3c_9)k_-k_z\\
                H^{kp}_{33}=&H^{kp}_{44}=a_1+2a_2-\frac{2\sqrt{3}}{3}a_4+d_3^{+}k_+k_-+d_4^+k_z^2\\
                H^{kp}_{34}=&-2\xi_+(b_2-b_4-b_5)k_-\\
                H^{kp}_{35}=&H^{kp*}_{46}=-\frac{2\sqrt{3}}{3}\xi_-c_4k_-^2\\
                H^{kp}_{36}=&-(4 b_2 - 3 b_3 - 2 b_4)\xi_-k_+-\frac{2\sqrt{3}}{3}i\xi_-c_7k_+k_z\\
                H^{kp}_{45}=&-(4 b_2 - 3 b_3 - 2 b_4)\xi_+k_--\frac{2\sqrt{3}}{3}i\xi_+c_7k_-k_z\\
                H^{kp}_{55}=&H^{kp}_{66}=a_1-2a_2-\frac{2\sqrt{3}}{3}a_4+d_3^-k_+k_-+d_4^-k_z^2\\
                &d_3^\pm=c_1\pm2c_2-\frac{2\sqrt{3}}{3}c_5,\\
                &d_4^\pm=c_{10}\pm2c_{11}-\frac{2\sqrt{3}}{3}c_{13}\\
                H^{kp}_{56}=&0\\
                &\text{with} \quad\xi_\pm=1\pm\sqrt{3}i
            \end{aligned}\right.
        \end{equation}
        
        The Zeeman's coupling of six valence bands of InAs at $ \Gamma $ can be expressed by
        \begin{equation}
            \begin{aligned}
                H^Z=&\frac{\mu_B}{2}\begin{pmatrix}
                    h_1B_z&h_3B_+&2g_7B_z&h_4B_-&0&h_5B_+\\
                    &-h_1B_z&h_4^*B_+&-2g_7B_z&h_5^*B_-&0\\
                    &&3g_9B_z&h_6B_-&0&h_7B_+\\
                    &&&-3g_9B_z&h_7^*B_-&0\\
                    &\dagger&&&h_2B_z&0\\&&&&&-h_2B_z
                \end{pmatrix}\\
                &h_1=2h_6+\frac{2\sqrt{3}}{3}g_8+3g_9\\
                &h_2=\frac{4\sqrt{3}}{3}g_8 + 3 g_9\\
                &h_3=-\frac{\sqrt{3}}{3}i\xi_+(2 g_1 + 4 g_2 - 3 g_3 - 2 g_4)\\
                &h_4=-\frac{2\sqrt{3}}{3}i\xi_+g_2\\
                &h_5=\frac{2\sqrt{3}}{3}i\xi_-(g_2-g_4)\\
                &h_6=\frac{2\sqrt{3}}{3}i\xi_+(g_2-g_4-g_5)\\
                &h_7=-\frac{\sqrt{3}}{3}i\xi_-(4g_2-3g_3-2g_4)\\
                &\text{with} \quad\xi_\pm=1\pm\sqrt{3}i
            \end{aligned}
        \end{equation}
        The values of the parameters $ \{a_i, b_i, c_i, g_i\} $ are presented in TABLE \ref{InAs-kp-par}.

        \begin{table}[H]
		\centering
            \caption{The computed values of parameters $ \{a_i, b_i, c_i, g_i\} $ for six valence states in InAs are obtained from the VASP calculations directly.}
		\begin{tabular}{p{1.9cm}p{1.9cm}p{1.9cm}p{1.9cm}}
			\hline\hline
			$a$ (eV)& $b$ (eV$\cdot$\AA) & $c$ (eV$\cdot$\AA$^2$)&$g$ \\
			\hline
			$a_1=3.91$&$b_1=0.13$&$c_1=-45.78$&$g_1=-3.28$\\
            $a_2=-0.09$&$b_2=0.08$&$c_2=12.94$&$g_2=-2.44$\\
            $a_3=0.02$&$b_3=0.11$&$c_3=-1.54$&$g_3=-3.16$\\
            $a_4=0.03$&$b_4=-0.09$&$c_4=-20.14$&$g_4=5.50$\\
            &$b_5=0.06$&$c_5=7.65$&$g_5=-2.76$\\
            &$b_6=0.21$&$c_6=-2.23$&$g_6=-11.20$\\
            &&$c_7=21.62$&$g_7=-6.90$\\
            &&$c_8=32.86$&$g_8=19.82$\\
            &&$c_9=-60.12$&$g_9=-3.88$\\
            &&$c_{10}=-45.68$&\\ &&$c_{11}=-7.45$&\\ &&$c_{12}=28.03$&\\ &&$c_{13}=-23.43$&\\
			\hline\hline
		\end{tabular}
		\label{InAs-kp-par}
	\end{table}

	\noindent

\begin{thebibliography}{60}
	\expandafter\ifx\csname natexlab\endcsname\relax\def\natexlab#1{#1}\fi
	\expandafter\ifx\csname bibnamefont\endcsname\relax
	  \def\bibnamefont#1{#1}\fi
	\expandafter\ifx\csname bibfnamefont\endcsname\relax
	  \def\bibfnamefont#1{#1}\fi
	\expandafter\ifx\csname citenamefont\endcsname\relax
	  \def\citenamefont#1{#1}\fi
	\expandafter\ifx\csname url\endcsname\relax
	  \def\url#1{\texttt{#1}}\fi
	\expandafter\ifx\csname urlprefix\endcsname\relax\def\urlprefix{URL }\fi
	\providecommand{\bibinfo}[2]{#2}
	\providecommand{\eprint}[2][]{\url{#2}}
	
	\bibitem[{\citenamefont{Hohenberg and Kohn}(1964)}]{PhysRev.136.B864}
	\bibinfo{author}{\bibfnamefont{P.}~\bibnamefont{Hohenberg}} \bibnamefont{and} \bibinfo{author}{\bibfnamefont{W.}~\bibnamefont{Kohn}}, \bibinfo{journal}{Phys. Rev.} \textbf{\bibinfo{volume}{136}}, \bibinfo{pages}{B864} (\bibinfo{year}{1964}), \urlprefix\url{https://link.aps.org/doi/10.1103/PhysRev.136.B864}.
	
	\bibitem[{\citenamefont{Kohn and Sham}(1965)}]{PhysRev.140.A1133}
	\bibinfo{author}{\bibfnamefont{W.}~\bibnamefont{Kohn}} \bibnamefont{and} \bibinfo{author}{\bibfnamefont{L.~J.} \bibnamefont{Sham}}, \bibinfo{journal}{Phys. Rev.} \textbf{\bibinfo{volume}{140}}, \bibinfo{pages}{A1133} (\bibinfo{year}{1965}), \urlprefix\url{https://link.aps.org/doi/10.1103/PhysRev.140.A1133}.
	
	\bibitem[{\citenamefont{Kresse and Furthm\"uller}(1996)}]{VASP1}
	\bibinfo{author}{\bibfnamefont{G.}~\bibnamefont{Kresse}} \bibnamefont{and} \bibinfo{author}{\bibfnamefont{J.}~\bibnamefont{Furthm\"uller}}, \bibinfo{journal}{Phys. Rev. B} \textbf{\bibinfo{volume}{54}}, \bibinfo{pages}{11169} (\bibinfo{year}{1996}), \urlprefix\url{https://link.aps.org/doi/10.1103/PhysRevB.54.11169}.
	
	\bibitem[{\citenamefont{Kresse and Furthmüller}(1996)}]{VASP2}
	\bibinfo{author}{\bibfnamefont{G.}~\bibnamefont{Kresse}} \bibnamefont{and} \bibinfo{author}{\bibfnamefont{J.}~\bibnamefont{Furthmüller}}, \bibinfo{journal}{Computational Materials Science} \textbf{\bibinfo{volume}{6}}, \bibinfo{pages}{15} (\bibinfo{year}{1996}), ISSN \bibinfo{issn}{0927-0256}, \urlprefix\url{https://www.sciencedirect.com/science/article/pii/0927025696000080}.
	
	\bibitem[{\citenamefont{Giannozzi et~al.}(2009)\citenamefont{Giannozzi, Baroni, Bonini, Calandra, Car, Cavazzoni, Ceresoli, Chiarotti, Cococcioni, Dabo et~al.}}]{Giannozzi_2009}
	\bibinfo{author}{\bibfnamefont{P.}~\bibnamefont{Giannozzi}}, \bibinfo{author}{\bibfnamefont{S.}~\bibnamefont{Baroni}}, \bibinfo{author}{\bibfnamefont{N.}~\bibnamefont{Bonini}}, \bibinfo{author}{\bibfnamefont{M.}~\bibnamefont{Calandra}}, \bibinfo{author}{\bibfnamefont{R.}~\bibnamefont{Car}}, \bibinfo{author}{\bibfnamefont{C.}~\bibnamefont{Cavazzoni}}, \bibinfo{author}{\bibfnamefont{D.}~\bibnamefont{Ceresoli}}, \bibinfo{author}{\bibfnamefont{G.~L.} \bibnamefont{Chiarotti}}, \bibinfo{author}{\bibfnamefont{M.}~\bibnamefont{Cococcioni}}, \bibinfo{author}{\bibfnamefont{I.}~\bibnamefont{Dabo}}, \bibnamefont{et~al.}, \bibinfo{journal}{Journal of Physics: Condensed Matter} \textbf{\bibinfo{volume}{21}}, \bibinfo{pages}{395502} (\bibinfo{year}{2009}), \urlprefix\url{https://dx.doi.org/10.1088/0953-8984/21/39/395502}.
	
	\bibitem[{\citenamefont{Giannozzi et~al.}(2017)\citenamefont{Giannozzi, Andreussi, Brumme, Bunau, Nardelli, Calandra, Car, Cavazzoni, Ceresoli, Cococcioni et~al.}}]{Giannozzi_2017}
	\bibinfo{author}{\bibfnamefont{P.}~\bibnamefont{Giannozzi}}, \bibinfo{author}{\bibfnamefont{O.}~\bibnamefont{Andreussi}}, \bibinfo{author}{\bibfnamefont{T.}~\bibnamefont{Brumme}}, \bibinfo{author}{\bibfnamefont{O.}~\bibnamefont{Bunau}}, \bibinfo{author}{\bibfnamefont{M.~B.} \bibnamefont{Nardelli}}, \bibinfo{author}{\bibfnamefont{M.}~\bibnamefont{Calandra}}, \bibinfo{author}{\bibfnamefont{R.}~\bibnamefont{Car}}, \bibinfo{author}{\bibfnamefont{C.}~\bibnamefont{Cavazzoni}}, \bibinfo{author}{\bibfnamefont{D.}~\bibnamefont{Ceresoli}}, \bibinfo{author}{\bibfnamefont{M.}~\bibnamefont{Cococcioni}}, \bibnamefont{et~al.}, \bibinfo{journal}{Journal of Physics: Condensed Matter} \textbf{\bibinfo{volume}{29}}, \bibinfo{pages}{465901} (\bibinfo{year}{2017}), \urlprefix\url{https://dx.doi.org/10.1088/1361-648X/aa8f79}.
	
	\bibitem[{\citenamefont{Clark et~al.}(2005)\citenamefont{Clark, Segall, Pickard, Hasnip, Probert, Refson, and Payne}}]{Clark}
	\bibinfo{author}{\bibfnamefont{S.~J.} \bibnamefont{Clark}}, \bibinfo{author}{\bibfnamefont{M.~D.} \bibnamefont{Segall}}, \bibinfo{author}{\bibfnamefont{C.~J.} \bibnamefont{Pickard}}, \bibinfo{author}{\bibfnamefont{P.~J.} \bibnamefont{Hasnip}}, \bibinfo{author}{\bibfnamefont{M.~I.~J.} \bibnamefont{Probert}}, \bibinfo{author}{\bibfnamefont{K.}~\bibnamefont{Refson}}, \bibnamefont{and} \bibinfo{author}{\bibfnamefont{M.~C.} \bibnamefont{Payne}}, \bibinfo{journal}{Zeitschrift für Kristallographie - Crystalline Materials} \textbf{\bibinfo{volume}{220}}, \bibinfo{pages}{567} (\bibinfo{year}{2005}), \urlprefix\url{https://doi.org/10.1524/zkri.220.5.567.65075}.
	
	\bibitem[{\citenamefont{Gonze et~al.}(2020)\citenamefont{Gonze, Amadon, Antonius, Arnardi, Baguet, Beuken, Bieder, Bottin, Bouchet, Bousquet et~al.}}]{Gonze2020}
	\bibinfo{author}{\bibfnamefont{X.}~\bibnamefont{Gonze}}, \bibinfo{author}{\bibfnamefont{B.}~\bibnamefont{Amadon}}, \bibinfo{author}{\bibfnamefont{G.}~\bibnamefont{Antonius}}, \bibinfo{author}{\bibfnamefont{F.}~\bibnamefont{Arnardi}}, \bibinfo{author}{\bibfnamefont{L.}~\bibnamefont{Baguet}}, \bibinfo{author}{\bibfnamefont{J.-M.} \bibnamefont{Beuken}}, \bibinfo{author}{\bibfnamefont{J.}~\bibnamefont{Bieder}}, \bibinfo{author}{\bibfnamefont{F.}~\bibnamefont{Bottin}}, \bibinfo{author}{\bibfnamefont{J.}~\bibnamefont{Bouchet}}, \bibinfo{author}{\bibfnamefont{E.}~\bibnamefont{Bousquet}}, \bibnamefont{et~al.}, \bibinfo{journal}{Comput. Phys. Commun.} \textbf{\bibinfo{volume}{248}}, \bibinfo{pages}{107042} (\bibinfo{year}{2020}), \urlprefix\url{https://doi.org/10.1016/j.cpc.2019.107042}.
	
	\bibitem[{\citenamefont{Romero et~al.}(2020)\citenamefont{Romero, Allan, Amadon, Antonius, Applencourt, Baguet, Bieder, Bottin, Bouchet, Bousquet et~al.}}]{Romero2020}
	\bibinfo{author}{\bibfnamefont{A.~H.} \bibnamefont{Romero}}, \bibinfo{author}{\bibfnamefont{D.~C.} \bibnamefont{Allan}}, \bibinfo{author}{\bibfnamefont{B.}~\bibnamefont{Amadon}}, \bibinfo{author}{\bibfnamefont{G.}~\bibnamefont{Antonius}}, \bibinfo{author}{\bibfnamefont{T.}~\bibnamefont{Applencourt}}, \bibinfo{author}{\bibfnamefont{L.}~\bibnamefont{Baguet}}, \bibinfo{author}{\bibfnamefont{J.}~\bibnamefont{Bieder}}, \bibinfo{author}{\bibfnamefont{F.}~\bibnamefont{Bottin}}, \bibinfo{author}{\bibfnamefont{J.}~\bibnamefont{Bouchet}}, \bibinfo{author}{\bibfnamefont{E.}~\bibnamefont{Bousquet}}, \bibnamefont{et~al.}, \bibinfo{journal}{J. Chem. Phys.} \textbf{\bibinfo{volume}{152}}, \bibinfo{pages}{124102} (\bibinfo{year}{2020}).
	
	\bibitem[{\citenamefont{Soler et~al.}(2002)\citenamefont{Soler, Artacho, Gale, García, Junquera, Ordejón, and Sánchez-Portal}}]{Jose_M_Soler_2002}
	\bibinfo{author}{\bibfnamefont{J.~M.} \bibnamefont{Soler}}, \bibinfo{author}{\bibfnamefont{E.}~\bibnamefont{Artacho}}, \bibinfo{author}{\bibfnamefont{J.~D.} \bibnamefont{Gale}}, \bibinfo{author}{\bibfnamefont{A.}~\bibnamefont{García}}, \bibinfo{author}{\bibfnamefont{J.}~\bibnamefont{Junquera}}, \bibinfo{author}{\bibfnamefont{P.}~\bibnamefont{Ordejón}}, \bibnamefont{and} \bibinfo{author}{\bibfnamefont{D.}~\bibnamefont{Sánchez-Portal}}, \bibinfo{journal}{Journal of Physics: Condensed Matter} \textbf{\bibinfo{volume}{14}}, \bibinfo{pages}{2745} (\bibinfo{year}{2002}), \urlprefix\url{https://dx.doi.org/10.1088/0953-8984/14/11/302}.
	
	\bibitem[{\citenamefont{García et~al.}(2020)\citenamefont{García, Papior, Akhtar, Artacho, Blum, Bosoni, Brandimarte, Brandbyge, Cerdá, Corsetti et~al.}}]{10.1063/5.0005077}
	\bibinfo{author}{\bibfnamefont{A.}~\bibnamefont{García}}, \bibinfo{author}{\bibfnamefont{N.}~\bibnamefont{Papior}}, \bibinfo{author}{\bibfnamefont{A.}~\bibnamefont{Akhtar}}, \bibinfo{author}{\bibfnamefont{E.}~\bibnamefont{Artacho}}, \bibinfo{author}{\bibfnamefont{V.}~\bibnamefont{Blum}}, \bibinfo{author}{\bibfnamefont{E.}~\bibnamefont{Bosoni}}, \bibinfo{author}{\bibfnamefont{P.}~\bibnamefont{Brandimarte}}, \bibinfo{author}{\bibfnamefont{M.}~\bibnamefont{Brandbyge}}, \bibinfo{author}{\bibfnamefont{J.~I.} \bibnamefont{Cerdá}}, \bibinfo{author}{\bibfnamefont{F.}~\bibnamefont{Corsetti}}, \bibnamefont{et~al.}, \bibinfo{journal}{The Journal of Chemical Physics} \textbf{\bibinfo{volume}{152}}, \bibinfo{pages}{204108} (\bibinfo{year}{2020}), ISSN \bibinfo{issn}{0021-9606}, \urlprefix\url{https://doi.org/10.1063/5.0005077}.
	
	\bibitem[{\citenamefont{Blaha et~al.}(2020)\citenamefont{Blaha, Schwarz, Tran, Laskowski, Madsen, and Marks}}]{10.1063/1.5143061}
	\bibinfo{author}{\bibfnamefont{P.}~\bibnamefont{Blaha}}, \bibinfo{author}{\bibfnamefont{K.}~\bibnamefont{Schwarz}}, \bibinfo{author}{\bibfnamefont{F.}~\bibnamefont{Tran}}, \bibinfo{author}{\bibfnamefont{R.}~\bibnamefont{Laskowski}}, \bibinfo{author}{\bibfnamefont{G.~K.~H.} \bibnamefont{Madsen}}, \bibnamefont{and} \bibinfo{author}{\bibfnamefont{L.~D.} \bibnamefont{Marks}}, \bibinfo{journal}{The Journal of Chemical Physics} \textbf{\bibinfo{volume}{152}}, \bibinfo{pages}{074101} (\bibinfo{year}{2020}), ISSN \bibinfo{issn}{0021-9606}, \urlprefix\url{https://doi.org/10.1063/1.5143061}.
	
	\bibitem[{\citenamefont{Hourahine et~al.}(2020)\citenamefont{Hourahine, Aradi, Blum, Bonafé, Buccheri, Camacho, Cevallos, Deshaye, Dumitrică, Dominguez et~al.}}]{10.1063/1.5143190}
	\bibinfo{author}{\bibfnamefont{B.}~\bibnamefont{Hourahine}}, \bibinfo{author}{\bibfnamefont{B.}~\bibnamefont{Aradi}}, \bibinfo{author}{\bibfnamefont{V.}~\bibnamefont{Blum}}, \bibinfo{author}{\bibfnamefont{F.}~\bibnamefont{Bonafé}}, \bibinfo{author}{\bibfnamefont{A.}~\bibnamefont{Buccheri}}, \bibinfo{author}{\bibfnamefont{C.}~\bibnamefont{Camacho}}, \bibinfo{author}{\bibfnamefont{C.}~\bibnamefont{Cevallos}}, \bibinfo{author}{\bibfnamefont{M.~Y.} \bibnamefont{Deshaye}}, \bibinfo{author}{\bibfnamefont{T.}~\bibnamefont{Dumitrică}}, \bibinfo{author}{\bibfnamefont{A.}~\bibnamefont{Dominguez}}, \bibnamefont{et~al.}, \bibinfo{journal}{The Journal of Chemical Physics} \textbf{\bibinfo{volume}{152}}, \bibinfo{pages}{124101} (\bibinfo{year}{2020}), ISSN \bibinfo{issn}{0021-9606}, \urlprefix\url{https://doi.org/10.1063/1.5143190}.
	
	\bibitem[{\citenamefont{Luque et~al.}(2015)\citenamefont{Luque, Panchak, Mellor, Vlasov, Martí, and Andreev}}]{LUQUE201582}
	\bibinfo{author}{\bibfnamefont{A.}~\bibnamefont{Luque}}, \bibinfo{author}{\bibfnamefont{A.}~\bibnamefont{Panchak}}, \bibinfo{author}{\bibfnamefont{A.}~\bibnamefont{Mellor}}, \bibinfo{author}{\bibfnamefont{A.}~\bibnamefont{Vlasov}}, \bibinfo{author}{\bibfnamefont{A.}~\bibnamefont{Martí}}, \bibnamefont{and} \bibinfo{author}{\bibfnamefont{V.}~\bibnamefont{Andreev}}, \bibinfo{journal}{Physica B: Condensed Matter} \textbf{\bibinfo{volume}{456}}, \bibinfo{pages}{82} (\bibinfo{year}{2015}), ISSN \bibinfo{issn}{0921-4526}, \urlprefix\url{https://www.sciencedirect.com/science/article/pii/S0921452614006929}.
	
	\bibitem[{\citenamefont{Gresch et~al.}(2017)\citenamefont{Gresch, Wu, Winkler, and Soluyanov}}]{Gresch_2017}
	\bibinfo{author}{\bibfnamefont{D.}~\bibnamefont{Gresch}}, \bibinfo{author}{\bibfnamefont{Q.}~\bibnamefont{Wu}}, \bibinfo{author}{\bibfnamefont{G.~W.} \bibnamefont{Winkler}}, \bibnamefont{and} \bibinfo{author}{\bibfnamefont{A.~A.} \bibnamefont{Soluyanov}}, \bibinfo{journal}{New Journal of Physics} \textbf{\bibinfo{volume}{19}}, \bibinfo{pages}{035001} (\bibinfo{year}{2017}), \urlprefix\url{https://dx.doi.org/10.1088/1367-2630/aa5de7}.
	
	\bibitem[{\citenamefont{Luttinger and Kohn}(1955)}]{PhysRev.97.869}
	\bibinfo{author}{\bibfnamefont{J.~M.} \bibnamefont{Luttinger}} \bibnamefont{and} \bibinfo{author}{\bibfnamefont{W.}~\bibnamefont{Kohn}}, \bibinfo{journal}{Phys. Rev.} \textbf{\bibinfo{volume}{97}}, \bibinfo{pages}{869} (\bibinfo{year}{1955}), \urlprefix\url{https://link.aps.org/doi/10.1103/PhysRev.97.869}.
	
	\bibitem[{\citenamefont{Marquardt et~al.}(2015)\citenamefont{Marquardt, Geelhaar, and Brandt}}]{doi:10.1021/acs.nanolett.5b00101}
	\bibinfo{author}{\bibfnamefont{O.}~\bibnamefont{Marquardt}}, \bibinfo{author}{\bibfnamefont{L.}~\bibnamefont{Geelhaar}}, \bibnamefont{and} \bibinfo{author}{\bibfnamefont{O.}~\bibnamefont{Brandt}}, \bibinfo{journal}{Nano Letters} \textbf{\bibinfo{volume}{15}}, \bibinfo{pages}{4289} (\bibinfo{year}{2015}), \bibinfo{note}{pMID: 26042638}, \urlprefix\url{https://doi.org/10.1021/acs.nanolett.5b00101}.
	
	\bibitem[{\citenamefont{Kane}(1957)}]{KANE1957249}
	\bibinfo{author}{\bibfnamefont{E.~O.} \bibnamefont{Kane}}, \bibinfo{journal}{Journal of Physics and Chemistry of Solids} \textbf{\bibinfo{volume}{1}}, \bibinfo{pages}{249} (\bibinfo{year}{1957}), ISSN \bibinfo{issn}{0022-3697}, \urlprefix\url{https://www.sciencedirect.com/science/article/pii/0022369757900136}.
	
	\bibitem[{\citenamefont{Zhang and Zhang}(2013)}]{HZhang}
	\bibinfo{author}{\bibfnamefont{H.}~\bibnamefont{Zhang}} \bibnamefont{and} \bibinfo{author}{\bibfnamefont{S.-C.} \bibnamefont{Zhang}}, \bibinfo{journal}{physica status solidi (RRL) – Rapid Research Letters} \textbf{\bibinfo{volume}{7}}, \bibinfo{pages}{72} (\bibinfo{year}{2013}), \urlprefix\url{https://onlinelibrary.wiley.com/doi/abs/10.1002/pssr.201206414}.
	
	\bibitem[{\citenamefont{Zhang et~al.}(2009)\citenamefont{Zhang, Liu, Qi, Dai, Fang, and Zhang}}]{zhang2009topological}
	\bibinfo{author}{\bibfnamefont{H.}~\bibnamefont{Zhang}}, \bibinfo{author}{\bibfnamefont{C.-X.} \bibnamefont{Liu}}, \bibinfo{author}{\bibfnamefont{X.-L.} \bibnamefont{Qi}}, \bibinfo{author}{\bibfnamefont{X.}~\bibnamefont{Dai}}, \bibinfo{author}{\bibfnamefont{Z.}~\bibnamefont{Fang}}, \bibnamefont{and} \bibinfo{author}{\bibfnamefont{S.-C.} \bibnamefont{Zhang}}, \bibinfo{journal}{Nature Physics} \textbf{\bibinfo{volume}{5}}, \bibinfo{pages}{438} (\bibinfo{year}{2009}).
	
	\bibitem[{\citenamefont{Fu}(2009)}]{PhysRevLett.103.266801}
	\bibinfo{author}{\bibfnamefont{L.}~\bibnamefont{Fu}}, \bibinfo{journal}{Phys. Rev. Lett.} \textbf{\bibinfo{volume}{103}}, \bibinfo{pages}{266801} (\bibinfo{year}{2009}), \urlprefix\url{https://link.aps.org/doi/10.1103/PhysRevLett.103.266801}.
	
	\bibitem[{\citenamefont{Xu et~al.}(2011)\citenamefont{Xu, Weng, Wang, Dai, and Fang}}]{PhysRevLett.107.186806}
	\bibinfo{author}{\bibfnamefont{G.}~\bibnamefont{Xu}}, \bibinfo{author}{\bibfnamefont{H.}~\bibnamefont{Weng}}, \bibinfo{author}{\bibfnamefont{Z.}~\bibnamefont{Wang}}, \bibinfo{author}{\bibfnamefont{X.}~\bibnamefont{Dai}}, \bibnamefont{and} \bibinfo{author}{\bibfnamefont{Z.}~\bibnamefont{Fang}}, \bibinfo{journal}{Phys. Rev. Lett.} \textbf{\bibinfo{volume}{107}}, \bibinfo{pages}{186806} (\bibinfo{year}{2011}), \urlprefix\url{https://link.aps.org/doi/10.1103/PhysRevLett.107.186806}.
	
	\bibitem[{\citenamefont{Faria~Junior et~al.}(2015)\citenamefont{Faria~Junior, Xu, Lee, Gerhardt, Sipahi, and \ifmmode \check{Z}\else \v{Z}\fi{}uti\ifmmode~\acute{c}\else \'{c}\fi{}}}]{PhysRevB.92.075311}
	\bibinfo{author}{\bibfnamefont{P.~E.} \bibnamefont{Faria~Junior}}, \bibinfo{author}{\bibfnamefont{G.}~\bibnamefont{Xu}}, \bibinfo{author}{\bibfnamefont{J.}~\bibnamefont{Lee}}, \bibinfo{author}{\bibfnamefont{N.~C.} \bibnamefont{Gerhardt}}, \bibinfo{author}{\bibfnamefont{G.~M.} \bibnamefont{Sipahi}}, \bibnamefont{and} \bibinfo{author}{\bibfnamefont{I.}~\bibnamefont{\ifmmode \check{Z}\else \v{Z}\fi{}uti\ifmmode~\acute{c}\else \'{c}\fi{}}}, \bibinfo{journal}{Phys. Rev. B} \textbf{\bibinfo{volume}{92}}, \bibinfo{pages}{075311} (\bibinfo{year}{2015}), \urlprefix\url{https://link.aps.org/doi/10.1103/PhysRevB.92.075311}.
	
	\bibitem[{\citenamefont{Holub and Jonker}(2011)}]{PhysRevB.83.125309}
	\bibinfo{author}{\bibfnamefont{M.}~\bibnamefont{Holub}} \bibnamefont{and} \bibinfo{author}{\bibfnamefont{B.~T.} \bibnamefont{Jonker}}, \bibinfo{journal}{Phys. Rev. B} \textbf{\bibinfo{volume}{83}}, \bibinfo{pages}{125309} (\bibinfo{year}{2011}), \urlprefix\url{https://link.aps.org/doi/10.1103/PhysRevB.83.125309}.
	
	\bibitem[{\citenamefont{Marquardt}(2021)}]{MARQUARDT2021110318}
	\bibinfo{author}{\bibfnamefont{O.}~\bibnamefont{Marquardt}}, \bibinfo{journal}{Computational Materials Science} \textbf{\bibinfo{volume}{194}}, \bibinfo{pages}{110318} (\bibinfo{year}{2021}), ISSN \bibinfo{issn}{0927-0256}, \urlprefix\url{https://www.sciencedirect.com/science/article/pii/S0927025621000434}.
	
	\bibitem[{\citenamefont{Faria~Junior et~al.}(2019)\citenamefont{Faria~Junior, Kurpas, Gmitra, and Fabian}}]{PhysRevB.100.115203}
	\bibinfo{author}{\bibfnamefont{P.~E.} \bibnamefont{Faria~Junior}}, \bibinfo{author}{\bibfnamefont{M.}~\bibnamefont{Kurpas}}, \bibinfo{author}{\bibfnamefont{M.}~\bibnamefont{Gmitra}}, \bibnamefont{and} \bibinfo{author}{\bibfnamefont{J.}~\bibnamefont{Fabian}}, \bibinfo{journal}{Phys. Rev. B} \textbf{\bibinfo{volume}{100}}, \bibinfo{pages}{115203} (\bibinfo{year}{2019}), \urlprefix\url{https://link.aps.org/doi/10.1103/PhysRevB.100.115203}.
	
	\bibitem[{\citenamefont{Kormányos et~al.}(2015)\citenamefont{Kormányos, Burkard, Gmitra, Fabian, Zólyomi, Drummond, and Fal’ko}}]{Kormanyos_2015}
	\bibinfo{author}{\bibfnamefont{A.}~\bibnamefont{Kormányos}}, \bibinfo{author}{\bibfnamefont{G.}~\bibnamefont{Burkard}}, \bibinfo{author}{\bibfnamefont{M.}~\bibnamefont{Gmitra}}, \bibinfo{author}{\bibfnamefont{J.}~\bibnamefont{Fabian}}, \bibinfo{author}{\bibfnamefont{V.}~\bibnamefont{Zólyomi}}, \bibinfo{author}{\bibfnamefont{N.~D.} \bibnamefont{Drummond}}, \bibnamefont{and} \bibinfo{author}{\bibfnamefont{V.}~\bibnamefont{Fal’ko}}, \bibinfo{journal}{2D Materials} \textbf{\bibinfo{volume}{2}}, \bibinfo{pages}{049501} (\bibinfo{year}{2015}), \urlprefix\url{https://dx.doi.org/10.1088/2053-1583/2/4/049501}.
	
	\bibitem[{\citenamefont{Deilmann et~al.}(2020)\citenamefont{Deilmann, Kr\"uger, and Rohlfing}}]{PhysRevLett.124.226402}
	\bibinfo{author}{\bibfnamefont{T.}~\bibnamefont{Deilmann}}, \bibinfo{author}{\bibfnamefont{P.}~\bibnamefont{Kr\"uger}}, \bibnamefont{and} \bibinfo{author}{\bibfnamefont{M.}~\bibnamefont{Rohlfing}}, \bibinfo{journal}{Phys. Rev. Lett.} \textbf{\bibinfo{volume}{124}}, \bibinfo{pages}{226402} (\bibinfo{year}{2020}), \urlprefix\url{https://link.aps.org/doi/10.1103/PhysRevLett.124.226402}.
	
	\bibitem[{\citenamefont{Faria~Junior and Sipahi}(2012)}]{Junior}
	\bibinfo{author}{\bibfnamefont{P.}~\bibnamefont{Faria~Junior}} \bibnamefont{and} \bibinfo{author}{\bibfnamefont{G.}~\bibnamefont{Sipahi}}, \bibinfo{journal}{Journal of Applied Physics} \textbf{\bibinfo{volume}{112}} (\bibinfo{year}{2012}).
	
	\bibitem[{\citenamefont{Xuan and Quek}(2020)}]{PhysRevResearch.2.033256}
	\bibinfo{author}{\bibfnamefont{F.}~\bibnamefont{Xuan}} \bibnamefont{and} \bibinfo{author}{\bibfnamefont{S.~Y.} \bibnamefont{Quek}}, \bibinfo{journal}{Phys. Rev. Res.} \textbf{\bibinfo{volume}{2}}, \bibinfo{pages}{033256} (\bibinfo{year}{2020}), \urlprefix\url{https://link.aps.org/doi/10.1103/PhysRevResearch.2.033256}.
	
	\bibitem[{\citenamefont{Climente et~al.}(2016)\citenamefont{Climente, Segarra, Rajadell, and Planelles}}]{10.1063/1.4945112}
	\bibinfo{author}{\bibfnamefont{J.~I.} \bibnamefont{Climente}}, \bibinfo{author}{\bibfnamefont{C.}~\bibnamefont{Segarra}}, \bibinfo{author}{\bibfnamefont{F.}~\bibnamefont{Rajadell}}, \bibnamefont{and} \bibinfo{author}{\bibfnamefont{J.}~\bibnamefont{Planelles}}, \bibinfo{journal}{Journal of Applied Physics} \textbf{\bibinfo{volume}{119}}, \bibinfo{pages}{125705} (\bibinfo{year}{2016}), ISSN \bibinfo{issn}{0021-8979}, \urlprefix\url{https://doi.org/10.1063/1.4945112}.
	
	\bibitem[{\citenamefont{Lucignano et~al.}(2007)\citenamefont{Lucignano, Giuliano, and Tagliacozzo}}]{PhysRevB.76.045324}
	\bibinfo{author}{\bibfnamefont{P.}~\bibnamefont{Lucignano}}, \bibinfo{author}{\bibfnamefont{D.}~\bibnamefont{Giuliano}}, \bibnamefont{and} \bibinfo{author}{\bibfnamefont{A.}~\bibnamefont{Tagliacozzo}}, \bibinfo{journal}{Phys. Rev. B} \textbf{\bibinfo{volume}{76}}, \bibinfo{pages}{045324} (\bibinfo{year}{2007}), \urlprefix\url{https://link.aps.org/doi/10.1103/PhysRevB.76.045324}.
	
	\bibitem[{\citenamefont{Zamani et~al.}(2017)\citenamefont{Zamani, Setareh, Azargoshasb, and Niknam}}]{ZAMANI2017243}
	\bibinfo{author}{\bibfnamefont{A.}~\bibnamefont{Zamani}}, \bibinfo{author}{\bibfnamefont{F.}~\bibnamefont{Setareh}}, \bibinfo{author}{\bibfnamefont{T.}~\bibnamefont{Azargoshasb}}, \bibnamefont{and} \bibinfo{author}{\bibfnamefont{E.}~\bibnamefont{Niknam}}, \bibinfo{journal}{Superlattices and Microstructures} \textbf{\bibinfo{volume}{110}}, \bibinfo{pages}{243} (\bibinfo{year}{2017}), ISSN \bibinfo{issn}{0749-6036}, \urlprefix\url{https://www.sciencedirect.com/science/article/pii/S0749603617314672}.
	
	\bibitem[{\citenamefont{León-González et~al.}(2023)\citenamefont{León-González, Toscano-Negrette, Morales, Vinasco, Yücel, Sari, Kasapoglu, Sakiroglu, Mora-Ramos, Restrepo et~al.}}]{nano13091461}
	\bibinfo{author}{\bibfnamefont{J.~C.} \bibnamefont{León-González}}, \bibinfo{author}{\bibfnamefont{R.~G.} \bibnamefont{Toscano-Negrette}}, \bibinfo{author}{\bibfnamefont{A.~L.} \bibnamefont{Morales}}, \bibinfo{author}{\bibfnamefont{J.~A.} \bibnamefont{Vinasco}}, \bibinfo{author}{\bibfnamefont{M.~B.} \bibnamefont{Yücel}}, \bibinfo{author}{\bibfnamefont{H.}~\bibnamefont{Sari}}, \bibinfo{author}{\bibfnamefont{E.}~\bibnamefont{Kasapoglu}}, \bibinfo{author}{\bibfnamefont{S.}~\bibnamefont{Sakiroglu}}, \bibinfo{author}{\bibfnamefont{M.~E.} \bibnamefont{Mora-Ramos}}, \bibinfo{author}{\bibfnamefont{R.~L.} \bibnamefont{Restrepo}}, \bibnamefont{et~al.}, \bibinfo{journal}{Nanomaterials} \textbf{\bibinfo{volume}{13}} (\bibinfo{year}{2023}), ISSN \bibinfo{issn}{2079-4991}, \urlprefix\url{https://www.mdpi.com/2079-4991/13/9/1461}.
	
	\bibitem[{\citenamefont{Zamani and Rezaei}(2018)}]{ZAMANI2018145}
	\bibinfo{author}{\bibfnamefont{A.}~\bibnamefont{Zamani}} \bibnamefont{and} \bibinfo{author}{\bibfnamefont{G.}~\bibnamefont{Rezaei}}, \bibinfo{journal}{Superlattices and Microstructures} \textbf{\bibinfo{volume}{124}}, \bibinfo{pages}{145} (\bibinfo{year}{2018}), ISSN \bibinfo{issn}{0749-6036}, \urlprefix\url{https://www.sciencedirect.com/science/article/pii/S0749603618316781}.
	
	\bibitem[{\citenamefont{Pryor and Flatt\'e}(2006)}]{PhysRevLett.96.026804}
	\bibinfo{author}{\bibfnamefont{C.~E.} \bibnamefont{Pryor}} \bibnamefont{and} \bibinfo{author}{\bibfnamefont{M.~E.} \bibnamefont{Flatt\'e}}, \bibinfo{journal}{Phys. Rev. Lett.} \textbf{\bibinfo{volume}{96}}, \bibinfo{pages}{026804} (\bibinfo{year}{2006}), \urlprefix\url{https://link.aps.org/doi/10.1103/PhysRevLett.96.026804}.
	
	\bibitem[{\citenamefont{Gharaati}(2017)}]{GHARAATI201717}
	\bibinfo{author}{\bibfnamefont{A.}~\bibnamefont{Gharaati}}, \bibinfo{journal}{Solid State Communications} \textbf{\bibinfo{volume}{258}}, \bibinfo{pages}{17} (\bibinfo{year}{2017}), ISSN \bibinfo{issn}{0038-1098}, \urlprefix\url{https://www.sciencedirect.com/science/article/pii/S0038109817301229}.
	
	\bibitem[{\citenamefont{Kotlyar et~al.}(2001)\citenamefont{Kotlyar, Reinecke, Bayer, and Forchel}}]{PhysRevB.63.085310}
	\bibinfo{author}{\bibfnamefont{R.}~\bibnamefont{Kotlyar}}, \bibinfo{author}{\bibfnamefont{T.~L.} \bibnamefont{Reinecke}}, \bibinfo{author}{\bibfnamefont{M.}~\bibnamefont{Bayer}}, \bibnamefont{and} \bibinfo{author}{\bibfnamefont{A.}~\bibnamefont{Forchel}}, \bibinfo{journal}{Phys. Rev. B} \textbf{\bibinfo{volume}{63}}, \bibinfo{pages}{085310} (\bibinfo{year}{2001}), \urlprefix\url{https://link.aps.org/doi/10.1103/PhysRevB.63.085310}.
	
	\bibitem[{\citenamefont{Kiselev et~al.}(1998)\citenamefont{Kiselev, Ivchenko, and R\"ossler}}]{PhysRevB.58.16353}
	\bibinfo{author}{\bibfnamefont{A.~A.} \bibnamefont{Kiselev}}, \bibinfo{author}{\bibfnamefont{E.~L.} \bibnamefont{Ivchenko}}, \bibnamefont{and} \bibinfo{author}{\bibfnamefont{U.}~\bibnamefont{R\"ossler}}, \bibinfo{journal}{Phys. Rev. B} \textbf{\bibinfo{volume}{58}}, \bibinfo{pages}{16353} (\bibinfo{year}{1998}), \urlprefix\url{https://link.aps.org/doi/10.1103/PhysRevB.58.16353}.
	
	\bibitem[{\citenamefont{Winkler et~al.}(2017)\citenamefont{Winkler, Varjas, Skolasinski, Soluyanov, Troyer, and Wimmer}}]{PhysRevLett.119.037701}
	\bibinfo{author}{\bibfnamefont{G.~W.} \bibnamefont{Winkler}}, \bibinfo{author}{\bibfnamefont{D.}~\bibnamefont{Varjas}}, \bibinfo{author}{\bibfnamefont{R.}~\bibnamefont{Skolasinski}}, \bibinfo{author}{\bibfnamefont{A.~A.} \bibnamefont{Soluyanov}}, \bibinfo{author}{\bibfnamefont{M.}~\bibnamefont{Troyer}}, \bibnamefont{and} \bibinfo{author}{\bibfnamefont{M.}~\bibnamefont{Wimmer}}, \bibinfo{journal}{Phys. Rev. Lett.} \textbf{\bibinfo{volume}{119}}, \bibinfo{pages}{037701} (\bibinfo{year}{2017}), \urlprefix\url{https://link.aps.org/doi/10.1103/PhysRevLett.119.037701}.
	
	\bibitem[{\citenamefont{Toloza~Sandoval et~al.}(2012)\citenamefont{Toloza~Sandoval, Ferreira~da Silva, de~Andrada~e Silva, and La~Rocca}}]{PhysRevB.86.195302}
	\bibinfo{author}{\bibfnamefont{M.~A.} \bibnamefont{Toloza~Sandoval}}, \bibinfo{author}{\bibfnamefont{A.}~\bibnamefont{Ferreira~da Silva}}, \bibinfo{author}{\bibfnamefont{E.~A.} \bibnamefont{de~Andrada~e Silva}}, \bibnamefont{and} \bibinfo{author}{\bibfnamefont{G.~C.} \bibnamefont{La~Rocca}}, \bibinfo{journal}{Phys. Rev. B} \textbf{\bibinfo{volume}{86}}, \bibinfo{pages}{195302} (\bibinfo{year}{2012}), \urlprefix\url{https://link.aps.org/doi/10.1103/PhysRevB.86.195302}.
	
	\bibitem[{\citenamefont{Alegre et~al.}(2006)\citenamefont{Alegre, Hern\'andez, Pereira, and Medeiros-Ribeiro}}]{PhysRevLett.97.236402}
	\bibinfo{author}{\bibfnamefont{T.~P.~M.} \bibnamefont{Alegre}}, \bibinfo{author}{\bibfnamefont{F.~G.~G.} \bibnamefont{Hern\'andez}}, \bibinfo{author}{\bibfnamefont{A.~L.~C.} \bibnamefont{Pereira}}, \bibnamefont{and} \bibinfo{author}{\bibfnamefont{G.}~\bibnamefont{Medeiros-Ribeiro}}, \bibinfo{journal}{Phys. Rev. Lett.} \textbf{\bibinfo{volume}{97}}, \bibinfo{pages}{236402} (\bibinfo{year}{2006}), \urlprefix\url{https://link.aps.org/doi/10.1103/PhysRevLett.97.236402}.
	
	\bibitem[{\citenamefont{Wang et~al.}(2015)\citenamefont{Wang, Yan, Zhang, Liao, Wu, and Yu}}]{C5NR05250E}
	\bibinfo{author}{\bibfnamefont{L.-X.} \bibnamefont{Wang}}, \bibinfo{author}{\bibfnamefont{Y.}~\bibnamefont{Yan}}, \bibinfo{author}{\bibfnamefont{L.}~\bibnamefont{Zhang}}, \bibinfo{author}{\bibfnamefont{Z.-M.} \bibnamefont{Liao}}, \bibinfo{author}{\bibfnamefont{H.-C.} \bibnamefont{Wu}}, \bibnamefont{and} \bibinfo{author}{\bibfnamefont{D.-P.} \bibnamefont{Yu}}, \bibinfo{journal}{Nanoscale} \textbf{\bibinfo{volume}{7}}, \bibinfo{pages}{16687} (\bibinfo{year}{2015}), \urlprefix\url{http://dx.doi.org/10.1039/C5NR05250E}.
	
	\bibitem[{\citenamefont{Liu et~al.}(2021)\citenamefont{Liu, Entin-Wohlman, Aharony, You, and Xu}}]{PhysRevB.104.085302}
	\bibinfo{author}{\bibfnamefont{Z.-H.} \bibnamefont{Liu}}, \bibinfo{author}{\bibfnamefont{O.}~\bibnamefont{Entin-Wohlman}}, \bibinfo{author}{\bibfnamefont{A.}~\bibnamefont{Aharony}}, \bibinfo{author}{\bibfnamefont{J.~Q.} \bibnamefont{You}}, \bibnamefont{and} \bibinfo{author}{\bibfnamefont{H.~Q.} \bibnamefont{Xu}}, \bibinfo{journal}{Phys. Rev. B} \textbf{\bibinfo{volume}{104}}, \bibinfo{pages}{085302} (\bibinfo{year}{2021}), \urlprefix\url{https://link.aps.org/doi/10.1103/PhysRevB.104.085302}.
	
	\bibitem[{\citenamefont{Xin and Reid}(2002)}]{10.1063/1.1423328}
	\bibinfo{author}{\bibfnamefont{J.}~\bibnamefont{Xin}} \bibnamefont{and} \bibinfo{author}{\bibfnamefont{S.~A.} \bibnamefont{Reid}}, \bibinfo{journal}{The Journal of Chemical Physics} \textbf{\bibinfo{volume}{116}}, \bibinfo{pages}{525} (\bibinfo{year}{2002}), ISSN \bibinfo{issn}{0021-9606}, \urlprefix\url{https://doi.org/10.1063/1.1423328}.
	
	\bibitem[{\citenamefont{Semenov et~al.}(2016)\citenamefont{Semenov, Yurchenko, and Tennyson}}]{SEMENOV201657}
	\bibinfo{author}{\bibfnamefont{M.}~\bibnamefont{Semenov}}, \bibinfo{author}{\bibfnamefont{S.~N.} \bibnamefont{Yurchenko}}, \bibnamefont{and} \bibinfo{author}{\bibfnamefont{J.}~\bibnamefont{Tennyson}}, \bibinfo{journal}{Journal of Molecular Spectroscopy} \textbf{\bibinfo{volume}{330}}, \bibinfo{pages}{57} (\bibinfo{year}{2016}), ISSN \bibinfo{issn}{0022-2852}, \bibinfo{note}{potentiology and Spectroscopy in Honor of Robert Le Roy}, \urlprefix\url{https://www.sciencedirect.com/science/article/pii/S0022285216303186}.
	
	\bibitem[{\citenamefont{Fischer and Jönsson}(2001)}]{FISCHER200155}
	\bibinfo{author}{\bibfnamefont{C.}~\bibnamefont{Fischer}} \bibnamefont{and} \bibinfo{author}{\bibfnamefont{P.}~\bibnamefont{Jönsson}}, \bibinfo{journal}{Journal of Molecular Structure: THEOCHEM} \textbf{\bibinfo{volume}{537}}, \bibinfo{pages}{55} (\bibinfo{year}{2001}), ISSN \bibinfo{issn}{0166-1280}, \urlprefix\url{https://www.sciencedirect.com/science/article/pii/S0166128000004607}.
	
	\bibitem[{\citenamefont{Gao et~al.}(2021)\citenamefont{Gao, Wu, Persson, and Wang}}]{GAO2021107760}
	\bibinfo{author}{\bibfnamefont{J.}~\bibnamefont{Gao}}, \bibinfo{author}{\bibfnamefont{Q.}~\bibnamefont{Wu}}, \bibinfo{author}{\bibfnamefont{C.}~\bibnamefont{Persson}}, \bibnamefont{and} \bibinfo{author}{\bibfnamefont{Z.}~\bibnamefont{Wang}}, \bibinfo{journal}{Computer Physics Communications} \textbf{\bibinfo{volume}{261}}, \bibinfo{pages}{107760} (\bibinfo{year}{2021}), ISSN \bibinfo{issn}{0010-4655}, \urlprefix\url{https://www.sciencedirect.com/science/article/pii/S0010465520303805}.
	
	\bibitem[{\citenamefont{Jiang et~al.}(2021)\citenamefont{Jiang, Fang, and Fang}}]{Jiang_2021}
	\bibinfo{author}{\bibfnamefont{Y.}~\bibnamefont{Jiang}}, \bibinfo{author}{\bibfnamefont{Z.}~\bibnamefont{Fang}}, \bibnamefont{and} \bibinfo{author}{\bibfnamefont{C.}~\bibnamefont{Fang}}, \bibinfo{journal}{Chinese Physics Letters} \textbf{\bibinfo{volume}{38}}, \bibinfo{pages}{077104} (\bibinfo{year}{2021}), \urlprefix\url{https://dx.doi.org/10.1088/0256-307X/38/7/077104}.
	
	\bibitem[{\citenamefont{Song et~al.}(2021)\citenamefont{Song, Sun, Xu, Nie, Weng, Fang, and Dai}}]{song-gfactor}
	\bibinfo{author}{\bibfnamefont{Z.}~\bibnamefont{Song}}, \bibinfo{author}{\bibfnamefont{S.}~\bibnamefont{Sun}}, \bibinfo{author}{\bibfnamefont{Y.}~\bibnamefont{Xu}}, \bibinfo{author}{\bibfnamefont{S.}~\bibnamefont{Nie}}, \bibinfo{author}{\bibfnamefont{H.}~\bibnamefont{Weng}}, \bibinfo{author}{\bibfnamefont{Z.}~\bibnamefont{Fang}}, \bibnamefont{and} \bibinfo{author}{\bibfnamefont{X.}~\bibnamefont{Dai}}, \emph{\bibinfo{title}{First Principle Calculation of the Effective Zeeman’s Couplings in Topological Materials}} (\bibinfo{publisher}{World Scientific}, \bibinfo{year}{2021}), chap. \bibinfo{chapter}{Chapter 11}, pp. \bibinfo{pages}{263--281}, \urlprefix\url{https://www.worldscientific.com/doi/abs/10.1142/9789811231711_0013}.
	
	\bibitem[{\citenamefont{Zhang et~al.}(2023{\natexlab{a}})\citenamefont{Zhang, Deng, Sun, Fang, Guo, and Wang}}]{RHZhang}
	\bibinfo{author}{\bibfnamefont{R.}~\bibnamefont{Zhang}}, \bibinfo{author}{\bibfnamefont{J.}~\bibnamefont{Deng}}, \bibinfo{author}{\bibfnamefont{Y.}~\bibnamefont{Sun}}, \bibinfo{author}{\bibfnamefont{Z.}~\bibnamefont{Fang}}, \bibinfo{author}{\bibfnamefont{Z.}~\bibnamefont{Guo}}, \bibnamefont{and} \bibinfo{author}{\bibfnamefont{Z.}~\bibnamefont{Wang}}, \bibinfo{journal}{Phys. Rev. Res.} \textbf{\bibinfo{volume}{5}}, \bibinfo{pages}{023142} (\bibinfo{year}{2023}{\natexlab{a}}), \urlprefix\url{https://link.aps.org/doi/10.1103/PhysRevResearch.5.023142}. The IR2PW code is available at \url{https://github.com/zjwang11/IR2PW}.
	
	\bibitem[{\citenamefont{Iraola et~al.}(2022)\citenamefont{Iraola, Mañes, Bradlyn, Horton, Neupert, Vergniory, and Tsirkin}}]{IeRep}
	\bibinfo{author}{\bibfnamefont{M.}~\bibnamefont{Iraola}}, \bibinfo{author}{\bibfnamefont{J.~L.} \bibnamefont{Mañes}}, \bibinfo{author}{\bibfnamefont{B.}~\bibnamefont{Bradlyn}}, \bibinfo{author}{\bibfnamefont{M.~K.} \bibnamefont{Horton}}, \bibinfo{author}{\bibfnamefont{T.}~\bibnamefont{Neupert}}, \bibinfo{author}{\bibfnamefont{M.~G.} \bibnamefont{Vergniory}}, \bibnamefont{and} \bibinfo{author}{\bibfnamefont{S.~S.} \bibnamefont{Tsirkin}}, \bibinfo{journal}{Computer Physics Communications} \textbf{\bibinfo{volume}{272}}, \bibinfo{pages}{108226} (\bibinfo{year}{2022}), ISSN \bibinfo{issn}{0010-4655}, \urlprefix\url{https://www.sciencedirect.com/science/article/pii/S0010465521003386}.
	
	\bibitem[{\citenamefont{Cassiano et~al.}(2023)\citenamefont{Cassiano, Ara{\'u}jo, Junior, and Ferreira}}]{cassiano2023dft2kp}
	\bibinfo{author}{\bibfnamefont{J.~V.~V.} \bibnamefont{Cassiano}}, \bibinfo{author}{\bibfnamefont{A.~L.} \bibnamefont{Ara{\'u}jo}}, \bibinfo{author}{\bibfnamefont{P.~E.~F.} \bibnamefont{Junior}}, \bibnamefont{and} \bibinfo{author}{\bibfnamefont{G.~J.} \bibnamefont{Ferreira}}, \bibinfo{journal}{arXiv preprint arXiv:2306.08554}  (\bibinfo{year}{2023}).
	
	\bibitem[{\citenamefont{Bl\"ochl}(1994)}]{PhysRevB.50.17953}
	\bibinfo{author}{\bibfnamefont{P.~E.} \bibnamefont{Bl\"ochl}}, \bibinfo{journal}{Phys. Rev. B} \textbf{\bibinfo{volume}{50}}, \bibinfo{pages}{17953} (\bibinfo{year}{1994}), \urlprefix\url{https://link.aps.org/doi/10.1103/PhysRevB.50.17953}.
	
	\bibitem[{\citenamefont{Bl{\"o}chl et~al.}(2003)\citenamefont{Bl{\"o}chl, F{\"o}rst, and Schimpl}}]{blochl2003projector}
	\bibinfo{author}{\bibfnamefont{P.~E.} \bibnamefont{Bl{\"o}chl}}, \bibinfo{author}{\bibfnamefont{C.~J.} \bibnamefont{F{\"o}rst}}, \bibnamefont{and} \bibinfo{author}{\bibfnamefont{J.}~\bibnamefont{Schimpl}}, \bibinfo{journal}{Bulletin of Materials Science} \textbf{\bibinfo{volume}{26}}, \bibinfo{pages}{33} (\bibinfo{year}{2003}).
	
	\bibitem[{\citenamefont{Albrecht et~al.}(2016)\citenamefont{Albrecht, Higginbotham, Madsen, Kuemmeth, Jespersen, Nygård, Krogstrup, and Marcus}}]{InAs-nature17162}
	\bibinfo{author}{\bibfnamefont{S.~M.} \bibnamefont{Albrecht}}, \bibinfo{author}{\bibfnamefont{A.~P.} \bibnamefont{Higginbotham}}, \bibinfo{author}{\bibfnamefont{M.}~\bibnamefont{Madsen}}, \bibinfo{author}{\bibfnamefont{F.}~\bibnamefont{Kuemmeth}}, \bibinfo{author}{\bibfnamefont{T.~S.} \bibnamefont{Jespersen}}, \bibinfo{author}{\bibfnamefont{J.}~\bibnamefont{Nygård}}, \bibinfo{author}{\bibfnamefont{P.}~\bibnamefont{Krogstrup}}, \bibnamefont{and} \bibinfo{author}{\bibfnamefont{C.~M.} \bibnamefont{Marcus}}, \bibinfo{journal}{Nature} \textbf{\bibinfo{volume}{531}}, \bibinfo{pages}{206–209} (\bibinfo{year}{2016}), \urlprefix\url{https://doi.org/10.1038/s41467-020-18521-6}.
	
	\bibitem[{\citenamefont{Köhler and Wöchner}(1975)}]{https://doi.org/10.1002/pssb.2220670229}
	\bibinfo{author}{\bibfnamefont{H.}~\bibnamefont{Köhler}} \bibnamefont{and} \bibinfo{author}{\bibfnamefont{E.}~\bibnamefont{Wöchner}}, \bibinfo{journal}{physica status solidi (b)} \textbf{\bibinfo{volume}{67}}, \bibinfo{pages}{665} (\bibinfo{year}{1975}), \urlprefix\url{https://onlinelibrary.wiley.com/doi/abs/10.1002/pssb.2220670229}.
	
	\bibitem[{\citenamefont{Bj\"ork et~al.}(2005)\citenamefont{Bj\"ork, Fuhrer, Hansen, Larsson, Fr\"oberg, and Samuelson}}]{WZ-InAs-exp}
	\bibinfo{author}{\bibfnamefont{M.~T.} \bibnamefont{Bj\"ork}}, \bibinfo{author}{\bibfnamefont{A.}~\bibnamefont{Fuhrer}}, \bibinfo{author}{\bibfnamefont{A.~E.} \bibnamefont{Hansen}}, \bibinfo{author}{\bibfnamefont{M.~W.} \bibnamefont{Larsson}}, \bibinfo{author}{\bibfnamefont{L.~E.} \bibnamefont{Fr\"oberg}}, \bibnamefont{and} \bibinfo{author}{\bibfnamefont{L.}~\bibnamefont{Samuelson}}, \bibinfo{journal}{Phys. Rev. B} \textbf{\bibinfo{volume}{72}}, \bibinfo{pages}{201307} (\bibinfo{year}{2005}), \urlprefix\url{https://link.aps.org/doi/10.1103/PhysRevB.72.201307}.
	
	\bibitem[{\citenamefont{F{\"o}rste et~al.}(2020)\citenamefont{F{\"o}rste, Tepliakov, Kruchinin, Lindlau, Funk, F{\"o}rg, Watanabe, Taniguchi, Baimuratov, and H{\"o}gele}}]{WSe2-exp}
	\bibinfo{author}{\bibfnamefont{J.}~\bibnamefont{F{\"o}rste}}, \bibinfo{author}{\bibfnamefont{N.~V.} \bibnamefont{Tepliakov}}, \bibinfo{author}{\bibfnamefont{S.~Y.} \bibnamefont{Kruchinin}}, \bibinfo{author}{\bibfnamefont{J.}~\bibnamefont{Lindlau}}, \bibinfo{author}{\bibfnamefont{V.}~\bibnamefont{Funk}}, \bibinfo{author}{\bibfnamefont{M.}~\bibnamefont{F{\"o}rg}}, \bibinfo{author}{\bibfnamefont{K.}~\bibnamefont{Watanabe}}, \bibinfo{author}{\bibfnamefont{T.}~\bibnamefont{Taniguchi}}, \bibinfo{author}{\bibfnamefont{A.~S.} \bibnamefont{Baimuratov}}, \bibnamefont{and} \bibinfo{author}{\bibfnamefont{A.}~\bibnamefont{H{\"o}gele}}, \bibinfo{journal}{Nature Communications} \textbf{\bibinfo{volume}{11}}, \bibinfo{pages}{4539} (\bibinfo{year}{2020}), \urlprefix\url{https://doi.org/10.1038/s41467-020-18019-1}.
	
	\bibitem[{\citenamefont{Zhang et~al.}(2023{\natexlab{b}})\citenamefont{Zhang, Sheng, Song, Liang, and Wang}}]{vasp2kp.com}
	\bibinfo{author}{\bibfnamefont{S.}~\bibnamefont{Zhang}}, \bibinfo{author}{\bibfnamefont{H.}~\bibnamefont{Sheng}}, \bibinfo{author}{\bibfnamefont{Z.-D.} \bibnamefont{Song}}, \bibinfo{author}{\bibfnamefont{C.}~\bibnamefont{Liang}}, \bibnamefont{and} \bibinfo{author}{\bibfnamefont{Z.}~\bibnamefont{Wang}}, \emph{\bibinfo{title}{VASP2KP}}, \bibinfo{howpublished}{\url{http://www.vasp2kp.com}} (\bibinfo{year}{2023}{\natexlab{b}}).
	
	\end{thebibliography}
\end{document}